\documentclass[5p,authoryear]{elsarticle}
\usepackage[colorlinks]{hyperref}
\usepackage[normalem]{ulem}
\usepackage{xcolor}
\usepackage{natbib,graphicx,aas_macros,amsmath,amssymb,xspace,multirow}

\newcommand{\Gaia}{\textit{Gaia}\xspace}
\newcommand{\vlos}{$v_\text{los}$\xspace}
\newcommand{\kms}{km\,s$^{-1}$\xspace}
\newcommand{\kmskpc}{km\,s$^{-1}$\hspace{0.5mm}kpc$^{-1}$\xspace}
\newcommand{\masyr}{mas\,yr$^{-1}$\xspace}
\newcommand{\zvz}{$z$--$v_z$\xspace}
\begin{document}

\begin{frontmatter}
\title{Milky Way dynamics in light of \textit{Gaia}}
\author{Jason A. S. Hunt}
\ead{j.a.hunt@surrey.ac.uk}
\author{Eugene Vasiliev}
\ead{eugvas@protonmail.com}
\affiliation{organization={University of Surrey}, city={Guildford}, postcode={GU2 7XH}, country={UK}}

\date{December 2024}
\journal{New Astronomy Reviews}

\begin{abstract}
The \Gaia mission has triggered major developments in the field of Galactic dynamics in recent years, which we discuss in this review. The structure and kinematics of all Galactic components -- disc, bar/bulge and halo -- are now mapped in great detail not only in the Solar neighbourhood, but across a large part of the Milky Way. The dramatic improvements in the coverage and precision of observations revealed various disequilibrium processes, such as perturbations in the Galactic disc and the deformations of the outer halo, which are partly attributed to the interaction with satellite galaxies. The knowledge of the gravitational potential at all scales has also advanced considerably, but we are still far from having a consistent view on the key properties of the Galaxy, such as the bar pattern speed or the mass profile and shape of the dark halo. The complexity and interplay of several dynamical processes makes the interpretation of observational data challenging, and it is fair to say that more theoretical effort is needed to fully reap the fruit of the \Gaia revolution.
\end{abstract}
\end{frontmatter}

\section{Introduction}  \label{sec:introduction}

One of the key science objectives of the \Gaia mission, as summarised by \citet{Gaia2016a}, is to advance the research of the structure, kinematics and dynamics of our Galaxy.
Indeed, this field has seen enormous progress in less than a decade since the comprehensive review by \citet{BlandHawthorn2016}, which appeared just before the first data release of \Gaia. While the global features of our Galaxy have stood the test of time, our understanding of the dynamical mechanisms relevant for its evolution has deepened significantly since then. A major theme emerging in recent years is the discovery and characterisation of various disequilibrium effects in the Milky Way system, made possible thanks to the vastly increased quantity and precision of observational measurements. Since this topic was virtually non-existent in the pre-\Gaia era, we devote a substantial fraction of our review to it, but we also discuss more traditional aspects of Galaxy dynamics that have also seen significant new developments.
We do not aim to present a full compendium of Galactic dynamics research, but focus on studies that appeared in the last few years, significantly rely on \Gaia data, and explore the structure and dynamics of the entire Milky Way or its major components.
Our discussion has some overlap with other reviews in this series covering the Galactic assembly history and evolution \citep{ArchaeologyReview} and stellar streams \citep{StreamsReview}, and can be complemented by reviews of dynamical modelling and mass measurement by \citet{Binney2013}, \citet{Wang2020a}, \citet{Gardner2021}, and \citet{deSalas2021}.

We start in Section~\ref{sec:observations} by introducing the currently available observational inventory: \Gaia and complementary major photometric and spectroscopic surveys; methods for measuring the distances, velocities and other stellar properties; and basic definitions for dynamical analysis.
The rest of the review splits the discussion both by structural components and by dynamical mechanisms. Section~\ref{sec:disc} focuses on the equilibrium structure and kinematics of the Galactic disc, radial and vertical gradients of velocity distributions, variations with ages and between different populations.
Section~\ref{sec:nonaxi_response} discusses major planar non-axisymmetric features that have been known from well before \Gaia, namely the Galactic bar and spiral arms, whose dynamical effects can be studied both `locally' (in the Solar neighbourhood) as well as `remotely' (using observational measurements in the corresponding spatial regions).
In Section~\ref{sec:perturbations}, we consider the vertical perturbations of the disc: the Galactic warp, waves and corrugations, and the major new area of research in recent years -- the \Gaia phase spiral.
We then turn to the structure and kinematics of the stellar halo in Section~\ref{sec:halo}, minimising the overlap in scope with the aforementioned reviews of Galactic archaeology and stellar streams. Section~\ref{sec:dynamics} discusses the mass modelling techniques and current constraints on the Milky Way mass distribution and shape. Finally, we summarise the key discoveries enabled by \Gaia and list the open problems that can be addressed in the near future in Section~\ref{sec:conclusions}.

\section{Observational inventory}  \label{sec:observations}

\begin{table*}
\caption{Large-scale photometric surveys of the Milky Way. Shown are the photometric bands and corresponding wavelength ranges in nm, approximate limiting magnitude (not directly comparable between different bands), sky coverage, and sizes of published catalogues.}  \label{tab:photometric_surveys}
\newcommand{\GaiaBands}{$G, G_\text{BP}, G_\text{RP}$}
\newcommand{\TwoMASS}{\hspace{-2mm}\raisebox{1mm}{\begin{tabular}{l}{Whipple (USA) 1.3m\!\!\!\!\!}\\[-2mm] {CTIO (Chile) 1.3m}\end{tabular}}}
\renewcommand{\arraystretch}{1.5}
\begin{tabular}{llp{32mm}crrrr}
survey   & date       & telescope         & bands     & wavelength & depth      & coverage & num.stars\\
\hline\vspace*{1mm}
Gaia     & 2014--now  & space (L2) 1.2m   & \GaiaBands & 330--1050 & $G\!<$21   & all sky & 1.8bn \\
SDSS     & 2000--2009 & Sloan (USA) 2.5m  & $u,g,r,i,z$& 320--1000 & $g\!<$22.2 & 1/3 sky & 0.47bn \\
PanSTARRS& 2010--2014 & PS1 (Hawaii) 1.8m & $g,r,i,z,y$& 400--1030 & $g\!<$23.3 & 3/4 sky & 3bn \\
Legacy   & 2013--2019 & Blanco (Chile) 4m & $g,r,z$    & 400--1000 & $g\!<$24.0 & 1/3 sky & 2bn \\
DECaPS   & 2016--2019 & Blanco (Chile) 4m & $g,r,i,z,Y$& 400--1030 & $g\!<$23.7 & 2\,700\ deg$^2$\!\! & 3.3bn \\
2MASS    & 1997--2001 & \TwoMASS          & $J,H,K_s$  &1100--2300 &$K_s\!<$15.3& all sky & 0.47bn\\
VVV(X)   & 2010--2023 & VISTA (Chile) 4m  & $J,H,K_s$  &1100--2300 &$K_s\!<$17.5& 1\,700\ deg$^2$\!\! & 1.5bn \\
WISE     & 2010--2018 & space (LEO) 0.4m  & $W_1,W_2$  &2800--5200 &$W_1\!<$18  & all sky & 2bn \\
\end{tabular}
\end{table*}

Before discussing the latest discoveries, it is worth reminding what kind of data we have to deal with, and how do we proceed from observational measurements to the determination of stellar phase-space coordinates and characterisation of their orbits.

\subsection{\Gaia}  \label{sec:observations_gaia}

As is well known, \Gaia catalogues contain well over a billion stars brighter than $G \approx 21$, which is still only $\sim 1$\% of the total number of stars in the Milky Way. The first data release (DR1, \citealt{Gaia2016b}) provided positions for all stars, plus a few million parallaxes and proper motions (PM) measurements for bright stars previously observed by the \textit{Hipparcos} satellite (the TGAS catalogue).
Starting from DR2 \citep{Gaia2018a}, the vast majority of stars have measured parallaxes $\varpi$, PM $\boldsymbol \mu$, and colours (BP and RP magnitudes). EDR3 \citep{Gaia2021a} significantly improved the astrometric precision (by a factor of two for PM), while DR3 \citep{Gaia2023a} added many new data products, including ultra-low-resolution BP/RP spectra for $2.2\times10^8$ objects brighter than $G=17.65$ \citep{deAngeli2023}, medium-resolution spectra from the RVS spectrograph for $10^6$ stars brighter than $G\approx 14$ (Seabroke et al., in prep.), as well as derived quantities, such as astrophysical parameters estimated from BP/RP spectra for $4.7\times10^8$ stars \citep{Andrae2023a} and from RVS spectra for $5.6\times10^6$ stars \citep{RecioBlanco2023}, a catalog of $9\times10^6$ variable stars split into 22 classes \citep{Rimoldini2023}, and a lot more. DR2 also provided line-of-sight velocities $v_\text{los}$ for over 7 million stars, and DR3 increased this sample to nearly 34 million \citep{Katz2023}, by far the largest catalogue to date.

\begin{table*}[t]
\caption{Large-scale spectroscopic surveys of the Milky Way. Shown are the wavelength range in nm (not necessarily contiguous), typical spectral resolution, number of fibres available in a single exposure, and the total number of unique stars in published catalogues; values prepended with $\sim$ refer to future surveys, and in brackets -- to those that have not published their data (H3) or published only a fraction of it (ARGOS, S$^5$, DESI).
}  \label{tab:spectroscopic_surveys}
\newcommand{\APOGEE}{\hspace{-2mm}\raisebox{1mm}{\begin{tabular}{l}{Sloan (USA) 2.5m}\\[-2mm] {du Pont (Chile) 2.5m}\end{tabular}}}
\newcommand{\mr}[1]{\multirow{2}{*}{#1}}
\newcommand{\p}{\makebox[0cm][l]{+}}
\renewcommand{\arraystretch}{1.5}
\begin{tabular}{llp{36mm}crrr}
survey    & date       & telescope               &wavelength&spec.res.& fibres & num.stars  \\
\hline\vspace*{1mm}
Gaia RVS  & 2014--now  & space (L2) 1.2m         & 845--872 & 11\,500 & --- & 33\,800\,000 \\
RAVE      & 2003--2013 & UKST (Australia) 1.2m   & 841--879 &  7\,500 & 120 & 518\,000 \\
ARGOS     & 2008--2011 & AAT (Australia) 4m      & 840--885 & 11\,000 & 350 & (28\,000) \\
SDSS/SEGUE& 2005--2009 & Sloan (USA) 2.5m        & 380--920 &  1\,800 & 640 & 510\,000 \\
APOGEE    & 2008--2021 & \APOGEE                 &1500--1700& 22\,500 & 300 & 657\,000 \\
Gaia--ESO & 2011--2018 & VLT (Chile) 8.2m        & 400--900 & 18\,000\p&150 & 114\,000 \\
GALAH     & 2014--now  & AAT (Australia) 4m      & 472--789 & 28\,000 & 390 & 918\,000 \\
\mr{LEGUE}& 2012--now  &\mr{LAMOST (China) 4m}   & 380--900 &  1\,500 &\mr{4000}& 4\,840\,000 \\[-2mm]
          & 2017--now  &                         & 495--680 &  7\,500 &     & 840\,000 \\
H3        & 2017--now  & MMT (USA) 6.5m          & 515--530 & 23\,000 & 240 & (300\,000) \\
S$^5$     & 2018--now  & AAT (Australia) 4m      & 842--882 & 10\,000 & 350 & (100\,000) \\
DESI      & 2021--now  & Mayall (USA) 4m         & 360--982 &  2\,500 & 5000& (7\,200\,000) \\
\hline
SDSS-V MWM& since 2022 & Sloan + du Pont 2.5m    &1500--1700& 22\,500 & 300 & $\sim$5\,000\,000\\
\mr{WEAVE}&\mr{since 2023}&\mr{WHT (Spain) 4m}   & 366--959 &  5\,000 &\mr{1000}& $\sim$3\,000\,000 \\[-2mm]
          &            &                         & 404--685 & 20\,000 &     & $\sim$2\,000\,000 \\
\mr{4MOST}&\mr{after 2025}&\mr{VISTA (Chile) 4m} & 370--950 &  6\,500 & 1600& $\sim$15\,000\,000 \\[-2mm]
          &            &                         & 393--679 & 20\,000 & 800 & $\sim$5\,000\,000 \\
MOONS     & after 2025 & VLT (Chile) 8.2m        & 645--1800&  5\,000\p&1000& $\sim$500\,000 \\
\hline
\end{tabular}
\end{table*}

\subsection{Photometric and astrometric surveys}  \label{sec:observations_photometric_surveys}

\begin{figure}
\centering
\includegraphics{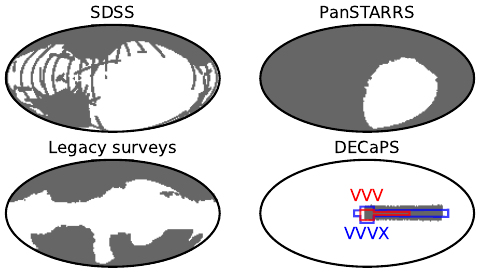}
\caption{Major photometric surveys of the Milky Way. Sky coverage of four optical surveys (SDSS, PanSTARRS, Legacy surveys comprising DECaLS, DES and a few others, and DECaPS) is shown as shaded regions in four panels. Footprints of near-IR surveys of the bulge and disc region (VVV and its extension VVVX) are shown by red and blue polygons in the last panel. Two other IR surveys (2MASS and WISE) have full-sky coverage and are not shown.}  \label{fig:photometric_surveys}
\end{figure}

\Gaia's $G$-band photometry is very precise and well calibrated, but the BP and RP detectors have smaller area and larger effective PSF, thus are prone to systematic errors in crowded areas (see \citealt{Riello2021} for discussion) and have large colour uncertainties for faint stars. It can be complemented by several large-area photometric surveys in the optical and near-IR bands, listed in Table~\ref{tab:photometric_surveys}, whose footprints are shown in Figure~\ref{fig:photometric_surveys}. The most widely used ground-based optical surveys are SDSS (final photometric catalogue published in DR12, \citealt{Alam2015}), PanSTARRS \citep{Chambers2016}, DESI Legacy Imaging Surveys \citep{Dey2019}, which include DECaLS, DES \citep{DES2016}, DELVE \citep{DrlicaWagner2022} and a number of other projects, and DECaPS \citep{Saydjari2023} survey of the inner Galaxy. 2MASS \citep{Skrutskie2006} and WISE \citep{Wright2010,Marocco2021} are two all-sky surveys in near-IR, and the bulge and inner disc region are covered by the VISTA Variables in the V\'\i a L\'actea survey (VVV; \citealt{Minniti2010}) and its extension VVVX \citep{Saito2024}.
These photometric surveys are valuable for determination of stellar parameters and photometric distance estimates (Section~\ref{sec:observations_distance_measurement}). Although the low-resolution BP/RP spectra in \Gaia DR3 can be used to reconstruct magnitudes in any photometric band within their wavelength range \citep{Gaia2023d}, these data are provided only for bright stars ($G<17.65$), whereas most ground-based surveys are significantly deeper, and some cover wavelength ranges inaccessible to \Gaia; in particular, the near-IR photometry from 2MASS and WISE improves the accuracy of extinction and effective temperature estimates \citep[e.g.,][]{Andrae2023b}. In addition, star counts from near-IR photometric surveys provide a more unbiased information about the structural properties of the high-extinction central region of the Galaxy \citep[e.g.,][]{McWilliam2010, Saito2011, Wegg2013, Ness2016b}.

Some of these surveys are multi-epoch and have sufficient astrometric precision for a meaningful determination of PM, either using the survey data alone, or in conjunction with \Gaia. For instance, the 10--15 year baseline between SDSS and \Gaia DR1 allowed \citet{Deason2017} and \citet{Belokurov2018} to study the halo kinematics before \Gaia DR2 became available (Section~\ref{sec:halo_kinematics}), while \citet{Altmann2017} and \citet{Tian2017} produced other combinations of first-epoch ground-based measurements with \Gaia DR1. Although ground-based astrometry has been superseded by \Gaia for most practical purposes, it is still relevant in some situations. For instance, the inner Galaxy region suffers from high extinction in optical passbands, but is more accessible in the near-IR. \citet{Smith2018} used many dozens of observational epochs from the VVV survey to construct the VIRAC catalogue of $\sim 3\times10^8$ relative PM in the bulge/bar region, which has been later crossmatched with \Gaia DR2 to study the kinematics of the bar (Section~\ref{sec:bar_continuity}); \citet{Smith2024} published the updated VIRAC2 catalogue calibrated against \Gaia DR3. Other examples involve faints stars, which are either entirely out of reach for \Gaia or have too poor PM precision, but can be studied with a combination of ground-based surveys (e.g., SDSS and Subaru HSC, \citealt{Qiu2021}), or with \Gaia and other space telescopes -- Hubble \citep{delPino2022,McKinnon2024}, Euclid \citep{Libralato2024}, and in the future Roman \citep{Sanderson2019}.

\subsection{Spectroscopic surveys}  \label{sec:observations_spectroscopic_surveys}

\begin{figure*}[t]
\includegraphics{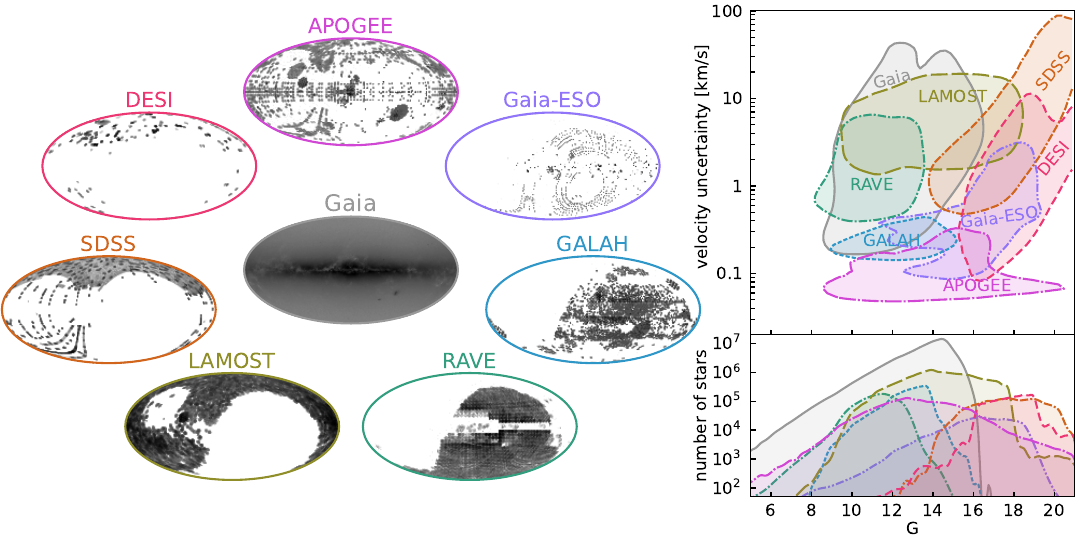}
\caption{Major spectroscopic surveys of the Milky Way. Left panels show the spatial coverage of each dataset in galactic coordinates; top right panel shows the velocity uncertainty as a function of $G$-band magnitude, and bottom right panel shows the distribution of magnitudes. }  \label{fig:spectroscopic_surveys}
\end{figure*}

In addition to \Gaia, a number of recent and ongoing ground-based spectroscopic surveys also feature prominently in the studies of Milky Way structure and dynamics (see Table~\ref{tab:spectroscopic_surveys}).
\begin{itemize}
\item RAVE \citep{Steinmetz2006,Steinmetz2020} was completed before the launch of \Gaia, and has a similar wavelength range and spectral resolution to \Gaia RVS, being limited to bright stars ($G\lesssim 13$--14) in the southern hemisphere ($\sim 5\times10^5$ objects). It is largely superseded by the more recent surveys, but played a historically important role in dynamical studies of the Solar neighbourhood \citep[e.g.,][]{Binney2014}.
\item Sloan Digital Sky Survey (SDSS) contained several Milky Way-related spectroscopic programmes. SEGUE-1 \citep{Yanny2009} and SEGUE-2 \citep{Rockosi2022} are low-resolution optical spectroscopic surveys focusing on faint stars at high Galactic latitudes, and together with stars observed in other SDSS programmes contain $\sim 5\times10^5$ stars.
\item APOGEE \citep{Majewski2017} is also part of SDSS, but uses a higher resolution IR spectrograph, facilitating the exploration of regions with high extinction in the Galactic bulge and disc, covering both hemispheres.
The final DR17 \citep{Abdurrouf2022} provides high-resolution spectra for $\sim 6.5\times10^5$ unique stars, most of them observed multiple times.
The analysis pipeline delivers high-precision (sub-\kms) line-of-sight velocities \vlos and abundances of $\sim 20$ chemical elements, making it a cornerstone for chemodynamical analysis of our Galaxy.
\item GALAH \citep{Buder2021,Buder2024} is similar in scope, but is conducted in the optical wavelength range, focuses on brighter and thus more nearby stars, and provides velocities and abundances for $\sim 9\times10^5$ objects mostly in the southern hemisphere (in their latest DR4), with an ultimate goal of $\sim10^6$ by the end of the survey \citep{DeSilva2015}.
\item \Gaia--ESO survey \citep{Gilmore2022, Randich2022} is the only one performed with an 8-metre class telescope, and focuses on accurate determination of stellar parameters, velocities and abundances using several cross-calibrated analysis pipelines. Its main targets are open clusters and stars in the Galactic disc.
\item LAMOST \citep{Zhao2012} consists of low- and medium-resolution surveys covering large areas in the northern hemisphere and containing respectively $\sim 5\times10^6$ and $\sim 8\times10^5$ unique stars in the current DR9 (by far the largest published ground-based catalogue).
\item DESI Milky Way Survey \citep{Cooper2023} has just started recently, and \citet{DESI2024} published an early data release containing the survey validation data ($\sim$1\% of the planned footprint, but still a respectable catalogue of $4\times10^5$ stars, see \citealt{Koposov2024} for details).
\end{itemize}

These surveys cover different parts of the Milky Way in terms of distance range, magnitude limits and structural components (see Figure~\ref{fig:spectroscopic_surveys}), although there is a sufficient number of stars in common to permit cross-calibration of line-of-sight velocities \citep{Tsantaki2022} and chemical abundances \citep[e.g.,][]{Soubiran2022,Hegedus2023,Thomas2024} between ground-based surveys, as well as to use them to calibrate stellar parameters determined from \Gaia BP/RP spectra \citep{Anders2023,Andrae2023b,Zhang2023b,Khalatyan2024}.

There are several other surveys focusing on specific components of the Milky Way: S$^5$ \citep{Li2019} is a dedicated survey of stellar streams, ARGOS \citep{Freeman2013} concentrates on the bulge (\citealt{Wylie2021} published a recalibrated subset of the original data), and H3 \citep{Conroy2019} explores the stellar halo, but the data from the latter survey are not publicly available, limiting its utility for the community. Future surveys, such as SDSS-V Milky Way Mapper \citep{Kollmeier2017} (a continuation of APOGEE), WEAVE \citep{Jin2024}, 4MOST \citep{deJong2019} and MOONS \citep{Gonzalez2020}, will push the magnitude limits to fainter and more distant stars and dramatically increase the catalogue sizes.

\subsection{Stellar ages and abundances}  \label{sec:observations_ages_and_chemistry}

Spectral analysis pipelines produce not only line-of-sight velocities, but also abundances of various chemical elements, as well as stellar parameters, such as effective temperature and surface gravity (a measure of the stellar radius), which together indicate the location of a star on the Herzsprung--Russell diagram and thus inform us about its evolutionary status. Among chemical elements, the abundance of iron [Fe/H] is most commonly measured, followed by one or more $\alpha$-elements (Mg, Ca, etc.), but high-resolution spectroscopic surveys such as APOGEE and \Gaia--ESO may provide dozens of elements, enabling a detailed characterisation of stellar populations. The key assumption is that stars inherit the abundance patterns from their birthplace environments, although this no longer holds for some elements at later evolutionary stages.

From the spectral stellar parameters, one can infer the physical properties of a star, such as mass, age and luminosity. Several groups have developed codes for performing this kind of inference, e.g., \texttt{The Cannon} \citep{Ness2016a}, \texttt{The Payne} \citep{Ting2019a}, \texttt{MADE} \citep{Das2019}, \texttt{StarHorse} \citep{Queiroz2018,Queiroz2020,Queiroz2023}, \texttt{AstroNN} \citep{Leung2019} and \texttt{DistMass} \citep{StoneMartinez2024}. The age distribution of an entire stellar population can also be constrained from photometry alone, fitting the observed colour--magnitude diagram with a linear combination of single-age, single-metallicity isochrones (e.g., \citealt{Gallart2005,Gallart2019}).

\subsection{Distance measurement}  \label{sec:observations_distance_measurement}

For most tasks in Galactic dynamical studies, one needs to know the distance to stars and other tracers -- it is used both for locating the object in the Galaxy and for determining the transverse velocity from PM. The most obvious method for computing the distance is to take the inverse of the parallax, but this gives sensible results only for high signal-to-noise measurements ($\varpi / \epsilon_\varpi \gtrsim 10$). Only $\sim 10^8$ stars in \Gaia DR3 satisfy this condition, and about twice as many pass a weaker cut $\varpi / \epsilon_\varpi \gtrsim 5$, which is borderline acceptable for subsequent analysis. The problem with making signal-to-noise cuts on parallax, highlighted by \citet{Luri2018}, is that the remaining sample becomes biased: the selection is made on the \textit{measured} value of $\varpi$, which is equally likely to be higher or lower than the \textit{true} value $\varpi^\text{(t)}$ (indeed, the distribution of measurement errors normalised by the estimated uncertainty $\epsilon_\varpi$ is close to standard normal for most stars with reliable astrometric solution, as quantified by the \texttt{ruwe} parameter). However, if $\varpi<\varpi^\text{(t)}$, the star may be excluded from the sample defined by a signal-to-noise cut, whereas it would have passed the cut if $\varpi>\varpi^\text{(t)}$; the true distance distribution of the resulting sample is therefore skewed to larger values than the distribution of $1/\varpi$.

The values of $\varpi$ listed in the \Gaia catalogue are known to contain some systematic bias at the level of a few tens $\mu$as, which depends on the magnitude, colour, sky position and other factors. \citet{Lindegren2021} provided a recipe for compensating the zero-point offset in EDR3, but subsequent studies generally found it to be overcorrecting (see figure 7 in \citealt{Khan2023} for a compilation of literature results).

The problem of inferring the distance to an individual star from its parallax (without a priori excluding it from the sample) becomes increasingly ill-defined for low signal-to-noise levels: without a meaningful prior, one cannot say much about the distance to a star whose measured value of $\varpi$ is negative (a quarter of the entire sample in DR3) or even consistent with zero. \citet{BailerJones2018} provide a recipe for computing the probability distribution for the distance, given the sky position, parallax and its uncertainty, as well as the catalogue of `geometric' distance estimates for nearly the entire \Gaia DR2 dataset. Of course, for low signal-to-noise ratio, the result heavily depends on the adopted prior for the Galactic density distribution. \citet{BailerJones2021} update the catalogue to EDR3, use a more realistic prior from a mock \Gaia catalogue by \citet{Rybizki2020}, and add a second method (`photogeometric'), adding information about colour and apparent magnitude in conjunction with a prior for the intrinsic colour--absolute magnitude diagram in each spatial region (taking into account spatially variable extinction, although rather crudely). \citet{BailerJones2023} introduced another modification to the method, adding the PM measurements as another constraint on the distance, again in conjunction with a prior derived from a mock \Gaia catalogue.

The use of photometry in conjunction with parallax should obviously improve the precision of distance estimates compared to parallax alone. For example, a star whose parallax is consistent with zero within uncertainty, and yet bright enough to be observed by \Gaia, is much more likely a luminous but distant giant rather than a nearby dwarf with the same colour (in the latter case it would have been much closer and thus its parallax would have been significantly different from zero); under this assumption, the distance could be inferred from the photometry using theoretical isochrones.
Unfortunately, the isochrones for giants depend quite strongly on metallicity, and moreover the colour and apparent magnitude are affected by extinction, adding another unknown parameter.
On the other hand, with the metallicity and absolute magnitude provided by spectroscopic analysis pipelines, one can measure both distance and extinction.
Moreover, the aforementioned \texttt{StarHorse} code has been adapted to work with photo-astrometric catalogues only, using parallaxes from \Gaia and multiband photometry from \Gaia, PanSTARRS, 2MASS and WISE; \citet{Anders2019} and \citet{Anders2022} provided estimates of stellar parameters (including distance) for 265 and 362 million stars from \Gaia DR2 and EDR3 respectively.

\Gaia DR3 analysis pipeline includes GSP-Phot \citep{Andrae2023a}, a module for determining stellar parameters from BP/RP spectra, $G$-band magnitude and parallax. The catalogue contains nearly half a billion stars brighter than $G<19$ with metallicity, effective temperature, surface gravity and distance estimates. However, as already mentioned in the \Gaia team papers and confirmed by independent analysis, the reported values are often significantly biased when compared to high-resolution spectroscopic measurements. In particular, distances are only reliable for stars with well-measured parallaxes ($\varpi/\epsilon_\varpi \gtrsim 20$), mostly within $\sim 2$~kpc, due to a rather strong prior on the extinction imposed in GSP-Phot. These shortcomings are largely caused by the deliberate choice to avoid using any external (non-\Gaia) data in the analysis pipeline -- this provides estimates fully independent of other (even higher-quality) datasets, but is vulnerable to imperfections in the theoretical models. On the other hand, nothing prevents the community from developing alternative catalogues of stellar parameters calibrated on high-resolution spectroscopic datasets and involving all available complementary information (e.g., near-IR photometry from WISE). Several groups provided such catalogues of metallicity, $\alpha$-abundances and other stellar parameter estimates from BP/RP spectra calibrated against APOGEE, GALAH and LAMOST \citep{Anders2023, Andrae2023b, Li2024a, Hattori2024, Fallows2024, Khalatyan2024}, and \citet{Zhang2023b} in addition provide distance estimates (however, these should be taken with a grain of salt -- for instance, distances to stars in the Magellanic Clouds are considerably underestimated).

Certain types of stars are much better suited for distance measurement by virtue of being standard(izable) candles. Blue horizontal branch (BHB) stars have the advantage of possessing unusually blue colours (bluer than the main-sequence turn-off for old stellar populations), and thus easily detectable in the halo, being good tracers of metal-poor populations. However, they are also easily confused with fainter blue stragglers, so additional $z$-band photometry is needed to separate these types (e.g., section 5.2 in \citealt{Li2019}). Red clump (RC) stars play a similar role for metal-rich populations, having absolute $G$-band magnitude $M_G\simeq 0.43$ and intrinsic scatter $\sim 0.12$ \citep{Chan2020, Khan2023}. However, their colours overlap with the more luminous red giant branch (RGB) stars, necessitating either additional photometry for an unambiguous selection or a probabilistic treatment of RC+RGB populations \citep[e.g.,][]{Sanders2019a}.

Variable stars can also be used for distance measurement through their period--luminosity relations (as usual, compounded by other factors, primarily metallicity). Among the most important types of variables are Cepheids and RR Lyrae. Classical Cepheids are young (few hundred Myr), massive (3--10\,$M_\odot$) and bright stars (detectable even in Andromeda), which trace metal-rich populations mainly in the disc, but they are fairly rare (a few thousand catalogued so far).
RR Lyrae are more numerous ($\gtrsim 10^5$), sit between BHB and RC stars in the colour--magnitude diagram, and are typical of old and metal-poor stellar populations mainly in the halo (although see \citealt{Prudil2020}, \citealt{Zinn2020}, \citealt{Iorio2021}, \citealt{Sarbadhicary2021}, \citealt{Bobrick2024} for counterexamples), with absolute magnitudes $m_G\simeq+0.6$ making them detectable out to nearly 100~kpc. Naturally, \Gaia is an excellent tool for identification of these variable stars across the entire sky. DR2 provided the first full-sky catalogue of RR Lyrae and Cepheids \citep{Clementini2019}, which was further updated in DR3 \citep{Clementini2023, Ripepi2023} and used by numerous studies to better calibrate the period--luminosity relations and distances (\citealt{Muraveva2018, Garofalo2022, Li2023a} for RR Lyrae, \citealt{Groenewegen2024} and references therein for Cepheids). It is worth mentioning that a significant fraction of RR Lyrae in the catalogue might be misclassified, but the sample can be neatly cleaned up by retaining only objects with \texttt{best\_class\_score} above 0.5 from the \texttt{vari\_classifier\_result} table (S.Ryan, priv.comm.)

\begin{figure}[t]
\includegraphics{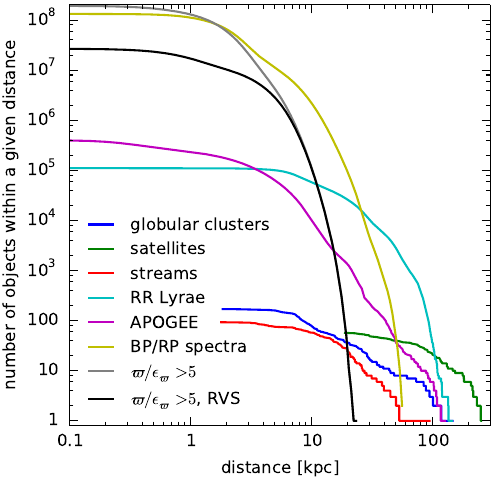}
\caption{Cumulative distributions of various kinematic tracers (number of objects with distances larger than the abscissa). Blue: globular clusters \citep{Baumgardt2021}; green: satellite galaxies \citep{Battaglia2022}; red: stellar streams \citep{Mateu2023}; cyan: RR Lyrae from \Gaia DR3 \citep{Li2023a}, excluding those in satellite galaxies; magenta: APOGEE DR17 \citep{Abdurrouf2022} with distances from \texttt{StarHorse} \citep{Queiroz2023} having relative uncertainty better than 20\%; yellow: distances from \Gaia BP/RP spectra calibrated on LAMOST \citep{Zhang2023b} with the same cut on uncertainty; grey: all stars with parallax uncertainty better than 20\%; black: subset of these stars with line-of-sight velocity measurements. }  \label{fig:distance_distribution}
\end{figure}

Figure~\ref{fig:distance_distribution} illustrates the distance distribution of various subsets of stars and other kinematic tracers. The distances computed from parallax (with a fairly liberal quality cut of 20\%) are largely limited to 10--15~kpc (grey and black curves, the latter showing the RVS subsample), whereas photometric estimates from \citet{Zhang2023b} with the same quality cut may be usable up to distances twice as large, and spectrophotometric distances for APOGEE stars from \citet{Queiroz2023} extend even further, although the entire catalogue is much smaller. RR Lyrae are observable by \Gaia up to 100~kpc and offer the largest sample of objects in the outer halo (beyond 40~kpc), whereas a dozen satellite galaxies are the only available tracers at even larger distances.

\subsection{Spatial coverage and selection biases}  \label{sec:observations_selection}

Although \Gaia is an all-sky survey, it certainly does not see every star in the Galaxy (in fact, the entire \Gaia catalogue contains only $\sim$1\% of all Milky Way stars). It misses both the (few) very bright stars and objects with apparent magnitudes fainter than $G\approx 21$, although the completeness of the catalogue gradually decreases towards the faint end, and in the areas of high source density this limit is considerably shallower. An increasing fraction of stars at the faint end have only the positional information, but lack parallax and PM measurements. The completeness of other subsets of the catalogue (colour information, BP/RP spectra, RVS velocity measurements, etc.) also varies with magnitude, colour and sky position (including effects of the scanning law and the source density). Ground-based spectroscopic surveys mentioned in Section~\ref{sec:observations_spectroscopic_surveys} have very patchy footprints (see Figure~\ref{fig:spectroscopic_surveys}) and even more dramatic variations in completeness, and their source list is usually assembled \textit{a priori} using some other catalogues and selection criteria (unlike \Gaia, which aims to record all objects that can possibly be observed).

All this complexity is encoded in the concept of \textit{selection function} $S(\boldsymbol q)$, which quantifies the fraction of stars in the entire population with properties $\boldsymbol q$ (including position, velocity, magnitude, colour, etc.) that appear in the observed catalogue (see \citealt{Rix2021} for a discussion). The \Gaia collaboration did not originally plan to provide the selection function of its catalogue, but following a `grassroots' initiative of \citet{Boubert2022} and associated papers, the \textit{Gaia Unlimited} project was set up to provide an official toolbox and pre-computed selection functions for various subsets of \Gaia data \citep[etc.]{CantatGaudin2023, CastroGinard2023}, as well as crossmatches between \Gaia and other surveys (e.g., APOGEE; \citealt{CantatGaudin2024}).

In most cases, the dominant factors in the selection function are the sky footprint and the apparent magnitude limits, which both affect the spatial coverage of the sample. More complicated cases include colour cuts, which introduces the dependence on the interstellar extinction and thus on the 3d dust distribution, and source lists for spectroscopic catalogues based on the \Gaia astrometry itself (e.g., H3 and S$^5$ surveys). Fortunately, apart from the latter case, the selection function seldom depends on the stellar velocity, i.e., the catalogues are usually kinematically unbiased (though there may be deliberate exceptions, e.g., the `reduced proper motion' approach of \citealt{Koppelman2021a,Koppelman2021b} and \citealt{Viswanathan2023}). Therefore, determining the kinematic properties of a given subset of stars is usually much easier than its spatial distribution. Nevertheless, when it comes to dynamical modelling (Section~\ref{sec:dynamics}), the inference on the gravitational potential depends on both spatial and kinematic properties of tracer populations.

\subsection{Velocity measurement}  \label{sec:observations_velocity_measurement}

\begin{figure}[t]
\includegraphics{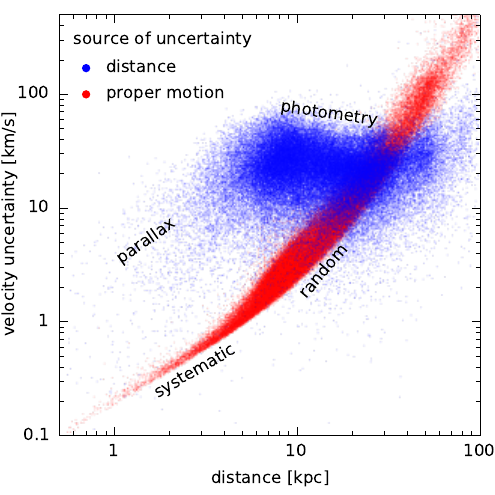}
\caption{Uncertainty in sky-plane velocity for RR Lyrae from \citet{Li2023a} as a function of distance. Blue dots show the contribution of distance uncertainty $\epsilon_v^{(D)} = v\,\epsilon_D/D \propto \mu\,\epsilon_D$, red -- the contribution of PM uncertainty $\epsilon_v^{(\mu)} \propto \epsilon_\mu\,D$. The first term dominates at distances $D\lesssim 30$~kpc, and the relative uncertainty $\epsilon_D/D$ is nearly constant at 0.1 when the distance is determined from photometry, whereas at even smaller distances $D\lesssim 5$~kpc it can be determined increasingly more precisely from parallax. In the second term, $\epsilon_\mu$ has a systematic floor around 0.035~\masyr (for both PM components combined) and rapidly increases for stars fainter than $G\simeq 15$, eventually dominating the total error budget. }  \label{fig:velocity_error_rrl}
\end{figure}

The line-of-sight component of velocity is measured from spectroscopy with precision ranging from $\sim 0.1$~\kms for high-resolution surveys such as APOGEE and GALAH to a few \kms for LAMOST and at the faint end of \Gaia RVS and DESI catalogues; faint stars in SDSS--SEGUE have even larger uncertainties up to a few tens \kms (right panel in Figure~\ref{fig:spectroscopic_surveys}). Except the latter case, all other measurement uncertainties are comfortably small for most galactic-dynamical applications (the velocity dispersion of Galactic populations ranges from 20--25~\kms in the thin disc to $\sim$100~\kms in the stellar halo), although for internal kinematics of star clusters and satellite galaxies one needs a precision $\lesssim 1$~\kms. However, all spectroscopic catalogues together contain $\lesssim 4\times10^7$ stars (with the vast majority contributed by \Gaia RVS).

On the other hand, the sky-plane velocity
\begin{equation*}
\frac{v}{\mbox{\kms}} = 4.74\, \frac{\mu}{\mbox{\masyr}}\, \frac{D}{\mbox{kpc}}
\end{equation*}
is available for all $1.5\times10^9$ stars with \Gaia astrometry, but of course it is only usable if the uncertainty is small enough. Its precision is determined by both the distance and PM uncertainties:
\begin{equation*}
\frac{\epsilon_v}{\mbox{\kms}} = \frac{\epsilon_D}{D}\, \frac{v}{\mbox{\kms}} + 4.74\, \frac{\epsilon_\mu}{\mbox{\masyr}}\, \frac{D}{\mbox{kpc}}.
\end{equation*}
When the distance can be determined from parallax, its relative uncertainty is $\epsilon_\varpi/\varpi$ and decreases towards smaller distances (the parallax uncertainty has a floor around 0.02 mas for $G\lesssim 14$). Beyond a few kpc, photometrically determined distances become more precise than parallax; a typical relative uncertainty is $\sim 0.1$ for RR Lyrae. Since the typical heliocentric velocities of halo stars are of order 200--300~\kms, the contribution of distance uncertainty to $\epsilon_v$ is of order 10\% of the velocity itself. However, at even larger distances, the second term in the above expression (PM uncertainty) becomes dominant (Figure~\ref{fig:velocity_error_rrl}), and as a result, $\epsilon_v$ exceeds the velocity dispersion in the halo beyond $\sim 50$~kpc. On the other hand, the PM uncertainty is expected to be reduced by a factor 2.5 in DR4 and 6 in DR5, leaving the distance uncertainty as the dominant term at all radii. The situation is similar for RGB stars, although their photometric distance uncertainties may be larger than 10\%, and PM uncertainties are smaller than those of RRL (since giant stars are brighter), hence the transition from distance- to PM-dominated regime occurs further out. In practice, this means that distance uncertainty is the limiting factor across almost entire Galaxy. In addition, the coupling between the distance errors and the magnitude-dependent selection function could bias the determination of the kinematic properties of a stellar population, if not accounted for.

\subsection{Coordinates}  \label{sec:observations_coordinates}

For studying the Milky Way, one needs to transform observations into the Galactocentric frame, or otherwise transform the model into the observable space (which is often preferred, as the measurement uncertainties are easier to deal with in their native space). The observational data are usually given by sky coordinates, parallax and two PM components (due to the way \Gaia scans the sky, their uncertainties have a non-trivial correlation matrix, which should be taken into account), and line-of-sight velocity%
\footnote{This quantity is often \textit{erroneously} called `radial velocity' in the astronomical literature; we reserve this term exclusively to refer to the component of velocity in the direction away from the centre of a given object (in this review, the Milky Way). The colloquial usage of `radial velocity' in place of the descriptive and unambiguous `line-of-sight velocity' essentially places the observer at the centre of the Universe, a view untenable since \cite{Copernicus1543}, although this usage may still be appropriate for cosmological applications.}; the latter is almost always given in the barycentric International Celestial Reference System (ICRS), i.e., compensated for the Earth motion around the Sun.

The first step is the rotational transformation of celestial coordinates from ICRS to Galactic, defined such that the equatorial plane \textit{approximately} coincides with the Galactic plane, and the origin \textit{approximately} coincides with the direction to the Galactic centre at a particular epoch. At the time when it was standardised \citep{Blaauw1960}, the precise location of Sgr~A$^\star$, which sits at the true centre of our Galaxy, was not known, and due to this tragic historical accident, it ended up being $\sim\!4'$ or 10~pc offset from $l=0,\, b=0$. If one uses a right-handed Cartesian rather than spherical heliocentric Galactic coordinate system, its $X$ axis and the corresponding velocity component (traditionally denoted as $U$) points \textit{towards} (not \textit{away from}) the Milky Way centre, the $Z$ axis and the velocity $W$ is perpendicular to the Galactic plane, and the $Y$ axis and the corresponding velocity component $V$ are defined to be in the direction of the disc rotation. Unfortunately, this choice of the $Y$ and $Z$ axes implies that the Galaxy rotates clockwise when viewed from the North pole (down the $Z$ axis), which is opposite to the Solar system, and corresponds to a negative angular momentum.

The final step is the transformation from the heliocentric Galactic coordinate system $XYZ$ to the Galactocentric coordinate system $xyz$. In the most common definition (e.g., in the \textit{de facto} standard \texttt{astropy.coordinates} module, \citealt{Astropy2022}), the orientations of Cartesian basis vectors in both systems \textit{nearly}%
\footnote{The difference mainly comes from the fact that the Sun is not exactly in the Galactic plane, but in addition to that, in \texttt{astropy} the projection of the origin of the Galactocentric system onto the celestial sphere differs from $l=0,\,b=0$ at a sub-arcsecond level without a good reason.}
coincide, but the origin of the heliocentric system is placed at a distance $R_0$ from the Galactic centre along the negative $x$ axis, at a height $z_0$ above its equatorial plane, and moving predominantly towards positive $y$ direction. Thus the angular momentum of the Sun and most disc stars is negative, which causes innumerable pains for astronomers; a less common but equally awkward alternative would be a left-handed Galactocentric coordinate system. For plotting purposes (e.g., the Lindblad diagram $E$ vs.\ $L_z$), one can either keep the disc stars on the left side, or put them on the right and flip the direction of the abscissa ($L_z$); here the community is more divided, with both conventions in widespread use (we note, however, that a leftward-pointing abscissa is quite common in astronomy, since this is the normal orientation of the longitude axis on the celestial sphere, although unfortunately it is shown incorrectly in some papers).

The heliocentric Galactic system moves with the Solar velocity $\boldsymbol v_\odot$ w.r.t.\ the Galactic centre. If one adopts Sgr A$^\star$ for the latter%
\footnote{The massive black hole is not exactly at rest, but the amplitude of its Brownian motion is estimated to be $\lesssim 1$~\kms \citep{Merritt2007, Reid2020}.}%
, its PM is measured very precisely from radio interferometry \citep{Reid2004,Reid2020}, but in order to translate it into actual velocity, one needs to know $R_0$. Most recent estimates are in the range 8.1--8.3~kpc, but even the very precise measurements by \citet{Gravity2018,Gravity2019,Gravity2022} differ by $\sim$0.15~kpc, well above their quoted uncertainties. Taking the value $R_0=8.122$~kpc from the first of these papers, \citet{Drimmel2018} determine the Solar velocity to be $v_{\odot}=\{12.9, 245.6, 7.8\}$~\kms. Together with $z_0=20.8$~pc estimated by \citet{Bennett2019}, these are currently the default parameters of the Galactocentric system in \texttt{astropy}. One should note that the value of $z_0$ is still not well constrained; for instance, \citet{Reid2019} find $z_0=7\pm2$~pc from the analysis of masers, which would align the $X$ and $x$ axes (while still placing Sgr A$^\star$ 7~pc below the plane of the Galactocentric system). The discrepancy between different studies likely arises from the different spatial coverage, coupled with the fact that the Galactic plane is not exactly flat (Section~\ref{sec:perturbations}, see also \citealt{Beane2019} for a simulations perspective). In addition, the line $b=0$ deviates from the actual midplane by $\sim 0.1^\circ$ (see \citealt{Anderson2019} for a discussion). The distance to the Galactic centre $R_0=8.2\pm0.1$~kpc quoted in the review by \citet[their table 3 and figure 4]{BlandHawthorn2016} is still our best estimate in the \Gaia era; see figure 1 in \citet{Leung2023} for a compilation of recent measurements.

The Solar velocity differs from the `local standard of rest' (LSR) by the peculiar velocity, which is measured to be 5--10 \kms in each coordinate \citep{Bobylev2018, Bobylev2022, Ding2019, Zbinden2019} depending on the sample of stars. In view of the obvious non-axisymmetries induced by the bar and spiral arms and various other perturbations discussed in Sections~\ref{sec:nonaxi_response} and \ref{sec:perturbations}, it is unlikely that the LSR can be nailed down to better than a few \kms. Nevertheless, the $y$-component of $v_\text{LSR}\simeq 235\pm 5$~\kms is a good proxy for the circular velocity at the Solar radius, formally defined as $v_\text{circ} \equiv \sqrt{R_0\,\partial\Phi^\text{(a)}/\partial R \big|_{R=R_0}}$ in the axisymmetrised Galactic potential $\Phi^\text{(a)}$. The transformation of velocities into the LSR frame is thus less well-defined than the transformation to the `Galactic standard of rest' (GSR; essentially the velocity in the Galactocentric system). Sometimes the line-of-sight velocity is quoted in the GSR frame, although unfortunately not always accompanied by the adopted value of $\boldsymbol v_\odot$. The difference between $v_\text{los}$ and $v_\text{GSR,los}$ is the projection of $\boldsymbol v_\odot$ onto the given line of sight, whereas applying a similar correction to the other two velocity components requires the knowledge of the distance, which is usually the dominant source of uncertainty, as discussed in the previous section.

\citet{Gaia2021c} were able to measure the Galactocentric acceleration of the Solar system directly from astrometry of $\sim10^6$ quasars. Their estimate translates to a circular velocity of $\sim 242\pm 8$~\kms, in good agreement with other methods, and the expected increase of precision in future data releases may make it the most accurate and most direct method for constraining the Galactic force field, albeit at only one location.

\subsection{Stellar orbits and integrals of motion}  \label{sec:observations_orbits_integrals}

In general, the instantaneous 6d phase-space coordinates of a star are not as interesting as its time-averaged orbital properties, given that the dynamical time (orbital period) is much smaller than the Hubble time, except in the outer parts of the Galaxy. To determine and characterise the orbit of an object, one needs to know the gravitational potential $\Phi(\boldsymbol x, t)$ in which it moves. As discussed in Section~\ref{sec:dynamics}, we now have a reasonably good understanding of the global properties of the Galactic potential, particularly in the Solar neighbourhood, where the uncertainties are at the level 10--20\%, although it is much less well known in the inner and outer parts of the Galaxy, and there is still no consensus about the properties of non-axisymmetric structures (bar and spiral arms). In addition, the time variation of the potential is virtually unconstrained over the entire Hubble time, except perhaps the last Gyr, where the dominant effect is the ongoing interaction with the Large Magellanic Cloud (LMC; see Section~\ref{sec:halo_perturbation}). Fortunately, the orbital properties of most stars are only weakly sensitive to the unknown details of the Galactic potential.

For any fiducial choice of the potential (even a time-varying one), one can numerically compute the trajectory of any object $\boldsymbol x(t), \boldsymbol v(t)$ from its current phase-space coordinates $\boldsymbol x(0), \boldsymbol v(0)$. Such a description is not very useful in practice, and to make further progress, one usually makes additional assumptions, for instance, that the potential is time-independent. In this case the orbit can be described in terms of its integrals of motion, in particular the energy $E$, and in the case of axisymmetric potentials, the $z$-components of the angular momentum $L_z$. For a given energy, $|L_z|$ cannot exceed the angular momentum of a circular orbit in the equatorial plane $L_\text{circ}(E)$, so the distribution of objects in the $E$--$L_z$ space is confined to a wedge-shaped region. Alternatively, one can characterise an orbit confined to the equatorial plane $z=0$ by its geometric properties: peri/apocentre radii $R_\text{p,a}$, defined as solutions of $\Phi(R) + L_z^2/(2R^2)=E$, or equivalently the eccentricity $e \equiv (R_\text{a}-R_\text{p}) / (R_\text{a}+R_\text{p})$ and the radius of the guiding centre $R_\text{g}$, defined as the solution of $R^3\;\partial \Phi(R)/\partial R = L_z^2$. One can generalise these definitions to non-planar orbits, adding the inclination $i\equiv \arccos L_z/L$ as the third variable (although $L$ is not strictly conserved in a general potential).

Finally, for some applications it is advantageous to choose the action variables $\boldsymbol J$ as the integrals of motion. They are defined only for bound and regular orbits (respecting at least three integrals of motion), and in practice are usually calculated approximately using the St\"ackel fudge method \citep{Binney2012}; see \citet{Sanders2016} for a detailed review and comparison of various methods of action computation. In an axisymmetric potential, the azimuthal action $J_\phi \equiv L_z$ uniquely maps to the guiding-centre radius $R_\text{g}$, and the radial and vertical actions $J_R, J_z$ quantify the extent of the orbit in the corresponding directions and generalise the concepts of eccentricity and inclination, respectively.
One of the benefits of actions $\boldsymbol J$ is that they are naturally complemented by the canonically conjugate angle variables $\boldsymbol \theta$, which describe the position of the star along its orbit and linearly change with time. By convention, $\theta_R=0,\pi$ correspond to peri/apocentres, $\theta_z=0,\pi$ -- to the upward/downward passages through the equatorial plane, and $\theta_\phi=0$ or $\pi$ usually corresponds to the azimuthal angle of the Sun (the latter two correspondences are approximate).

The various ways of describing the orbital properties can be summarised as follows:\\[2mm]
\mbox{}\!\!\!\!\begin{tabular}{cc}
6d phase-space coordinates $\boldsymbol x,\boldsymbol v$ \\
$\Downarrow$ & potential $\Phi(\boldsymbol x, t)$ \\
orbit $\boldsymbol x(t), \boldsymbol v(t)$\\
$\Downarrow$ & \makebox[35mm][r]{stationary potential $\Phi(\boldsymbol x)$}\\
integrals of motion $E\, [, L_z, \dots]$; & [, axisymmetry] \\
orbital parameters $R_\text{g}, e, i$ \\
$\Downarrow$ & integrable potential, \\
actions $J_R, J_\phi, J_z$ & bound orbit
\end{tabular}

\subsection{Galactic dynamics software}  \label{sec:observations_software}

In addition to the already mentioned \texttt{astropy}, which contains many general-purpose modules for dealing with coordinates, units, catalogues, etc., several publicly available software packages have been developed in the last decade specifically for studies in galactic dynamics. Namely, \texttt{galpy} \citep{Bovy2015}, \texttt{gala} \citep{PriceWhelan2017} and \texttt{agama} \citep{Vasiliev2019} are three widely used packages that provide largely similar functionality: various potential and density models, routines for numerical orbit integration, transformation between position--velocity and action--angle variables, distribution functions, generation of tidal streams, etc. Besides dynamics, other packages such as \texttt{dustmaps} \citep{Green2019} providing a common interface to most published studies of interstellar extinction, or the selection function toolbox developed by the \textit{Gaia Unlimited} team (Section~\ref{sec:observations_selection}), assist in analysis of observational data.

\section{Equilibrium structure and kinematics of the Ga\-lac\-tic disc}  \label{sec:disc}

In this Section, we discuss studies that explored the kinematic properties of the Galactic disc, but did not address the gravitational potential -- this topic is deferred to Section~\ref{sec:dynamics_disc}.

\subsection{Nomenclature and historical perspective}  \label{sec:disc_definition}

It is well established that stars in the Galactic disc belong to multiple populations, which differ in ages, chemistry, spatial distribution and kinematics.
\citet{Gilmore1983} found that the vertical distribution of stars in the Solar neighbourhood can be described by a sum of two exponential profiles $\rho \propto \exp\big(-|z|/h\big)$ with scale heights $h$ of order 300 pc and 1400 pc, which came to be known as `thin' and `thick' discs. The vertical velocity dispersion $\sigma_z$ is linked to the scale height by the [total] gravitational potential, so the thicker the disc, the higher is its dispersion. Given that the dispersion increases with age, the thick disc naturally contains the oldest stellar population with age $\tau \simeq 12$~Gyr, while the thin disc can be further decomposed into several age groups with different dispersions; this approach is used, e.g., in the Besan\c con Galaxy model \citep{Robin2003}.

The decomposition of stellar density profile into thin and thick discs stood the test of time, but became available over a range of Galactocentric radii with the advent of deeper photometric surveys, such as SDSS. Based on it, \citet{Juric2008} performed fits of two disc profiles plus a stellar halo, and determined the scale heights to be $\sim$300 and $\sim$900 pc for the two discs. Assuming that the radial profile of surface density also has an exponential form, $\Sigma \propto \exp(-R/L)$, they determined the scale lengths $L=2.6$ and $3.6$ kpc for the thin and thick discs, respectively.

Around the same time, it became clear that the chemical properties of thin and thick discs are also different. Both span a broad range of metallicities, but the thick disc has a higher abundance of $\alpha$ elements ([$\alpha$/Fe]$>$0), with Mg often used as a proxy for all $\alpha$ elements. Thus from the analysis of the Solar neighbourhood, thin and thick discs became synonymous with $\alpha$-poor and $\alpha$-rich populations. However, this distinction does not necessarily apply throughout the entire Galaxy. Using APOGEE DR12, \citet{Hayden2015} mapped the chemical compositions of disc stars across a wide range of radii (3--15 kpc) and demonstrated that the $\alpha$-rich population practically disappears beyond 11 kpc, even at high $|z|$ traditionally associated with the thick disc. More recently, \citet{Imig2023} confirmed these findings with the final DR17 of APOGEE, and by adding the age measurements, established that the $\alpha$-rich component has little radial variation in age and metallicity, while the $\alpha$-poor component tends to be more metal-poor and younger further out, while its scale height also increases with radius (i.e., the disc is flared).

\citet{Minchev2015} demonstrated that mono-age populations (and by consequence, chemically defined components) tend to flare, but their superposition may still be morphologically decomposed into thin and thick structures, with the latter having a longer scale length (echoing the conclusion of \citealt{Juric2008}), even though the oldest and most $\alpha$-rich population itself has a shorter scale length. This highlights the danger of using `thin' and `thick' labels when referring to chemically defined populations, which is unfortunately widespread in the literature and leads to much confusion.

\subsection{Spatial structure}  \label{sec:disc_spatial_structure}

As discussed in Section~\ref{sec:observations_selection}, determining the true spatial distribution of stars from the observed sample is difficult and requires a detailed knowledge of the survey selection function. An additional complication is the dust extinction at low Galactic latitudes ($b\lesssim 20^\circ$), which varies significantly on small angular scales and along the line of sight.
Several studies provided large catalogues of stellar parameters, including the reddening and attenuation, using \Gaia photometry and parallaxes combined with other photometric surveys, such as 2MASS, WISE and Pan\-STARRS (e.g., \citealt{Anders2019,Anders2022}, \citealt{Fouesneau2022}, \citealt{Speagle2024}). However, these fits are performed on a per-star basis, and there is no built-in constraint that the extinction amplitude monotonically increases with distance along a given line of sight. Other works build upon these individual estimates and impose some form of spatially correlated prior to produce 3d maps of dust distribution up to a few kpc from the Sun \citep[e.g.,][]{Green2019,Lallement2019,Lallement2022,Gontcharov2023,Dharmawardena2024,Edenhofer2024}. The density distribution of stars usually comes as a fixed prior in these analyses. Ideally, one would need to constrain the 3d density profile of both stars and dust, as well as the distances to individual stars, in the course of a single massive statistical inference procedure, but this appears to be computationally infeasible at present. Moreover, the addition of kinematic information and dynamical constraints (i.e., linking the density, kinematics and potential) should aid in breaking the degeneracies between stellar and dust distribution.

The problem of density profile measurement is somewhat simplified if one limits the analysis to high Galactic latitudes, where the dust extinction is negligible. \citet{Everall2022a,Everall2022b} fitted the magnitude and parallax distribution of $\sim 6\times10^5$ stars within $10^\circ$ of each Galactic pole by a sum of two exponential disc components and a power-law stellar halo. They determined the scale heights of thin and thick components to be 0.26 and 0.7 kpc, not too dissimilar to the SDSS-based study of \citet{Juric2008}. \citet{Vieira2023} obtained similar values (0.28 and 0.8 kpc) for a sample of main-sequence stars within $15^\circ$ of Galactic poles (although they used a sech$^2$ instead of an exponential profile).
Using the \textit{Gaia Unlimited} toolbox for dealing with selection functions, \citet{Khanna2024} fitted the density profile of RC stars by a sum of two exponential profiles with scale lengths of 3.6 and 2.6 kpc and scale heights 0.17 and 0.45 kpc at the Solar radius, although the thin component had a strong flare and reaches a scale height of 0.8~kpc at $R=12$~kpc (this is compounded by the onset of the disc warp, as discussed in Section~\ref{sec:perturbations_warp}). Notably, the thick component dominated the total mass outside 3~kpc, echoing similar findings of \citet{Vieira2023} and \citet{Everall2022b}, but in contrast to the pre-\Gaia literature-averaged value of $\Sigma_\text{thick}/\Sigma_\text{total} = 12$\% quoted in \citet{BlandHawthorn2016}.
\citet{Lian2022} used APOGEE spectroscopy to distinguish between $\alpha$-poor and $\alpha$-rich populations, and determined their scale heights as 0.39 and 0.85 kpc at the Solar radius, although both values increased with radius (i.e., the discs are flared). They also found that the surface density of these populations are not well described by a simple exponential profile in radius (see also \citealt{Lian2024}), and that the $\alpha$-rich component contributes about 1/4 of the surface density at the Solar radius. \citet{Xiang2022} and \citet{Xiang2024} used APOGEE and LAMOST chemistry and ages, and estimate the mass of the $\alpha$-rich disc as $2\times10^{10}\,M_\odot$ (including $2\times10^9\,M_\odot$ in the oldest disc-like population with age $\gtrsim 13$~Gyr), which is a substantial fraction of the total stellar mass of the Milky Way, variously estimated as (4--6)${}\times10^{10}\,M_\odot$.
Thus the emerging picture is that regardless of chemical or geometric definition, the older/thicker/$\alpha$-rich disc is much more prominent than previously assumed.

\subsection{Velocity dispersions for different populations}  \label{sec:disc_velocity_dispersions}

One of the most fundamental characteristics of a stellar population is the velocity dispersion tensor, and its variation with radius and stellar age.

\citet{Sanders2018} used the catalogue of stellar ages, distances and velocities determined from several spectroscopic surveys by \citet{Das2019}, in combination with PM from \Gaia DR2, to study the dependence of kinematic properties of disc populations on age and Galactocentric radius in the range 3--15 kpc. They found that both radial and vertical dispersions decrease quasi-exponentially with radius inside the Solar circle, but this trend slows down (for $\sigma_R$) or is even reversed (for $\sigma_z$) further out. The age ($\tau$) dependence of dispersions is approximately power-law $\sigma\propto \tau^\beta$ with an exponent $\beta\simeq 0.3$--0.4, comparable to the exponent inferred from the kinematics of the Solar neighbourhood in the pre-\Gaia era \citep[e.g.,][]{Aumer2009}.

\citet{Mackereth2019} combined \Gaia DR2 PM with line-of-sight velocities, spectrophotometric distances and ages from APOGEE determined using \texttt{AstroNN} \citep{Leung2019}, and explored the chemo-kinematic trends in the Galactic disc in the radial range 4--13~kpc and up to $|z|<2$~kpc. They split their dataset into multiple subsamples in narrow bins of age and metallicity, as well as the low- and high-$\alpha$ sequences, and fitted parameterised models with an exponential dependence of $\sigma_{R,z}$ on $R$ and a quadratic dependence on $z$ to each population. Since each population occupies a relatively small range of radii, the overall velocity dispersion profiles are not necessarily exponential or even monotonic with radius. They find that the scale radii are large ($L_\sigma\gtrsim 15$~kpc) and poorly constrained, and both radial and vertical dispersions vary little with $z$ (i.e., the distributions are nearly isothermal). In the $\alpha$-poor disc, $\sigma_z$ was found to increase with radius, contrary to the usual expectations for a disc with a constant scale height (see their section 5.1 for discussion), but in agreement with the study of \citet{Sanders2018}. Finally, the power-law indices of the age dependence of velocity dispersions were in the range 0.4--0.5 for $\beta_z$ and 0.2--0.4 for $\beta_R$ (increasing towards smaller radii). By contrast, the high-$\alpha$ disc appeared to have little variation in the shape of velocity ellipsoid and a flat age--velocity dispersion relation.

\citet{Sharma2021} used MSTO and RGB stars from LAMOST and GALAH surveys, in combination with \Gaia DR2 PM, to explore the multivariate dependence of velocity dispersions on age, metallicity and Galactic position. They assumed a separable parametric function with individual factors depending on $L_z$, $|z|$, age and metallicity, and fitted the four subsets (two surveys and two stellar types) independently, finding no systematic differences between the subsets. They confirm the flattening of velocity dispersion profiles at $L_z \gtrsim L_{z,\odot}$ (beyond the Solar radius), but contrary to \citet{Mackereth2019}, find little difference between $\alpha$-rich and $\alpha$-poor populations, after all other factors (chiefly, that the $\alpha$-rich stars have generally lower $L_z$) have been accounted for. This confirms the earlier findings of the same group \citep{Hayden2020} based on the local ($\lesssim 0.5$~kpc) sample of GALAH stars. The power-law indices for the age--velocity dispersion relations were determined as $\beta_z\simeq 0.45$ and $\beta_R\simeq 0.25$, in agreement with other studies.

\citet{Ting2019b} quantified the vertical heating in the disc expressed in terms of time evolution of the mean vertical action $\widehat J_z$ (instead of the more commonly used velocity dispersion $\sigma_z$), since it is better preserved during radial migration. They also relied on the combination of \Gaia DR2 PM with photometric distances, line-of-sight velocities and ages from APOGEE for a sample of $\sim 20\,000$ RC stars in the radial range 3 to 14 kpc and ages $\tau \lesssim 8$~Gyr. The inferred power-law index for $\widehat J_z(\tau)$ was around 1, corresponding to $\beta_z=0.5$ in agreement with other estimates, and consistent with the scattering by giant molecular clouds (GMCs) as the heating mechanism, rather than any major global dynamical perturbations in the disc. \citet{Garzon2024} conducted a similar analysis for a sample of young stars from \Gaia DR3 $\times$ LAMOST, also finding a nearly linear dependence of $\widehat J_z$ on age; however, they argue that the heating by GMCs is only one of several possibly important mechanisms responsible for this trend.

\citet{Frankel2020} used the same sample of RC stars from APOGEE and \Gaia DR2 to explore the time evolution of angular momentum $L_z\equiv J_\phi$ and radial action $J_R$, which together describe the radial migration of stars across the Galactic disc and the broadening of their eccentricity distribution (labelled respectively as `churning' and `blurring' by \citealt{Schoenrich2009}, although the former process may be more descriptively named `cold torquing', as suggested by \citealt{Daniel2019}). They found that the rms changes in $L_z$ are $\sim10\times$ larger than in $J_R$, suggesting the radial migration at corotation resonances of non-axisymmetric perturbations (such as spiral arms, e.g., \citealt{Sellwood2002}) as the driving mechanism. As discussed by \citet{Hamilton2024}, this finding places strong constraints on the possible mechanisms of migration.

\citet{Sun2024} used a larger sample of RC stars from LAMOST ($\sim 140\,000$) in combination with the latest \Gaia DR3 PM to map out the disc kinematics in the range 4--15 kpc. They split the sample by age and chemical abundances ($\alpha$ and metallicity) and considered not only the velocity dispersions, but also mean velocities, which are non-zero due to various perturbations discussed in Section~\ref{sec:perturbations}. They confirmed a nearly flat $\sigma_z$ profile outside the Solar radius and a heating law mostly consistent with previous studies that suggest GMCs as the driving mechanism, but also found a step-like increase in the dispersion at ages older than 7--9~Gyr, which may be caused by more violent dynamical perturbations, although they could not rule out a possible age underestimation bias of old and metal-poor stars.
Using the same sample of LAMOST RC stars, but selecting only the $\alpha$-poor ones, \citet{Das2024} found a jump in $\sigma_z$ beyond $\sim 10$~kpc for stars older than 6~Gyr, which they interpreted as a signature of heating by a passing satellite (tentatively the Sagittarius dSph, discussed in detail in Section~\ref{sec:perturbations_phase_spiral_origins}). This scenario is supported by the analysis of star formation history in the Solar neighbourhood \citep{RuizLara2020}, although it is not clear if Sagittarius orbit could affect stars so far inside the Galactic disc (the pericentre radius of its current orbit is $\gtrsim 15$~kpc).

From the 6d \Gaia DR3 RVS sample, \citet{Chandra2024a} selected $\sim$10$^7$ RGB stars with metallicity and $\alpha$ estimates from \Gaia BP/RP spectra \citep{Andrae2023b,Li2024a} and identified three main populations, clearly visible in the 2d distribution of stars in circularity $\eta\equiv L_z/L_\text{circ}(E)$ vs.\ metallicity (see their figures 5--7). Stars with halo-like kinematics ($\eta\simeq 0$) and low [$\alpha$/Fe] are dominant at [Fe/H]$<-1.2$, and in the range of metallicities from $-1.2$ to $-0.9$ they overlap with the $\alpha$-rich disc, which gradually increases its circularity from $\sim$0.8 to $>$0.9 as [Fe/H] increases. The $\alpha$-poor disc population starts to dominate at [Fe/H]$\gtrsim -0.6$, and its circularity is even closer to 1 (i.e., it is kinematically colder, as appropriate for the thin disc). Although in general metallicity is not monotonically increasing with time, these three populations largely form an evolutionary sequence from the proto-galaxy with little evidence for ordered rotation (previously identified by \citealt{Belokurov2022} under the name `Aurora'), through a spin-up phase of a still $\alpha$-rich population, and eventually to the $\alpha$-poor, metal-rich cold (thin) disc. A major merger that likely occurred around 8--10~Gyr ago (see Section~\ref{sec:halo_kinematics}) also heated part of the forming disc and created a `Splash' population with halo-like kinematics but disc-like chemistry (\citealt{Belokurov2020}, although alternative interpretations exist, e.g.\ \citealt{Amarante2020}); see section 4 in the review by \citet{ArchaeologyReview} for a detailed discussion. Thus the simple picture in which the velocity dispersions smoothly increase with time for all disc stars is certainly not valid for the oldest Milky Way components.

Going beyond the first two moments, \citet{Anguiano2020} studied the full velocity distribution (or rather, its projections onto each axis of the cylindrical coordinate system), using \Gaia DR2 RVS subset with well-measured parallaxes and augmenting it with the chemical classification based on APOGEE. They find that $\alpha$-poor disc and halo velocity distributions have a nearly Gaussian shape, but the $v_\phi$ distribution of the $\alpha$-rich disc is asymmetric with a sharper cutoff at high velocity and a long tail towards low or even retrograde rotation velocity.

\subsection{Orientation of velocity ellipsoid}  \label{sec:disc_velocity_ellipsoid_orientation}

The velocity dispersion is described by a symmetric $3\times3$ tensor, which can be diagonalised and represented by three principal values and three orientation angles. Under the assumption of axisymmetry, common to most kinematic and dynamical studies, it has only four independent parameters: one of the principal components ($\sigma_\phi$) must be aligned with the azimuthal direction, whereas the velocity ellipsoid in the meridional plane is parameterised by $\sigma_R^2$, $\sigma_z^2$ and the cross-term $\sigma_{Rz}^2$. Equivalently, one can express the orientation of this ellipsoid using the tilt angle $\alpha\equiv \frac{1}{2} \arctan \big[ 2\sigma_{Rz}^2 / (\sigma_R^2 - \sigma_z^2) \big]$. A zero tilt means that the velocity ellipsoid is aligned with cylindrical coordinates, whereas a tilt angle equal to $\arctan z/R$ means a spherical alignment. The knowledge of the tilt angle is important for dynamical modelling methods discussed in Section~\ref{sec:dynamics_disc}, particularly the Jeans equations. Although the velocity dispersions vary across different stellar populations, the tilt angle depends primarily on the structure of the underlying gravitational potential (see \citealt{Evans2016} for a discussion).

\citet{Posti2018} examined the orientation of the velocity ellipsoid for kinematically selected halo population in the \Gaia DR1+RAVE sample within a few kpc, and found a nearly spherical alignment. Using RR Lyrae from \Gaia DR2, \citet{Wegg2019} also found a nearly spherical alignment for this predominantly halo-like population, but in a larger range of Galactocentric radii (1.5--20 kpc). This was confirmed by \citet{Lancaster2019} for the BHB population in the halo. \citet{Hagen2019} used the DR2 RVS subset within 4 kpc, not separating it into disc-like and halo-like populations, and inferred a nearly spherical alignment in the inner Galaxy, gradually transitioning to a more cylindrical alignment outside the Solar radius. They did not find significant differences between tilt angles of disc and halo populations separated using metallicity information from LAMOST, but cautioned that the results could be affected by systematics in the distance measurement (i.e., adopted parallax zero-point). Using largely the same data, \citet{Everall2019} found a nearly spherical alignment everywhere; they attribute the disagreement with \citet{Hagen2019} to the use of inverse parallax as a distance proxy by the latter study -- a practice discouraged for the reason explained in Section~\ref{sec:observations_distance_measurement}.
More recently, \citet{Ding2021} and \citet{Sun2023b} combined \Gaia EDR3 PM with velocity and metallicity measurements from LAMOST, selecting respectively K giants and RC stars with photometrically inferred distances. Both papers focused on disc stars and inferred intermediate orientations of the velocity ellipsoid, closer to spherical than cylindrical, but in substantial disagreement with the above mentioned studies.

\subsection{Galactic models based on distribution functions}  \label{sec:disc_df}

There are several approaches for constructing global Galactic models combining stellar populations with different age and/or metallicity, spatial distributions and kinematics, and optionally incorporating dynamical self-\-consis\-tency constraints.

The most well-known is the Besan\c con model \citep{Robin2003}, which features several age groups for the thin disc, plus thick disc, bulge and stellar halo. Its parameters were originally optimised to fit the data from the Hipparcos mission \citep{Hipparcos}, and successive versions of the model were adapted to the \Gaia DR1 (TGAS) PM + RAVE line-of-sight velocities \citep{Robin2017}, and \Gaia EDR3 \citep{Robin2022}. The first of these papers fitted the age--velocity dispersion relation in the Solar neighbourhood only, because of the limited data coverage. The most recent version of the model is fitted to the joint distribution of parallaxes and PM (but not line-of-sight velocities) in 26 `deep' fields, as well as the local sample of stars within 100 pc \citep{Gaia2021b} and various external constraints on the potential. It also achieves dynamical self-consistency (in the axisymmetric approximation, \citealt{Bienayme2018}) by recomputing the potential using the density profile generated by the stellar DFs , and adding the contributions of the dark halo, whose parameters are also optimised during the fit.

An earlier version of the Besan\c con model was incorporated into the Galaxia code \citep{Sharma2011} for generating mock observational datasets, and combined with the Galactic extinction map from \citet{Green2019}, was used to produce the mock catalogues for \Gaia DR2 and EDR3 \citep{Rybizki2018,Rybizki2020}. These catalogues, in turn, were used as spatial priors for the distance estimation in \citet{BailerJones2018,BailerJones2021}. Thus the distance inference is based on pre-\Gaia Galactic models, but ideally, should be self-calibrating (updating the model in the course of the fit).

Another modelling framework was developed by \citet{Just2010} and focuses on the reconstruction of age-dependent vertical density and kinematic profiles. It also features a self-consistent computation of the axisymmetric gravitational potential, but only under the assumption of separability of radial and vertical directions. \citet{Sysoliatina2021} applied the method to the \Gaia DR2 RVS sample in the Solar neighbourhood, and \citet{Sysoliatina2022} extended it to a large range of radii (4--14 kpc), although still retaining the assumption of separability in $R$ and $z$. They determined the power-law index of the age--vertical velocity dispersion relation as $\beta_z\approx 0.4$, in agreement with other studies.

When the model is specified by a parametric distribution function (DF), one is not limited to datasets with precisely measured 6d phase-space coordinates. Instead, the likelihood of each observed star in the model is given by the convolution of the model DF with the uncertainty distribution of this star's measured coordinates (including the case of entirely missing dimensions, such as line-of-sight velocity). From Bayes's theorem, the likelihood of model parameters, given the data, is proportional to the product of likelihoods of all stars in the model with the given parameters, so maximising this likelihood is the way to find best-fit parameters. The model DF is typically specified in terms of integrals of motion, usually computed in the St\"ackel approximation for an axisymmetric potential $\Phi$ \citep{Binney2012,Bienayme2015}. This approach, developed in \citet{McMillan2013}, \citet{Ting2013}, \citet{Trick2016}, can be used to fit the DF parameters in the fixed potential, or even to determine the potential together with the DF.
The likelihood of finding a star with the given phase-space coordinates in the observational dataset is a product of the DF and the selection function of the survey, which usually does not depend on kinematics, but strongly depends on position (including the indirect dependence caused by magnitude selection). Therefore, a direct application of this approach necessitates a precise knowledge of the selection function, which may present significant practical challenges.

\citet{Binney2023} applied the modelling approach based on action-space DFs to the \Gaia DR2 RVS dataset, augmented with the vertical density profile from SDSS \citep{Juric2008}. In this approach, the potential is linked to the density generated by the DFs, thus the models are dynamically self-consistent and can be used to constrain the mass distribution in the Galaxy. \citet{Binney2024} extended this approach into chemical space, with each of the several disc and stellar halo DFs having action-dependent (and hence spatially varying) metallicity and $\alpha$-abundance profiles. The models were fitted to the combination of \Gaia DR3 and APOGEE DR17, and recover important features, such as the distinction between geometrically and chemically defined thick discs, although still not perfectly. In both papers, the relatively large (few $\times10^5$ to few $\times10^6$) observational sample with complete 6d phase-space coordinates was spatially binned, and the models fitted to the velocity histograms, rather than individual stellar velocities and PMs. This avoided the need to specify the spatial selection function explicitly, since the normalisation of the velocity distribution is rescaled to match the actual number of stars in the given bin (though this still assumes that the selection function is nearly constant inside each bin).

\citet{Li2022b} constructed DF-based models for the youngest stellar disc (OB stars), which are strongly concentrated near the equatorial plane and thus most affected by the dust extinction. They used stars with PM from \Gaia EDR3, but not necessarily having line-of-sight velocities (i.e., the DF was marginalised over the missing dimension), thereby considerably expanding the sample size. Their best-fit DF matches the kinematics of observed stars, but does not describe their spatial distribution well, which they attribute to the imperfection of extinction models, which ultimately determine the spatial selection function. The application of the DF-fitting approach to halo stars is discussed in Section~\ref{sec:halo}.

\subsection{Summary of disc properties}  \label{sec:disc_summary}

It is common to represent the stellar disc of the Milky Way as a sum of two components with constant scale heights and exponentially declining surface density and velocity dispersion profiles: the dominant, $\alpha$-poor, geometrically thin disc with a broad distribution of stellar ages and power-law relations between age and velocity dispersion, and a universally old, $\alpha$-rich, geometrically thick disc.
As could be seen from the above discussion, none of these assumptions is particularly good:
\begin{itemize}
    \item One should clearly distinguish between geometric (thick/thin) and chemical ($\alpha$-rich/poor) definitions of the two components. The $\alpha$-rich disc is indeed thick, but it ends rather sharply slightly outside the Solar circle; by contrast, the $\alpha$-poor disc is thinner in the Solar neighbourhood, but becomes thick (flared) in the outer Galaxy \citep{Imig2023,Binney2024}.
    \item Regardless of the definition, the $\alpha$-rich or geometrically thick population is much more prominent than assumed in many earlier studies, perhaps contributing $\gtrsim 1/4$ or even dominating the total stellar mass \citep[e.g.][]{Lian2022,Vieira2023,Khanna2024,Xiang2024}.
    \item The geometric structure of the disc deviates from a simple radially-exponential profile with a fixed scale height \citep[e.g.][]{Lian2024}.
    \item Velocity dispersions generally increase with stellar age, but there are distinct features on top of smooth trends, which are possibly associated with some dynamical perturbations at particular times (e.g., passages of a massive satellite galaxy through the disc; \citealt{Das2024}).
    \item Likewise, velocity dispersions generally decrease with radius, but these trends become shallower or even reverse in the outer Galaxy \citep{Sanders2018,Mackereth2019,Sharma2021}.
\end{itemize}
This complexity certainly reflects various mechanisms and events in the Milky Way evolution, but is difficult to capture in global models of the Galactic disc \citep[e.g.][]{Robin2022,Binney2024}. Nevertheless, if such models adequately describe mean trends, it becomes easier to study deviations from these trends and associate them with particular aspects of Galactic dynamics. Of particular importance are various disequilibrium features discussed in the next two Sections.

\section{Non-axisymmetric structure and the dynamical response}  \label{sec:nonaxi_response}

In this Section we discuss studies of non-axisymmetric structure in the Galactic disc, such as the Galactic bar (Section~\ref{sec:bar}) and spiral arms (Section~\ref{sec:spiralarm}) and the effect of these structures on the stellar kinematics of the disc. We focus primarily on studies in the \Gaia era, with some historical context. Large scale vertical disequilibrium such as the Galactic warp and corrugations are discussed separately in Section~\ref{sec:perturbations}.

\subsection{Large-scale mapping of the disc plane}  \label{sec:maps}

With \Gaia DR2 our view of the Milky Way disc expanded to provide all sky, 6D maps of Milky Way structure and dynamics across several kpc from the Sun. \cite{Gaia2018b} showed clear and significant kinematic substructure over $d\sim$ 2 kpc using the DR2 RVS sample, with large-scale streaming motion present in the radial, azimuthal and vertical velocities of stars near the Sun.

As discussed above, combining \Gaia DR2 and later releases with complementary spectroscopic surveys, most commonly APOGEE, and using spectrophotometric distance estimates, it becomes possible to study large-scale kinematics across the entire disc. These hybrid maps reach well into the Galactic centre, and in some cases beyond to the far side of the disc. For example, one can now make `face-on' ($x$--$y$) maps of a significant fraction of the Milky Way disc, and visualise the Galactic bar in both raw number counts \citep[e.g.][]{Anders2019,Queiroz2021,Zhang2024b}, and via its kinematics and chemistry \citep[e.g.][]{Bovy2019,Eilers2022,Gaia2023b,Leung2023,Hey2023,Zhang2024b}.

\begin{figure*}
\includegraphics[width=\linewidth]{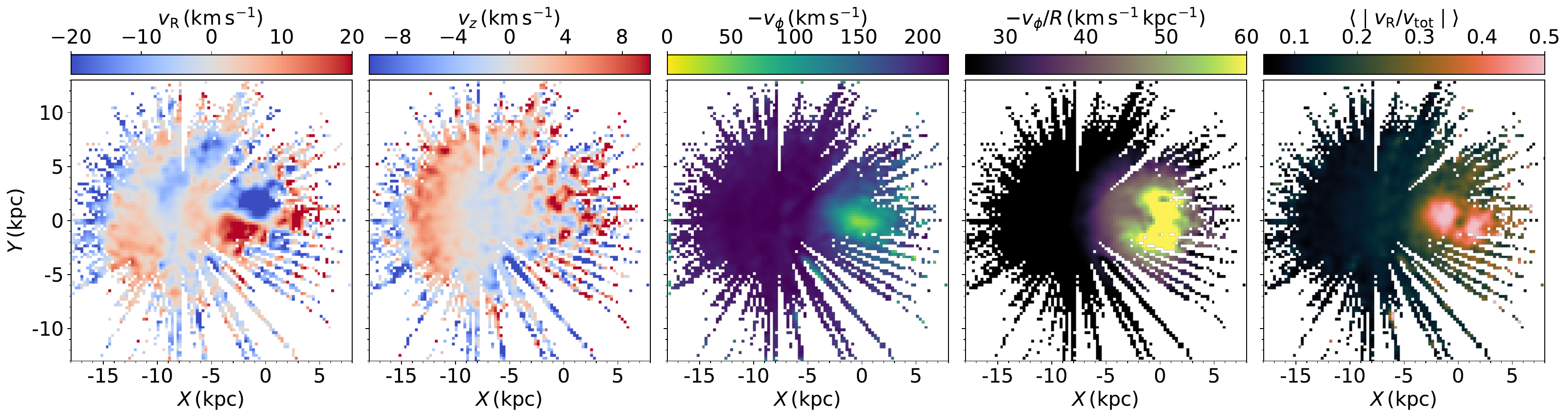}
\caption{$v_{\mathrm{R}}$ (left), $v_z$ (middle left), $v_{\phi}$ (middle), $v_\phi/R$ (middle right) and $\langle|v_{\mathrm{R}}/v_{\mathrm{tot}}|\rangle$ (right) maps across the Galactic disc using distances determined with \texttt{AstroNN}. Reconstructed following similar maps presented in \cite{Bovy2019,Carrillo2019,Eilers2020,Eilers2022,Gaia2023b,Leung2023,Hey2023,Zhang2024b}.
}  \label{fig:disc_kinematics}
\end{figure*}

Figure~\ref{fig:disc_kinematics} shows a reproduction of the $v_{\mathrm{R}}$, $v_{z}$, $v_{\phi}$ and $v_{\phi}/R$ maps from \cite{Bovy2019,Eilers2020,Leung2023} and the $\langle|v_{\mathrm{R}}/v_{\mathrm{tot}}|\rangle$ map from \cite{Zhang2024b} using \texttt{AstroNN} distance estimates. The characteristic radial velocity quadrupole appears clearly in the left panel, and the bar also appears as a function of mean $v_{\phi}$ and $v_\phi/R$, matching expectations for its kinematic signature \citep[see also][for similar maps and comparison to simulations]{Carrillo2019,Eilers2020,Eilers2022,Gaia2023b,Hey2023,Zhang2024b}. Note that while the angle of the kinematic features, such as the $v_{\mathrm{R}}$ quadrupole, should match the major axis of the bar, distance errors systematically `flatten' the feature to lie closer to the $y=0$ axis, making it non-trivial to measure the bar angle (see Section~\ref{sec:bar_angle} for further discussion).

It is possible to use such kinematic maps of the inner Galaxy to measure the distance to the Galactic centre. For example, the centre of the $v_{\mathrm{R}}$ quadrupole and the minimum of $v_\phi$ should lie at the barycentre of the bar+disc system. \cite{Leung2023} use maps from \texttt{AstroNN} (such as those shown in Figure~\ref{fig:disc_kinematics}) to fit for the distance to the Galactic centre. They rely on the minimum of the $v_\phi$ signature, claiming it to be less affected by observational errors than the $v_\mathrm{R}$ quadrupole, and find $R_0=8.23\pm0.12$ kpc. This purely dynamical measurement is in good agreement with other recent measurements discussed in Section~\ref{sec:observations_coordinates}.

Alongside the clear bar signature, \cite{Eilers2020} discovered a large-scale spiral-shaped pattern in the map of mean radial velocities (seen in the left panel of Figure~\ref{fig:disc_kinematics}). They linked this to a spiral arm, as discussed further in Sections~\ref{sec:spiralarm_response} and \ref{fig:rvphi}. They also showed the similar spiral-like arc in the vertical velocities. Both are discussed more in the context of external perturbations in Section~\ref{sec:perturbations_feathers}.

While this review is primarily focused on Galactic dynamics, the combination of \Gaia and complementary spectroscopic surveys have also made possible to map stellar chemistry across the disc, improving measurements of the radial and vertical gradients in several individual elements, as well as bulk metallicity and $\alpha$ elements \citep[e.g.][and references within]{Eilers2022,Hawkins2023,Hackshaw2024}. We discuss deviations from the smooth radial gradient, which likely arise from dynamical processes, in Section~\ref{sec:spiralarm_chemistry}.

\subsection{Structure and kinematics of the inner Galaxy}  \label{sec:inner_galaxy}

The structure and kinematics of the inner Galaxy are dominated by the Galactic bar, as discussed in the next Section. Observations of the inner Galaxy have to contend with large observational uncertainties, various biases, and a complex selection function arising from high levels of dust extinction towards the Galactic centre.

As discussed in Section~\ref{sec:observations_distance_measurement}, the majority of \Gaia parallaxes will not be reliable in the bar region, and studies of the Galactic centre that require distances must use either the `geometric' or `photogeometric' distance estimates of \cite{BailerJones2018,BailerJones2021}, which rely on some adopted prior, or catalogues such as \texttt{StarHorse}, \texttt{AstroNN} and others, which make use of data from complementary photometric and/or spectroscopic surveys, or limit the sample to standard candles, such as RC stars.

Such surveys have also revealed an old `knot' of stars in the Galactic centre \citep[][]{Horta2024b,Rix2024}. Using stars from \Gaia DR3 combined with APOGEE, \cite{Horta2024b} split the inner Galaxy into three populations, the disc, the bar, and a previously unknown `knot', which is approximately spherical and dynamically hot. \cite{Rix2024} found the same spherical knot structure in \Gaia data using metallicity from the BP/RP spectra for a large range of stellar ages ($3 \lesssim \tau \lesssim 10$ Gyr), but only in extremely metal-rich stars. These metal-rich knots are seen in Milky Way analogues in the TNG50 cosmological simulations, with similar ages and kinematics, yet the fraction of stellar mass is much lower in the observations ($\sim$0.1\%, although this is likely a lower limit owing to extinction) compared to the simulations (5--10\%). Interestingly, the central region also contains some of the most metal-poor stars in the Galaxy \citep{Belokurov2022,Rix2022,Arentsen2024}.

\subsection{The Galactic bar}  \label{sec:bar}

The Milky Way has been shown to contain a central bar through a combination of IR photometry from COBE and modelling of gas kinematics in the inner Galaxy \citep{Binney1991,Blitz1991,Weinberg1992,Weiland1994,Dwek1995}. Since then, the view on the central region has become increasingly clear with the advent of other state-of-the-art surveys, particularly in near-IR bands, which are less affected by dust extinction in the inner Milky Way. The near end of the bar lies at positive Galactic longitude ($l>0$) and appears more vertically extended in images of the inner Galaxy, owing to our perspective \citep[e.g.][]{Blitz1991}.

This apparent vertical extension of the bar has been shown to be arise from a boxy/peanut or X-shaped bulge, as seen from a viewing angle of approximately $20^{\circ}\lesssim\alpha_{\mathrm{b}}\lesssim30^{\circ}$ (see Section~\ref{sec:bar_angle} for a discussion of the bar angle, $\alpha_{\mathrm{b}}$) through a combination of near-IR survey data and comparison to simulations \citep[e.g.][]{McWilliam2010,Wegg2013,Wegg2015}. For instance, \cite{Wegg2015} examine a combination of data from 2MASS, UKIDSS, VVV, and GLIMPSE, finding a continuous transition from the boxy/peanut bulge down to a planar bar that extends out to $l\sim30^{\circ}$, or $R_{\mathrm{b}}\sim5$ kpc. This matches expectation from observations of external galaxies and $N$-body simulations. The edge on X-shape is also shown extremely clearly and conclusively by \cite{Ness2016b} using data from WISE. When assuming $R_0=8.3$ kpc and $\alpha_{\mathrm{b}}=27^\circ$, they estimate the radial extent of the X-shaped component to be $R_X\sim2.4$ kpc.

While the existence and general shape of the bar are now well known, its detailed dynamical properties, i.e.\ length $R_{\mathrm{b}}$, pattern speed $\Omega_{\mathrm{b}}$, angle with respect to the Sun $\alpha_{\mathrm{b}}$, radial and vertical density profile etc., have been difficult to pin down below the $\sim20\%$ level, with different methods used and different tracers being studied leading to a wide range of derived values in the decades since the discovery of the bar. Our review primarily focuses on the measurements made from \Gaia data specifically, both alone and in combination with ground-based surveys. For a more detailed discussion of the bar and its parameters from other complementary surveys, as well as historical context and additional references, see section 4 in \cite{BlandHawthorn2016} and the review of \cite{Shen2020}.

Even post-\Gaia there remains a disagreement between proponents of a `short fast' bar or a `long slow' bar as discussed below, although the \Gaia era has generally seen a shift in the community consensus to favour a longer, slower bar. Note that this discussion of a `slow' or a `fast' bar is related to a dichotomy of pattern speed estimates in the literature, where a `fast' bar typically associates the Hercules moving group with the Outer Lindblad Resonance (OLR) of a bar with $\Omega_{\mathrm{b}}\sim50$--60 \kmskpc, and a `slow' bar typically associates Hercules with the Corotation Resonance (CR) of a bar with $\Omega_{\mathrm{b}}\sim30$--40 \kmskpc, as discussed further in Section~\ref{sec:bar_inference}. This is not to be confused with the dynamically fast/slow classification of whether a bar extends close to its corotation radius\footnote{defined as the radius at which $\partial \Phi/\partial R = \Omega_\mathrm{b}^2\,R$.}, i.e.\ with the dimensionless ratio $\mathcal{R}=R_{\mathrm{corot}}/R_{\mathrm{b}}$, which is also discussed in various studies.

Methods for determining the Galactic bar parameters from \Gaia data (alone or combined with complementary ground based survey data) can be loosely split into two categories: methods that make use of data in the Galactic centre, directly analysing the distribution and kinematics of stars in the bar region (as discussed in this Section), or alternately methods that attempt to constrain the properties of the bar by modelling resonance features in the kinematics of stars in the Solar neighbourhood and across the disc, outside of the bar region (as discussed in Section~\ref{sec:bar_inference}).

\subsubsection{The angle of the Galactic bar; $\alpha_\mathrm{b}$}  \label{sec:bar_angle}

One may expect that large-scale maps similar to Figure~\ref{fig:disc_kinematics} with a clear bar signature in the inner Galaxy can be used to measure the angle of the Galactic bar with respect to the Sun--Galactic centre line. In the absence of observational uncertainties this would become trivial. However, multiple studies have shown that distance errors in the inner Galaxy can bias the apparent angle of both the density enhancement and the kinematic signature of the bar \citep[e.g.][]{Gaia2023b,Leung2023,Hey2023,Vislosky2024}, such that it cannot be immediately read off from Figure~\ref{fig:disc_kinematics}, or similar maps in the literature, without accounting for such biases.

For example, the apparent angle of the stellar density enhancement in the maps of \cite{Anders2019}, which are constructed from \texttt{StarHorse} distances calculated from a combination of \Gaia parallaxes and photometric data from Pan-STARRS, 2MASS and AllWISE, is found to be $\alpha_{\mathrm{b}}\sim40$--$45^\circ$. In contrast, \cite{Queiroz2021} find an apparent bar angle of $\alpha_{\mathrm{b}}\sim20^\circ$ by fitting an ellipse to the observed density enhancement using updated \texttt{StarHorse} distances calculated from a combination of APOGEE and \Gaia EDR3. They claim the discrepancy originates from high errors in the photometric distance estimates of the earlier catalogue, and an underestimation of extinction in the Galactic plane.

\cite{Bovy2019} find an apparent bar angle of $\alpha_{\mathrm{b}}=25^\circ$ after applying a correction for the distance biases based on a maximum-likelihood based comparison of \texttt{AstroNN} distance estimates derived from the APOGEE spectra, and the \Gaia DR2 parallaxes. \cite{Zhang2024b} also measure an apparent angle of $\alpha_{\mathrm{b}}=25^\circ$ from maps of the mean radial fraction of the total velocity, $\langle| v_{\mathrm{R}}/v_{\mathrm{tot}} |\rangle \equiv (1/N)\;\sum_{j=1}^N |v_{\mathrm{R},j}| / v_{\mathrm{tot},j}$, from a sample of low-amplitude long-period variable stars from \Gaia DR3. They show the angle of the $\langle |v_{\mathrm{R}}/v_{\mathrm{tot}} |\rangle$ feature is more robust to distance uncertainties than maps of $v_{\mathrm{R}}$ alone. In fact, a simple visual estimation of the apparent angle of the $v_{\mathrm{R}}$ quadrupole in the left panel of our Figure~\ref{fig:disc_kinematics} (uncorrected for distance biases) is $\sim20^{\circ}$, while the apparent angle of the $\langle| v_{\mathrm{R}}/v_{\mathrm{tot}} |\rangle$ feature in the right panel is $\alpha_{\mathrm{b}}\sim25^{\circ}$, matching that of \cite{Zhang2024b}.

\cite{Gaia2023b} find a bar angle of $\alpha_\mathrm{b}=19.2^\circ\pm1.5^\circ$ by modelling the bi-symmetry in disc kinematics from \Gaia DR3. They find the angle of the bar as estimated from azimuthal velocities to be more robust in the presence of observational uncertainties than estimates made from the radial velocities (as justified by comparison to a numerical simulation with observational errors imposed). The phase of the bi-symmetry also provides an estimation of the corotation radius of the bar, allowing them to estimate a bar pattern speed of $\Omega_{\mathrm{b}}=38.1^{+2.6}_{-2.0}$ \kmskpc when also assuming an angular velocity curve.

\cite{Simion2021} use a combination of \Gaia DR2 and ARGOS data to constrain the angle of the bar with respect to the Sun, by comparing bulge kinematics with a model from \cite{Shen2010}. They find $\alpha_{\mathrm{b}}=29\pm3^{\circ}$.

As such, there is significant scatter in \Gaia era estimates of the bar angle, although the discrepancy can be reasonably explained by distance uncertainties for stars in the inner Galaxy. The pre-\Gaia review of \cite{BlandHawthorn2016} recommends $\alpha_{\mathrm{b}}=27^\circ\pm2$ for the $X$-shaped bar, and $\alpha_{\mathrm{b}}=28$--$33^\circ$ for the long bar. More recent studies have reconciled the `short' and `long' bars to be a single structure with a single angle \citep[see the review of][and references within]{Shen2020}, but there remains uncertainty over approximately $\alpha_{\mathrm{b}}\approx25\pm5^{\circ}$.

\subsubsection{Bar pattern speed, $\Omega_{\mathrm{b}}$, from inner Galaxy kinematics and the continuity equation}  \label{sec:bar_continuity}

In this Section we discuss estimations of the bar pattern speed from the kinematics of stars in the inner Galaxy. See Section~\ref{sec:bar_summary} for a summary of bar pattern speed when also taking into account stellar kinematics across the wider disc.

With sufficient data coverage in the bar region, one can use the continuity equation (the basis for the well-known \citealt{Tremaine1984} method) to measure bar pattern speed from the 6D map (although the measurement may be bias in the presence of dust; e.g.\ \citealt{Gerssen2007}). For example, \cite{Bovy2019} apply their variant of the continuity equation method to \texttt{AstroNN} (\Gaia+APOGEE) data and measure $\Omega_{\mathrm{b}}=41\pm3$ \kmskpc (later updated to $\Omega_\mathrm{b} = 40.08\pm 1.78$ \kmskpc in \citealt{Leung2023} with a newer version of the catalogue).

\cite{Sanders2019b} independently apply their variant of the continuity equation method to a combination of data from \Gaia DR2 and the VIRAC catalogue of relative PM from the VVV survey \citep{Smith2018}, and also recover a bar pattern speed of $\Omega_{\mathrm{b}}=41\pm3$ \kmskpc. Concurrently, \cite{Clarke2019} also make use of a combination of VVV and \Gaia DR2 data to estimate a pattern speed $\Omega_\mathrm{b} = 37.5$ \kmskpc by \textit{qualitatively} comparing the M2M model of \cite{Portail2017} to the PM distribution in the inner disc. They then follow this up with a \textit{quantitative} fit, and find a lower pattern speed of $\Omega _\mathrm{b} = 33.29\pm1.81$ \kmskpc \citep{Clarke2022}.

\cite{Zhang2024b} explore a sample of low-amplitude long-period variable (LPV) stars from \Gaia DR3, whose distances can be measured with a $\sim10$\% error, and reveal the bar signature both in the density contrast and in its kinematic imprint on the far side of the Galactic centre. They apply the continuity equation to the LPV data and find $\Omega _\mathrm{b} = 34.1\pm2.4$ \kmskpc. They also analyse the orbit structure and make a dynamical measurement of the bar length of $\sim 4$ kpc.

\cite{Horta2024b} measure the pattern speed $\Omega_{\mathrm{b}} = 24\pm3$ \kmskpc by fitting the distribution of stars belonging to the bar in the $R^2$--$L_z$ space, which corresponds to a corotation radius of $R_{\mathrm{CR}}\sim9.6$ kpc and $\mathcal{R}\sim1.9$. This pattern speed is significantly slower than found by the other studies of inner-Galaxy dynamics discussed here, and also slower than studies of kinematics across the wider disc (see Section~\ref{sec:bar_inference}). \cite{Horta2024b} argue that this discrepancy is due to the contamination of the sample by disc stars in other studies. Further exploration is needed, but for now this measurement remains an outlier in the literature.

The application of the continuity equation and studies of PM in the Galactic centre consistently favour a `slow bar', with values converging to a pattern speed of $\sim34$--41 \kmskpc \citep[other than][]{Horta2024b}, a value which is also favoured by some studies of gas kinematics in the bar region (e.g. 40--42 \kmskpc; \citealt{Weiner1999,Sormani2015}). However, spiral arms can complicate the picture.

\subsubsection{Bar formation time; $\tau_{\mathrm{b}}$}  \label{sec:bar_formation}

As well as the current state of the Milky Way's bar, \Gaia and complementary surveys have also been used to constrain the bar formation time, $\tau_{\mathrm{b}}$.

\cite{Bovy2019} find the age of stars in the bar is consistent with the expected early formation of the thin disc, when the turbulent gas cooled enough to allow a bar-unstable thin disc to form, giving $\tau_{\mathrm{b}}\approx8$ Gyr.

One method of constraining the bar formation time is looking at the ages of stars in the nuclear stellar disc. The argument is that when the bar first forms, gas is funnelled in to the Galactic centre, creating a star-forming nuclear disc. Thus, the oldest stars in the nuclear stellar disc give an estimate of $\tau_{\mathrm{b}}$ \citep{Gadotti2015}. \cite{Sanders2024} examine the ages of Mira variable stars in the nuclear stellar disc using data from VVV, calibrated by \Gaia \citep{Sanders2022}. They estimate the nuclear stellar disc formed during a significant burst of star formation around $8\pm1$ Gyr ago, which also implies an early formation of the bar of $\tau_{\mathrm{b}}\gtrsim8$ Gyr.

\cite{Haywood2024} also estimate a bar formation time of $8\lesssim\tau_{\mathrm{b}}\lesssim10$ Gyr owing to a burst of radial migration seen in stars with ages $\tau\gtrsim7$--8 Gyr (from the \texttt{AstroNN} catalogue). This burst of migration is expected to occur during the early slowdown of the bar, lasting $\sim2$ Gyr after bar formation \citep[based on the $N$-body simulations from][]{Khoperskov2020}.

These estimates are consistent within error ($\tau_\mathrm{b}\gtrsim8$ Gyr), despite different techniques and arguments used. This value is also consistent with the possibility that the Milky Way's bar may have been induced by the last major merger, known as GSE (see Section~\ref{sec:halo_kinematics}), as discussed e.g.\ in \citet{Merrow2024}, although the Milky Way disc may also have been bar-unstable before the GSE merger, and further work on both the formation time of the bar and the dynamical state of the Milky Way's disc at early times is required before drawing firm conclusions.

\subsubsection{Bar--Spiral connection}  \label{sec:bar_spiral_connection}

\cite{Hilmi2020} show that an instantaneous measurement of bar length and pattern speed can be significantly biased when the bar is connected to a spiral arm. They argue that the Scutum--Centaurus spiral arm is likely currently connected to the Milky Way's bar, and thus estimates of its length and pattern speed may be biased by 1--1.5 kpc and 5--10 \kmskpc.

In general, estimating bar `length' in both external galaxies and the Milky Way is a surprisingly non-trivial question. A detailed review of bar length and measurement methods is beyond the scope of this review, but there can be a significant difference in length between `ellipse fitting' and similar methods that measure the `visual extent' of a bar, and the `dynamical length' of trapped orbits that make up the bar \citep[e.g.][]{Petersen2024,Ghosh2024}. For example, \cite{Lucey2023} compare a range of bar models to a combination of \Gaia and APOGEE data, and find the best-fitting model has a dynamical length of 3.5 kpc, while the visual extent of non-trapped orbits is approximately 5 kpc. Similarly, \cite{Vislosky2024} find that the radial velocity field from \Gaia DR3 in the inner Galaxy is consistent with both a short bar ($R_{\mathrm{b}}\sim3.6$ kpc) with moderately strong spiral arm attached, or a long bar ($R_{\mathrm{b}}\sim5.2$ kpc) with weaker spiral structure.

While the majority of the community appear to favour the long slow bar explanation, this will likely not be resolved until we can fully map the Milky Way's spiral structure into the inner Galaxy.

\subsection{Spiral arms}  \label{sec:spiralarm}

The Milky Way has long been known to be a spiral galaxy. Yet as with other aspects of Milky Way cartography, our location within this disc has made it difficult to reach a consensus on the number, location and nature of the Milky Way's spiral arms. A full review of the current state of the field with regards to the Milky Way's spiral structure across all tracers is beyond the scope of this review \citep[see instead prior reviews such as][]{Vallee2017a,Vallee2017b,Shen2020}. In this Section we focus solely on the detection of spiral structure in the Milky Way or estimation of its parameters from the stellar catalogue from \Gaia alone or combined with complementary spectroscopic surveys, and neglect much of the work involving other tracers, such as masers \citep[e.g.][and references within]{Reid2019} or gas \citep[e.g.][]{Soding2024}. For a general review of the nature of spiral structure in disc galaxies see \cite{Dobbs2014}.

\subsubsection{Direct stellar density mapping of spiral structure}  \label{sec:spiralarm_density}

\begin{figure*}
\includegraphics[width=\linewidth]{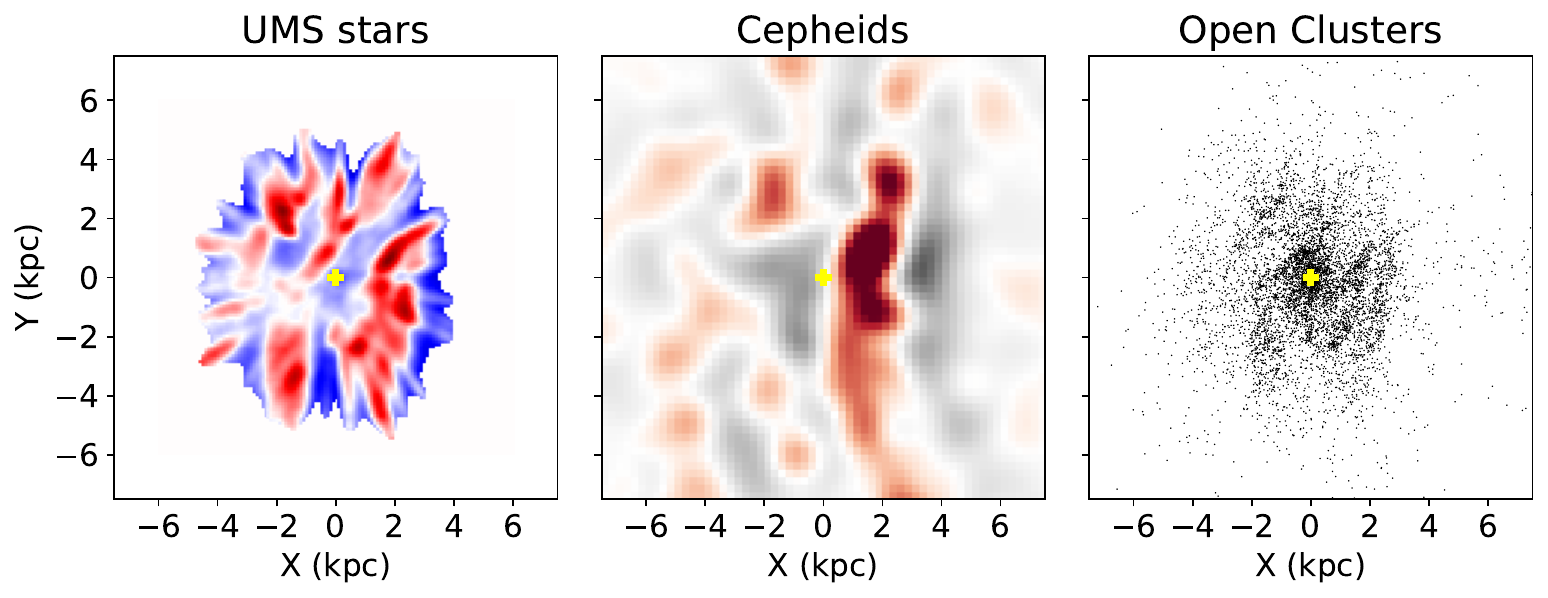}
\caption{Milky Way spiral structure seen in \Gaia EDR3, shown in density contrast for a sample of Upper Main Sequence stars (left) and using a wavelet analysis on a sample of Milky Way Cepheid stars (middle), reproducing the figures of \cite{Poggio2021b} from data available within. The right panel shows the distribution of open clusters from data available in \cite{Hunt2023}.}
\label{fig:Poggio2021spiralarms}
\end{figure*}

With \Gaia EDR3, \cite{Poggio2021b} map segments of the Milky Way's spiral arms directly in the number density of Upper Main Sequence (UMS) stars, with Cepheid variables. Figure~\ref{fig:Poggio2021spiralarms} shows a map of relative density in the UMS sample (left panel) and the Cepheid map following a wavelet transform (middle panel)\footnote{Recreated from \cite{Poggio2021b} using their notebook, available at \url{https://github.com/epoggio/Spiral_arms_EDR3}}. The UMS map reveals overdensities associated with the Perseus and Local arms, and some combination of the Sagittarius and Scutum arms. They show that the Perseus arm overdensity correlates well with the Perseus arm seen in the neutral Hydrogen \citep{Levine2006}, but is generally not well matched to the well-known model of star forming regions using masers \citep[e.g.][]{Reid2019}, though we direct the reader to \cite{Poggio2021b} for a direct comparison of the UMS sample with other models of the Milky Way spiral structure from the literature. The comparatively local UMS maps of \cite{Poggio2021b} have also now been remade in multiple follow-up studies with different tracers, e.g.\ OB stars \citep[e.g.][]{Gaia2023b,Xu2023b,Ge2024}, A stars \citep{Ardevol2023} and RC stars \citep[e.g.][]{Lin2022}, which are broadly in agreement with the maps of \cite{Poggio2021b}.

Cepheids are also useful for mapping the spiral structure across the disc, thanks to their precisely measured distances from the well-known period-luminosity relationship \citep[e.g.][]{Skowron2019a,Skowron2024}. Several studies examine the distribution of Cepheids in the disc plane, revealing faint spiral-like arcs over a significant fraction of the Galactic disc, as seen for the Cepheid sample of \cite{Poggio2021b} in the middle panel of Figure~\ref{fig:Poggio2021spiralarms} (see also \citealt{Lemasle2022,Gaia2023b,Drimmel2024,Funakoshi2024}).

The right panel shows the distribution of Open Clusters (OCs) from the catalogue of \cite{Hunt2023}. Studies of OCs in the era of \Gaia are discussed in another review in this series \citep{ClustersReview} and thus we do not go into detail here. But in brief, detections of spiral structure are also claimed in the distribution of OCs \citep[e.g.][and references within]{CastroGinard2021,Hao2021,Alfaro2022,Gaia2023b,Hunt2023,Joshi2023}. The distribution of OCs appears to favour more flocculent spiral structure over a grand design pattern, and the lack of an age gradient across the OCs (predicted e.g. in \citealt{Dobbs2010}) argues against a density wave or bar-driven nature.

\subsubsection{Mapping spiral structure by their imprint on stellar dynamics}  \label{sec:spiralarm_response}

As already mentioned in Section~\ref{sec:maps}, maps of the Milky Way disc plane cover a significant fraction of the disc. In this Section, we discuss large-scale kinematic patterns in the $x$--$y$ plane, which are thought to be the dynamical response to the Milky Way's spiral structure. Spiral structure also leaves an imprint in the stellar kinematics in other projections, which are discussed in their own Sections, such as the $U$--$V$ plane (Section~\ref{sec:bar_uv_plane}), the $R$--$v_\phi$ plane (Section~\ref{sec:bar_rphi}), and in action space (Section~\ref{sec:bar_action}).

\cite{Gaia2023b} compare the disc velocity field with the spiral-like overdensities in their OB star sample (for $|z|<0.3$ kpc, reaching approximately $d\lesssim2$ kpc), and find significant correlations between the density maps and the mean $v_{\mathrm{R}}$, $v_\phi$ and $v_z$ fields. Extending the analysis to their RGB sample highlights larger-scale coherent velocity patterns, but less obvious connection to spiral structure than the OB sample.

\cite{Eilers2020} show a large-scale radial velocity feature that appears to follow multiple wraps of a logarithmic spiral pattern using data from \Gaia DR2, APOGEE, 2MASS and WISE. They compare the velocity pattern with a toy model and find a best fit for $\alpha_{\mathrm{sp}}=12^{\circ}$, $\Omega_{\mathrm{sp}}=12$ \kmskpc, and an arm to inter-arm density contrast of $\sim10\%$ at the Solar radius, which increases towards the outer Galaxy.

A series of papers \citep{Fedorov2021,Fedorov2023,Dmytrenko2023,Denyshchenko2024} map the stellar spiral structure across the disc with \Gaia EDR3 and DR3 data using the Ogorodnikov--Milne model. They show that $\partial v_R/\partial R$ should be close to zero at the location of the spiral structure, and \cite{Denyshchenko2024} recover a four-armed global logarithmic spiral pattern, consisting of the Scutum--Centaurus, Sagittarius--Carina, Perseus and Norma--Outer arms, with a pitch angle of approximately $12^{\circ}$. This is interesting because it appears to match relatively well the four-armed logarithmic spiral pattern favoured by maser data \citep[e.g.][]{Reid2019}, in contrast to other stellar-dynamical studies. However, there are also numerous other spur-like or fragmented features in the map, which may also suggest some flocculency.

\cite{Palicio2023} show that plotting the disc plane as a function of radial action $J_{\mathrm{R}}$ (calculated in an axisymmetric potential) highlights multiple spiral-like features with low $J_{\mathrm{R}}$. However, \cite{Debattista2024} show that calculating instantaneous $J_{\mathrm{R}}$ in an axisymmetric potential leads to error on the order of $\lesssim100$ kpc\,\kms, and the actions should be averaged over the lifetime of the spiral arms to recover true $J_{\mathrm{R}}$ (which is of course impossible in the Milky Way). Nevertheless, they demonstrate with $N$-body simulations that while the true radial actions are lower in the spiral arms, the bias in instantaneous actions makes them appear higher at the location of the spiral density enhancement, meaning that even if the absolute values of $J_{\mathrm{R}}$ are wrong, they may still be useful for locating the spiral arm.

As such, the innermost arc of $J_{\mathrm{R}}$ in \cite{Palicio2023} correlates well with a combination of the expected location of the Sagittarius and Scutum arms. The local and Perseus arms are somewhat coincident with arcs of low $J_{\mathrm{R}}$ \citep[in contrast with expectation from][]{Debattista2024}, yet the variation with Galactic azimuth $\phi$ (or their pitch angles) are inconsistent with other studies. While it is not immediately clear whether these spiral-like regions of high $J_{\mathrm{R}}$ are a dynamical response to spiral structure, or otherwise related to the Solar neighbourhood moving groups, this is another potential signature for future studies to examine.

\cite{Widmark2024} make a detection of the Local arm overdensity ($\sim20\%$) by comparing the vertical potential in the midplane and at $z=400$ pc, using the vertical Jeans equation (as detailed in Section~\ref{sec:dynamics_disc_vertical}). When taking the residuals of $\Phi(400\ \mathrm{pc})-\Phi(0\ \mathrm{pc})$, there are clear spiral arm-like features, which match the location of the local arm in the UMS sample of \cite{Poggio2021b}, and are somewhat consistent with the location of the Sagittarius and Scutum arms.

\subsubsection{The chemical signature of the Milky Way's spiral structure?}  \label{sec:spiralarm_chemistry}

Recent work has also shown azimuthal variation in the chemistry of stars across the disc. Simulations show that such variations are predicted to arise both through bar-driven radial rearrangement of the disc \citep{Filion2023}, in isolated spiral galaxies \citep[e.g.][]{Debattista2024}, and in systems interacting with a satellite \citep[e.g.][]{Carr2022}. They are also predicted from chemical evolution models that take transient spiral structure into account \citep[e.g.][]{Spitoni2019,Spitoni2023}.This is to be expected, as in the presence of a radial metallicity gradient any perturbation will cause gradients in azimuth, and as a function of stellar kinematics (see \citealt{Frankel2024} for discussion of this effect).

Some weak variation may been seen in large-scale abundance maps across the disc \citep[e.g. with APOGEE DR16 in][]{Eilers2022}, but they are difficult to distinguish from noise and are consistent with a smooth and monotonic increase with $R$. When looking at a more local sample with \Gaia DR3 data, \cite{Poggio2022} find spiral-like features in [M/H] in a sample of bright giants, which correlates with the overdensity of UMS stars around the Sagittarius--Carina, Local, and Perseus spiral arms from \cite{Poggio2021b}. The signal in [M/H] is relatively weak, yet when studying the residuals deviating from the radial metallicity gradient of the Milky Way's disc, they become clear. \cite{Poggio2022} calculate the radial metallicity gradient\footnote{\cite{Poggio2022} find an azimuthally varying radial gradient from $\partial$[M/H]$/\partial R\sim-0.054$ dex kpc$^{-1}$ at $\phi=-20^{\circ}$ to $\partial$[M/H]$/\partial R\sim-0.036$ dex kpc$^{-1}$ at $\phi=20^{\circ}$.} and subtract it to reveal clear spiral-like features in the residuals, $\Delta$[M/H], which follow the stellar overdensity from their UMS sample.

\cite{Hawkins2023} calculate the radial and vertical metallicity gradient\footnote{\cite{Hawkins2023} find $\partial$[Fe/H]$/\partial R\sim-0.078\pm0.001$ dex kpc$^{-1}$, and $\partial$[Fe/H]$/\partial z\sim-0.15\pm0.01$ dex kpc$^{-1}$.} from data from LAMOST and \Gaia DR3, and subtract the radial gradient to reveal clear spiral-like features in the residuals. These residuals, $\Delta$[Fe/H] track azimuthal variations in the stellar metallicities, and they appear to correlate with the stellar overdensity in the UMS sample of \cite{Poggio2021b} for a \Gaia DR3 sample of giant stars, but not the LAMOST sample of O, B, A \& F stars, highlighting differences between different tracer populations.

In a follow-up study, \cite{Hackshaw2024} recalculate the gradients\footnote{\cite{Hackshaw2024} find $\partial$[Fe/H]$/\partial R\sim-0.066\pm0.0004$ dex kpc$^{-1}$, and $\partial$[Fe/H]$/\partial z\sim-0.164\pm0.001$ dex kpc$^{-1}$.} with the \texttt{AstroNN} catalogue, subtract them, and map the azimuthal variations in multiple elements. While [Fe/H] is strongest, the spiral-like azimuthal variations also appear in the $\alpha$-elements (O, Mg, Si, S) at the $\sim0.1$ dex level. They also find differences in the azimuthal variations with $J_{\mathrm{R}}$, eccentricity, and stellar ages.

This shows that they are dynamical features, potentially linked to the kinematic response to the spiral arms, rather than some chemical enhancement from star formation in the arms themselves. This is explained in \cite{Frankel2024}, who show how dynamical perturbations combine with an existing gradient to produce spiral structure in the metallicity residuals (see also Section~\ref{sec:perturbations_phase_spiral_chemistry}). However, \cite{Hackshaw2024} find less of a correlation between the spiral-like features in the chemistry and the stellar density maps \citep[such as from][]{Poggio2021b}. In addition, the azimuthal dependence on the radial gradient found in \cite{Poggio2022} may reflect the effect of the bar on the large-scale metallicity field \citep[see][]{Filion2023}, leaving open questions for future work.

\subsubsection{Growth and disruption of Milky Way spiral structure}  \label{sec:spiralarm_disruption}

\textit{If} spiral structure is transient, the effect it has on stellar kinematics is expected to differ in the growth and disruption phases of the spiral arms \citep[e.g.][]{Baba2013,Grand2014}.

\cite{Asano2024} detect a compression breathing mode in the Solar neighbourhood, which correlates with the location of the Local arm \citep[see also][]{Widmark2022b}. This is consistent with the expectation of a growing transient spiral arm, shown by comparison to a $N$-body simulation. They also make tentative detections of compression and expansion breathing modes around the Outer and Perseus arms, implying that the Perseus arm is disrupting, and the Outer arm is growing.

This is supported by Cepheid kinematics. For example, \cite{Baba2018} use a sample of Cepheids from \Gaia DR1 to show that the kinematics of the Cepheids are consistent with the Perseus arm being in the disruption phase. \cite{Funakoshi2024} follow this up and show that the Cepheids around the Perseus and Outer arms show different kinematic trends in the vertex deviation of the $v_\mathrm{R}$--$v_\phi$ planes either side of the two spiral arms. The vertex deviations are positive around the Perseus arm, and negative around the Outer arm, matching predictions from a simulation for disrupting and growing arms, respectively.

\subsection{Inferring dynamics from disc kinematics}  \label{sec:bar_inference}

While the bar itself is confined to the inner Galaxy, its influence is felt far across the Galactic disc, including the Solar neighbourhood. Similarly, spiral arms, whether quasi-stationary density waves or material winding transients, will also create non-axisymmetric motion across the disc. For example, such kinematic substructure can arise through resonant interaction or phase mixing following some perturbation.

It is known that some stars in the disc will interact with the Galactic bar through a variety of resonances such as the 1:1 Corotation resonance (CR), the inner and outer Lindblad resonances (OLR) and other higher order resonances such as the Ultraharmonic (UHR) resonances, and more. In addition, dynamical theory is able to make clear predictions of how bar resonances should shape kinematic substructure as a function of both bar parameters and position in the Galactic disc \citep[e.g.][]{Contopoulos1989,Fux2001,Bovy2010,Trick2021,Trick2022}, which can then be compared with the data from \Gaia and other stellar surveys. If a kinematic feature is known to arise from a given resonance, one can make precise measurements of the bar parameters.

The main advantage of inferring bar parameters from such local features in the stellar kinematics is that there is a much larger amount of high-quality data in the Solar neighbourhood and nearby disc (especially in the era of \Gaia) than there is available in the Galactic centre, which suffers both from high levels of extinction as a result of the dust, and large uncertainties at high distances.

However, the Galactic disc is not shaped by the bar alone, and other non-axisymmetric structures, such as the spiral arms, can produce qualitatively similar features in the local velocity distribution \citep[e.g.][]{Quillen2011,Monari2016,Hunt2018b,Sellwood2019,Hattori2019} or periodically destroy resonance features \citep[e.g.][]{Fujii2019,Hunt2019}, or create chaotic regions where the bar and spiral resonances overlap \citep[e.g.][]{Quillen2003}. In addition, as discussed in Section~\ref{sec:perturbations}, the Galactic disc appears to be responding to some external perturbation(s), which can also leave a significant imprint in the planar kinematics across the disc.
These additional perturbations make it non-trivial to conclusively associate a resonance to a specific feature, and an incorrect attribution of given feature to a resonance will lead to incorrect measurements of the various bar parameters.

\subsubsection{The $U$--$V$ plane}  \label{sec:bar_uv_plane}

This Section will focus on studies of the $U$--$V$ and $v_{\mathrm{R}}$--$v_\phi$ planes with \Gaia data, while subsequent Sections will extend the discussion across Galactocentric radius, and azimuth (see Section~\ref{sec:bar_rphi}), and into action--angle space (see Section~\ref{sec:bar_action}). As discussed in Section~\ref{sec:observations_coordinates}, the heliocentric velocity $\{U, V, W\}$ differ from the Galactocentric velocity $\boldsymbol v_\text{GSR}$ by the Solar velocity $\boldsymbol v_\odot$, and in the Solar neighbourhood $\boldsymbol v_\text{GSR}$ can be expressed interchangeably in Cartesian or cylindrical coordinates (e.g., for our chosen coordinate system with the Sun being located along the negative $x$ axis, $v_R=-v_x$ is \textit{the} radial velocity, and $v_\phi=-v_y$ is the velocity in the direction opposite to Galactic rotation). However, in more distant regions of the Milky Way the Galactocentric velocity in cylindrical coordinates is preferred, since $U,V$ are defined in the Cartesian frame fixed to the Solar neighbourhood.

The $U$--$V$ plane contains significant kinematic substructure. Since the early studies of Olin Eggen \citep[see][and references within]{Eggen1996}, it has been clear that there exist groups of stars on similar orbits in the Solar neighbourhood. These became known as the `moving groups', as they initially appeared to be a relatively small number of stars with similar kinematics. While initial theories suggested they may be dissolving clusters, several properties (a lack of chemical homogeneity, the sheer quantity of stars in such features, and the fact that one can now trace these groups for several kpc across the disc) have demonstrated conclusively that they have a dynamical origin\footnote{Note that some of the moving groups share their name with star clusters, e.g. Hyades, Pleiades, etc. owing to the historical naming convention. They are not equivalent and should not be confused.}. As such, understanding these substructures can potentially reveal the non-axisymmetric structure of the disc.

\begin{figure*}[t]
\includegraphics[trim={0 7cm 0 0.6cm},clip,width=\linewidth]{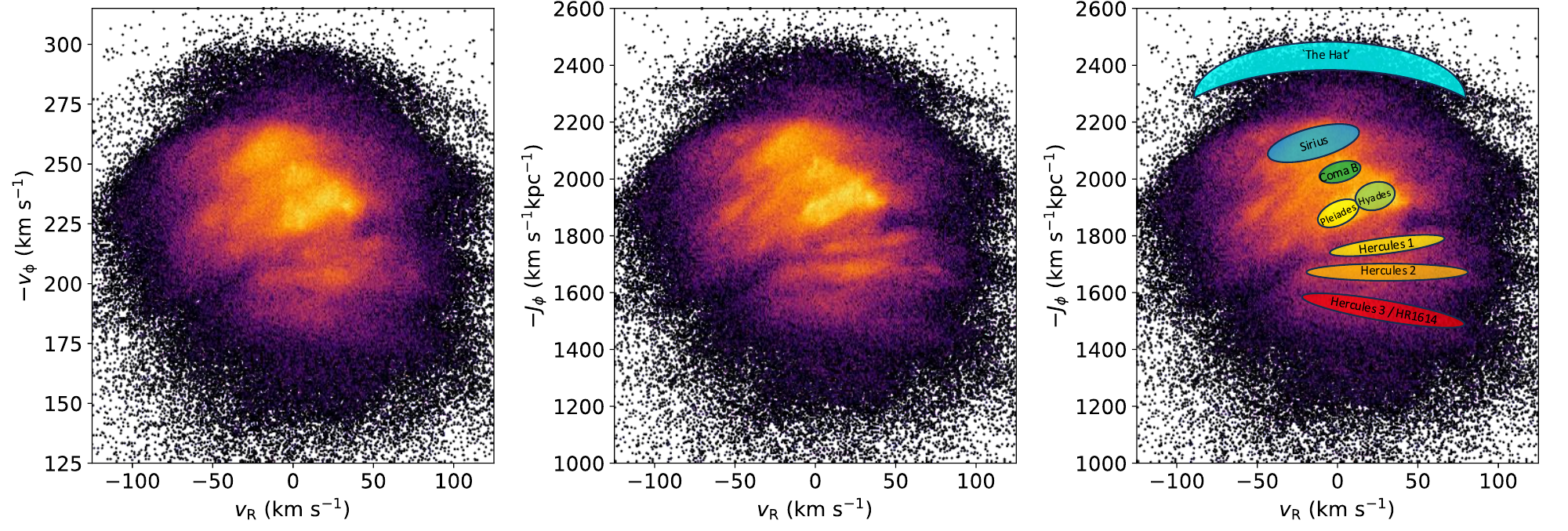}
\caption{$v_{\mathrm{R}}$--$v_\phi$ plane (left), $v_{\mathrm{R}}$--$L_z$ plane (middle, where $L_z\equiv R\, v_\phi$), and the same plane annotated with the locations of the classical moving groups and the `Hat' (right), for \Gaia DR3 stars with $d<200$ pc, and a fractional parallax error of $<10\%$.}
\label{fig:UVplane}
\end{figure*}

Figure~\ref{fig:UVplane} shows the $v_{\mathrm{R}}$--$v_\phi$ plane (left) for \Gaia DR3 stars with $d<200$ pc, and a fractional parallax error of $<10\%$. The middle panel shows $v_{\mathrm{R}}$--$L_z$ plane, and the right panel annotates it with the locations of the classical moving groups and the `Hat'; these are approximate `by hand' selections purely to aid the reader in following discussion. Note that the moving groups are `sharper' as a function of $L_z$, which is an orbit label as opposed to the instantaneous velocity, $v_\phi$. Note also that with \Gaia DR2 and beyond, there are many smaller moving groups visible which have been selected and mapped in other studies \citep[e.g.][and references within]{Gaia2018a,Ramos2018,Bernet2022,Lucchini2023,Medan2023}. It also becomes clear that the classical moving group known as Hercules is comprised of multiple sub-features which were not resolvable in earlier data. We mark them as Hercules 1 and 2, and Hercules 3 / HR1614 to be consistent with choices in the literature. What is not yet conclusive is the assignment of a specific dynamical process to each moving group. Dynamical explanations include trapping by bar resonances, trapping by spiral arm resonances, perturbation from transient spiral structure, or phase mixing following some perturbation.

The orbital dynamics of stars trapped around bar resonances are well known in the literature \citep[see e.g.][]{Contopoulos1989}. These orbit families have some characteristic orientation in the Solar neighbourhood, which is dependent on the bar angle with respect to the Sun--Galactic Centre line. \cite{Dehnen2000} showed that the Outer Lindblad Resonance (OLR) of a bar with pattern speed $\Omega_{\mathrm{b}}=1.85\pm0.15$ times the local circular frequency reproduces major features in data from \Gaia's predecessor, \textit{Hipparcos} \citep{Hipparcos}. This model qualitatively reproduces the Hercules moving group which is comprised of orbits belonging to the $x_1$(1) orbit family, the elongated inwards moving feature sometimes referred to as `the horn', which is comprised of the $x_1$(2) orbit family, and the gap between them \citep[see][for various illustrations of the orbit families]{Contopoulos1989,Fragkoudi2019,Hunt2019,Trick2021}.

This started a trend of attempting to measure bar parameters from the Hercules stream (and other moving groups) which persists into the \Gaia era. Initial studies (pre- and post-\Gaia DR1) predominantly followed \cite{Dehnen2000} in assuming that Hercules results from trapping around the OLR\footnote{See also \citealt{Fux2001} for an explanation involving chaos around the OLR}, supported by contemporary measurements of a fast pattern speed from gas kinematics, \citep{Englmaier1999} and thus measure similar `fast' pattern speeds with some small scatter \citep[e.g.][]{Antoja2014,Monari2017b}.

Alternatively, \cite{PerezVillegas2017} showed that the Hercules moving group as seen in a combination of \Gaia TGAS, RAVE and LAMOST could also be created by the corotation resonance (CR) of a long slow bar with pattern speed $\Omega_{\mathrm{b}}=39$ \kmskpc, where stars librating around the L4 Lagrange point in an $N$-body/M2M model from \cite{Portail2017} cause a kinematic overdensity similar to the Hercules moving group in the Solar neighbourhood \citep[see also][]{DOnghia2020}. However, this model did not reproduce the underdensity (or gap) between Hercules and the Hyades/Pleiades moving groups, and subsequent models using perturbation theory or the `Backwards integration technique' of \cite{Dehnen2000} found that the CR of a steadily rotating, purely quadrupole bar model produced a weaker feature than seen in the \Gaia data \citep[e.g.][]{Monari2017a,Binney2018,Hunt2018a}.

This was subsequently reconciled with the data either by: i) the addition of spiral structure which can couple with the CR to produce a distinct `gap' \citep[e.g.][]{Hunt2018b,Michtchenko2018b,Hattori2019}, ii) by including higher order bar components and resonances \citep[e.g.][]{Hunt2018a,Monari2019a,Asano2020,Khalil2024}, or iii) by including the slowdown of the Galactic bar \citep{Chiba2021a,Chiba2021b}.

Galactic bars are expected to slow down through transfer of angular momentum to the dark matter halo owing to dynamical friction \citep{Debattista1998,Athanassoula2003,Sellwood2014a}. The pioneering study of \cite{Fux2001} demonstrated that a slowing bar leaves a different imprint in local kinematics compared to a rigidly rotating potential in the context of the OLR. \cite{Chiba2021a} showed that the CR of a slowing bar naturally reproduces the kinematic overdensity of the Hercules stream and the gap between Hercules and Hyades/Pleiades significantly better than the CR of a steadily rotating bar when taking into account this expected dynamical behaviour (although they only consider a simple quadrupolar bar).

\cite{Chiba2021a} also show that the rate of slowing affects the amplitude of the feature, enabling them to make an estimate of the current slowing rate $\dot{\Omega }_{\text p}= (-4.5 \pm 1.4)$\, \kmskpc\,Gyr$^{-1}$ when comparing their models to the \Gaia data (although they note there is likely an additional systematic uncertainty from the influence of spiral structure, which is not modelled). \cite{Chiba2021b} follow this up by investigating the chemistry of stars within the Hercules stream, and comparing them to models of the decelerating bar, which predicts a `tree ring' structure, where the more metal-rich stars transported from further within the Galaxy will lie at the centre of the trapped region. They find a best-fitting pattern speed of $\Omega_{\mathrm{b}}=35.5\pm0.8$ \kmskpc, which is consistent with some other independent estimates across a variety of methods (see Figure~\ref{fig:BarSpeedsSummary}).

The inclusion of the bar slowdown is an important improvement on prior studies, which assume a fixed pattern speed, demonstrating that the inclusion of expected dynamical behaviour leads to better agreement with observed kinematics. It is also expected that the rate of slowdown decreases with time, and that the presence of gas inflow to the bar can reduce the rate with which a bar slows, as the gas imparts angular momentum to the bar \citep[e.g.][]{Beane2023}. Thus, more complex models, including self-gravity, spiral structure, and gas physics, are needed in order to determine whether the specific measurement of the slowdown rate is robust in the presence of these other complicating factors.

\Gaia DR2 and later releases significantly increased the resolution of the $U$--$V$ plane map, such that many new moving groups were discovered \citep[e.g.][]{Ramos2018,Lucchini2023,Mikkola2023}. Some but not all of these are `arches' of constant energy in the $U$--$V$ plane which follow expectations of phase mixing following some perturbation \citep[as long known from e.g.][]{Minchev2009,Gomez2012a}.

The Hercules stream was also shown to consist of multiple branches, which was not clear in older survey data. This becomes more clear when examining the local kinematics in action space as discussed in Section~\ref{sec:bar_action}. \cite{Asano2020} find that this multicomponent Hercules stream is reproduced in a very high resolution $N$-body barred galaxy simulation \citep[MWa from][]{Fujii2019} through a combination of the 4:1 and 5:1 resonance \citep[see also][for a model reproducing a Hercules-like feature with the 4:1 UHR]{Hunt2018a}. If this is the case for the Milky Way, it would correspond to a pattern speed of around $\Omega_{\mathrm{b}}\approx40$--45 \kmskpc. These higher order resonances can have a significant effect on the $U$--$V$ plane in general, explaining other moving groups beyond Hercules \citep[e.g.][]{Monari2019a} for a bar with structure beyond the quadrupole.

Spiral structure is also expected to have a significant effect on the $U$--$V$ plane, as shown in various pre-\Gaia studies \citep[e.g.][]{Fux2001,Quillen2003,DeSimone2004,Chakrabarty2007}. It should not be neglected in the modelling of local kinematics, and various models of spiral structure can potentially explain various aspects of the local kinematic substructure. For example, \cite{Quillen2018} link the boundaries in the kinematics between the moving groups in \Gaia DR2 to separate stars that have recently crossed a spiral arm, and thus experienced a different perturbation, from those that have not. They argue that the high degree of substructure is consistent with the spiral structure of the Milky Way disc being flocculent rather than grand design, containing multiple spiral patterns with different pattern speeds.

\cite{Hunt2018b} show similarly that the highly structured nature of the $U$--$V$ plane is well explained by transient spiral structure, using the `Backwards Integration' method of \cite{Dehnen2000}. Recent transient spiral arms with a lifetime of $\lesssim250$ Myr create both the division between moving groups, the vertex deviation of the $U$--$V$ plane, and some of the `ripple' or `arch' features which result from phase mixing of the stars back towards equilibrium once the arm disrupts. This process is described analytically in \cite{Binney2020c}. By combining the transient spiral arm model with a long slow bar ($R_{\mathrm{b}}$=5 kpc, $\Omega_{\mathrm{b}}$=35.75 \kmskpc), they reproduce the $U$--$V$ plane across 16 spatial bins within 1 kpc from the Sun for \Gaia DR2, including the multi-modal nature of the Hercules stream.

Conversely, a series of papers from \cite{Michtchenko2018a,Michtchenko2018b}, \cite{Barros2020} find that long-lived density wave-like spiral arms also reproduce well the $U$--$V$ plane. \cite{Michtchenko2018a} show the analytic prediction for resonance features in the $U$--$V$ plane arising from density wave-like spiral arms, before comparing them to the \Gaia data in \cite{Michtchenko2018b}. They find a good qualitative fit to the moving groups in the Solar neighbourhood for a $N$=4 spiral pattern with $\Omega_{\mathrm{sp}}$=28.5 \kmskpc. \cite{Barros2020} then verify this solution with a test-particle model. They initialise their model with stars `born' in a $N$=4 spiral pattern, and find that the $U$--$V$ plane is reproduced well within 1~Gyr of evolution, by trapping stars around spiral arm resonances. As with the transient arm model, the density wave-like spirals (where $\Omega_{\mathrm{sp}}=28.5$ \kmskpc) reproduce the divisions between moving groups, the overall vertex deviation and the multi-modal nature of the Hercules stream which is created through a combination of higher order resonances (mainly the 8:1 and 12:1 ILR). Notably, the model of \cite{Barros2020} does not contain a bar, yet still reproduces clear `Hercules-like' and `Hat-like' features often used to measure the bar pattern speed in the data.

Alongside the dynamics, the combination of \Gaia with complementary spectroscopic surveys makes it possible to explore the stellar label dependence of the moving groups. This is discussed more for the $R$--$v_\phi$ plane in Section~\ref{sec:bar_rphi} below. \cite{Medan2023} explore the $U$--$X_\mathrm{mix}$ plane (where $X_{\mathrm{mix}}=R_{\mathrm{g}}\cos{\phi}$) with a sample of $2.5\times10^{5}$ dwarf stars in the Solar neighbourhood with data from \Gaia DR3 and GALAH, showing distinct chemistry and ages for the moving groups compared to field stars.

While most studies of the $U$--$V$ plane are restricted to samples of stars with 6D phase space information, \cite{Mikkola2023} extend the sample to stretch further from the Solar neighbourhood by estimating the 3D velocities for stars with 5D phase space information in \Gaia DR3 using a maximum penalised-likelihood algorithm \cite[as described in][]{Mikkola2022}. This increases the number of stars present in previously low number moving groups, and also reveals new features.

\subsubsection{Kinematic substructure across Galactic radius R and azimuth $\phi$}  \label{sec:bar_rphi}

The increased spatial coverage of the sample of stars with 6D phase space information in \Gaia allowed studies to start examining and mapping kinematic features in 6D as a function of Galactocentric radius $R$, angular momentum $J_\phi$ and Galactic azimuth $\phi$.

While DR2 and beyond were revolutionary for studies of kinematics across the disc, already the DR1 catalogue made it possible to start to explore comparatively local kinematics just beyond the Solar neighbourhood. For example, \cite{Monari2017b} used a combination of \Gaia DR1 (TGAS) and LAMOST DR2 to track the Hercules stream for $8.1\lesssim R\lesssim8.8$ kpc, and show that its behaviour in $v_\phi$ as a function of $R$ matches predictions for the OLR of the Galactic bar. \cite{Hunt2017} analysed the distribution of $R$--$v_l$\footnote{$v_l=4.7047\mu_{l*}/\varpi$, where $\mu_{l*}$ is the PM in the longitude direction in true arc} for stars in TGAS, where $v_l$ is a proxy for $v_\phi$ along the $l=0^{\circ}$ and $l=180^{\circ}$ lines-of-sight (see Section~\ref{sec:observations_velocity_measurement}). They discovered a `fast moving' group of stars which has since become known as the `Hat'. \cite{Hunt2017} linked it to the influence of the Perseus spiral arm, although much of the literature now considers the `Hat' to arise from the OLR of a long slow bar.

\begin{figure}[t]
\includegraphics[width=\linewidth]{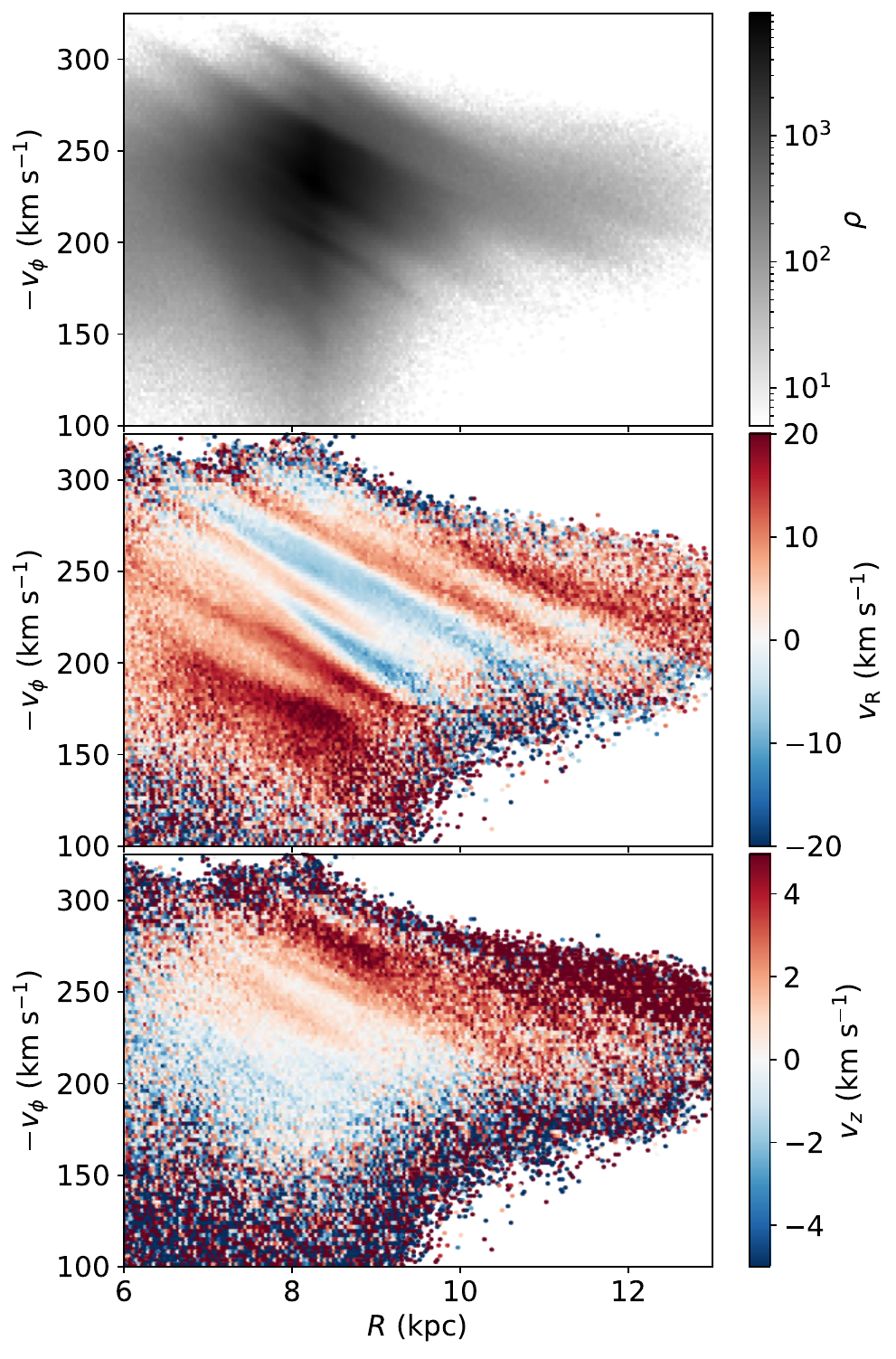}
\caption{$R$--$v_{\phi}$ plane from \Gaia DR3 in in number density (upper panel), mean radial velocity (middle panel) and mean vertical velocity (lower panel), for a sample of stars with fractional parallax error $<10\%$ and $|\phi|<0.1$ rad from the Sun Galactic centre line. Ridges are clearly visible in the density in the top panel, with coherent velocity signatures in the middle and lower panel. Reproduced following many papers in Section~\ref{sec:bar_rphi}.} 
\label{fig:rvphi}
\end{figure}

\Gaia DR2 conclusively confirmed that the classical moving groups are not local features, but instead stretch across the Galactic disc over several kpc \citep{Antoja2018,Kawata2018}. Both \cite{Antoja2018} and \cite{Kawata2018} show the moving groups create ridges or ripples in the stellar number density in $R$--$v_\phi$, over approximately $4\lesssim R\lesssim12$ kpc. Figure~\ref{fig:rvphi} shows the distribution of stars from \Gaia DR3 in $R$--$v_\phi$ in number density (upper panel), mean radial velocity (middle panel) and mean vertical velocity (lower panel), with ridges clearly visible in each panel. Both studies point out that some of these ridges likely originate from resonant interaction with the Galactic bar. However, \cite{Antoja2018} suggest that the majority of them may arise from phase mixing following the impact of the Sagittarius dwarf galaxy (see also Section~\ref{sec:perturbations}), while \cite{Kawata2018} suggest that many of them may arise from the motion of stars around transient spiral structure. \citet{Gaia2021d} then extended the detection of the ridges out to $R\approx18$ kpc with \Gaia EDR3, again using the relation that $v_l\sim v_{\phi}$ along the $l=180^{\circ}$ line-of-sight (the anticentre), detecting two new ridges in the outer disc.

\cite{Fragkoudi2019} showed that the OLR of the galactic bar creates the `longest ridge' in the $R$--$v_\phi$ plane of an $N$-body simulation, and that the OLR also creates a `flip' between positive and negative radial velocity in the ridge owing to the transition between the orientation of the $x_1$(1) and $x_1$(2) orbits (see \citealt{Contopoulos1989} for definition). Both these features may help identify the OLR in the data. \cite{Fragkoudi2019} link this to the Hercules stream in the Solar neighbourhood, with orbits librating around the $x_1$(1) and $x_1$(2) orbital families creating the Hercules stream, while orbits librating around the $x_1$(1) family also create the `horn' as seen in the $U$--$V$ plane. They state that the CR of the bar \textit{alone} cannot reproduce the Hercules/Horn split seen in the DR2 data, but that spiral structure or an external perturbation in concert with a long slow bar may be required in order to fit observations.

\cite{Monari2019a} show that the long slow bar model from \cite{Portail2017} creates numerous ridges in $R$--$v_\phi$ owing to the higher-order resonances. These higher-order resonances can then be linked to multiple moving groups, although the model of \cite{Monari2019a} does not match the `radial velocity flip' from positive to negative $v_{\mathrm{R}}$ seen in the data. \cite{Asano2020} reproduce a multicomponent `Hercules-like' ridge from the combination of the 4:1 and 5:1 resonances, while the 2:1 OLR then creates the `hat' at higher angular momentum. These ridges from higher-order resonances exist only for a bar with higher-order structure rather than a simple quadrupole ($m=2$) used in some other works. While the detailed structure of the Milky Way bar is still not firmly established, it is expected to contain such higher-order components \citep[as seen for example in the model of][fitted to IR survey data in the inner Galaxy]{Portail2017}, although there is variation in amplitude of the individual harmonics in the observed population of barred galaxies \citep[e.g.][]{Buta2006}.

\cite{Hunt2018b} show that transient recurrent spiral structure naturally creates clear ridges and ripples in the $R$--$v_\phi$ plane, which qualitatively matches the \Gaia data using the backwards integration method to construct the $R$--$v_\phi$ plane from $4<R<12$ kpc. The inclusion of a long slow bar then reproduces both the multi-modal Hercules stream via a combination of the bar CR and the spiral perturbation, and the OLR is associated with the `longest ridge' as shown in \cite{Fragkoudi2019}, corresponding to the `hat' feature in the $U$--$V$ plane. \cite{Hunt2019} extend this to compare eight models of bars and spirals, alone or combined, and find that both a long slow bar and a short fast bar can qualitatively reproduce the moving groups in the $R$--$v_\phi$ plane, while the `bar only' and `bar + density wave spiral arms' are a poor match to the \Gaia DR2 data.

The best-fitting model from \cite{Hunt2019} consists of a long slow bar ($R_{\mathrm{b}}=5$ kpc, $\Omega_{\mathrm{b}}=35.75$ \kmskpc) combined with a series of recurrent transient spiral arms. However, the important result is not the best fit, but rather that the combination of bar resonances and transient spiral structure does an excellent job of reproducing the overall shape and amplitude of the ridges and ripples, while the time-varying nature of the spiral arm potential and the non-linear dynamics in regions of interaction make it very difficult to get a \textit{quantitative} best fit. The shape, amplitude and radial velocity dependence of the moving groups vary on the order of a few Myr, and the perturbation from a passing transient spiral arm can temporarily `shift' the velocity signature expected from a resonance away from the expected location in kinematic space. \cite{MartinezMedina2019} also show how the combination of a bar and transient spiral arms naturally lead to ridges in ripples in the $R$--$v_\phi$ plane in $N$-body simulation, and that these in turn influence estimates of the rotation curve, in the Milky Way and external galaxies. \citet{Khalil2024} used the backward integration method and tuned the parameters of the bar and spiral arms to fit the local $U$--$V$ plane as well as other kinematic features across the disc. Their fiducial model has $\Omega_{\mathrm{b}}=37$ \kmskpc and two sets of spiral modes with slower pattern speeds.

\cite{Khanna2019} also map the $R$--$v_\phi$ plane and show that the ridges are clearly visible in density, $\overline{v}_\mathrm{R}$, $\overline{v}_z$, $\overline{z}$. They compare the data to a toy phase mixing simulation, an isolated $N$-body galaxy, and two $N$-body simulations with a $5\times10^{10}$ and a $10^{11}$ $M_{\odot}$ perturber. They show the toy phase mixing simulation reproduces the ridges (and arches in the $U$--$V$ plane; see Section~\ref{sec:bar_uv_plane}), as does the more realistic isolated $N$-body simulation, which causes phase mixing following the transient spiral arm perturbations. \cite{Khanna2019} and \cite{Hunt2019} find the same response in the $R$--$v_\phi$ plane to transient spiral structure, despite the different techniques used. The two merger simulations of \cite{Khanna2019} also create ridges in $R$--$v_{\phi}$, with their amplitude linked to the mass of the perturber.

\cite{Laporte2019b} examine the ridges in $R$--$v_\phi$ in the \Gaia DR2 data, in overdensity, $\overline{v}_\mathrm{R}$ and $\overline{v}_z$. They compare the data to an $N$-body simulation of a Sagittarius-like merger \citep[Model L2 from][]{Laporte2018} and show that the merger naturally reproduces ridges and ripples in $R$--$v_\phi$. \cite{Laporte2020b} then extend the detection of ridges to lower $v_{\phi}$ with \Gaia DR2 and the \cite{Sanders2018} catalogue, finding they are regularly separated by $\sim18$--20 \kms, matching predictions of phase mixing \citep[e.g.][]{Minchev2009}. The explanation that some of the ridges are caused by an external perturbation is supported by the finding that those further out in the disc show coherent signatures in $\overline{z}$, $\overline{v}_z$ and $\sigma_z$ \citep[e.g.][]{Laporte2019b,Khanna2019,Gaia2021d,McMillan2022}. External perturbations to the disc are discussed in more detail in Section~\ref{sec:perturbations}.

\cite{Antoja2022} examine the ridges in $R$--$v_\phi$ plane as seen by astrometry from \Gaia DR3 and distances from \texttt{StarHorse}. They compare the data with a series of models of tidally induced spiral structure, and again show that they create a series of ridges in $R$--$v_\phi$ in both density and $\overline{v}_\mathrm{R}$. The test particle model shows a very distinctive `saw-tooth' pattern as the tidal arms phase mix (which can be used to time the perturbation, as discussed further in Section~\ref{sec:perturbations_feathers}), but becomes messier in the more realistic $N$-body simulations.

Similar to the $U$--$V$ plane, \cite{Michtchenko2018a,Michtchenko2018b} and \cite{Barros2020} show that density wave-like spiral arms also reproduce ridges in the $R$--$v_\phi$ plane, which are consistent with those seen in \Gaia DR2. The higher-order resonances from the fixed spiral pattern create several ridges across the disc, both alone \citep{Barros2020} and in combination with a bar model \citep{Michtchenko2018b,Khalil2024}.

Along with density and radial velocity, the ridges are visible in stellar labels, such as chemistry and stellar ages. \cite{Khanna2019} show that the ridges are clearly visible in [$\alpha$/Fe] and [Fe/H] using \Gaia DR2 in combination with GALAH, while \cite{Wang2020c} and \cite{Wheeler2022} show the same for {\Gaia}+LAMOST DR4 and DR5 respectively. \cite{Wang2020c} map the ridges in $R$--$v_{\phi}$ as a function of stellar age $\tau$, and show that they are present across all ages. However, while some ridges maintain their shape and location in $L_z,R,v_\phi$, some do not, implying two different dynamical effects in place. In particular, the angular momentum of the Sirius moving group appears to change with age. While other studies have already shown that Sirius is not constant in angular momentum \citep[e.g.][]{Ramos2018}, future models of its origin may also benefit from including the age dependence. \cite{Laporte2020b} show that the ridges also appear coherent when plotted directly as a function of $\tau$ and $\Delta\tau$, and that Hercules, Sirius and the `hat' are predominantly younger than the field stars.

Thus, several studies show that ridges in $R$--$v_\phi$ naturally arise from several dynamical mechanisms; the bar, long-lived density wave \textit{or} transient winding spiral arms, or through the influence of an external perturbation.

Following such qualitative explanations, numerous groups have applied various techniques to track the moving groups, and quantitatively determine their origin. In an ideal case, it is expected that resonant kinematic structure should follow lines of constant angular momentum, and that phase-mixing structures should follow lines of constant energy in $R$--$v_\phi$ (although these are dependent on the choice of galactic potential). Resonant structures should also follow specific gradients in azimuth, so that one can theoretically distinguish between different resonant origins for moving groups by mapping them across the disc in both $R$--$v_\phi$ and $\phi$--$v_\phi$. Mapping methods range from simply fitting a line to the visual slope of a group in some parameter space \citep[e.g.][]{Monari2019b}, to performing wavelet transforms on the overdensities \citep{Ramos2018,Bernet2022,Lucchini2023,Lucchini2024,Bernet2024}, or instead tracking the gaps between the moving groups \citep{Contardo2022}.

\cite{Monari2019b} show the moving groups in $L_z$--$\phi$ in \Gaia DR2 as a function of $v_{\mathrm{R}}$, for $1200\lesssim L_z\lesssim2500$ kpc\,\kms and $-15^\circ\lesssim\phi\lesssim15^\circ$. They show how a negative gradient is expected for the CR, while the OLR should be flat in this projection, for a fixed $R$. They then fit the apparent gradient of the Hercules stream in this space, and find a negative slope of $-8$ kpc\,\kms\,deg$^{-1}$, providing evidence for a long slow bar. They also note that other ridges have significant slopes, with possible links to higher order resonances, or links to spiral structure, or an external perturbation to the disc, and that further coverage in Galactic azimuth is required.

\cite{Ramos2018} use wavelet transforms to map the ridges in $R$--$v_\phi$ and $\phi$--$v_\phi$ over $6\lesssim R\lesssim11$ kpc and $-15^\circ\lesssim\phi\lesssim15^\circ$ in \Gaia DR2, and show that some follow lines close to constant angular momentum (e.g. Hercules), while others follow lines of constant energy (e.g. Sirius). This suggests that Hercules would arise from a bar resonance, while Sirius would arise from phase mixing. They favour a short fast bar model linking Hercules to the OLR, estimating a pattern speed of $\Omega_{\mathrm{b}}=54$ \kmskpc.

\cite{Lucchini2023} also use wavelets to track the ridges in $R$--$v_\phi$ for $6.5\lesssim R\lesssim10$ kpc and $-15^\circ\lesssim\phi\lesssim15^\circ$. They also find numerous new structures in kinematic space (see within for a list), and find that the data is most \textit{qualitatively} consistent with a long slow bar, where Hercules originates from the CR, on account of the change in the number of stars in Hercules with $R$. This is extended in \cite{Lucchini2024} to map the change in Hercules with azimuth, where they again find in favour of a long slow bar ($\Omega_{\mathrm{b}}\sim40$ \kmskpc) owing to the change in strength of Hercules with $\phi$.

\cite{Bernet2022,Bernet2024} use the wavelet transforms on the \Gaia EDR3 and DR3 data respectively, tracking the moving groups across $6\lesssim R\lesssim14$ kpc and $-40^\circ\lesssim\phi\lesssim40^\circ$. In both works they find that the ridges in $R$--$v_\phi$ do not exactly follow lines of constant angular momentum or energy over large distances. They show instead that the spatial evolution of the moving groups in the data is complex, requiring more advanced theoretical predictions. They also show the same in simulations, where the kinematic ridges deviate from the expected lines. \cite{Bernet2024} find that the data are inconsistent at $3\sigma$ with simple predictions from both a long slow bar model and a short fast bar model, implying that more complex dynamics from spiral structure, bar spiral coupling, or incomplete phase mixing following some other perturbation need to be taken into account in future modelling.

Alternately, one may track the gaps between moving groups, instead of the groups themselves. \cite{Contardo2022} introduce a method to measure the under-densities in a data space
, and illustrate its use on the \Gaia DR2 data. While they primarily illustrate the gaps in the $U$--$V$ space, they are able to trace how those gaps change with $R$ and $\phi$, both via a series of individual slices, or 3D data sets. The decrease in $v_\phi$ with increasing $R$ is clear as expected, while the change in $v_\phi$ with $\phi$ shows non-linear features. For instance, the `gap' below the Sirius moving group is almost constant in $v_\phi$ for $v_{\mathrm{R}}<0$ \kms, yet changes significantly for $v_{\mathrm{R}}>0$ \kms, where the location of the gap decreases in both $v_\phi$ and $v_{\mathrm{R}}$ with increasing $\phi$. Smaller fluctuations are visible in the gaps between other moving groups.

Coma Berenices alone among the classical moving groups shows a significant vertical asymmetry \citep[e.g.][]{Monari2018,Quillen2018,Bernet2022,Mikkola2023} and coherent vertical motion \citep[e.g.][]{Khanna2019,Bernet2022}, although the groups, or ridges further out in the disc, all appear to have coherent vertical motion. This hints at an origin linked to the vertical perturbations discussed in Section~\ref{sec:perturbations}. While the wavelet analysis of \cite{Lucchini2023} find Coma Berenices extends for less than 0.5 kpc in $R$, \cite{Bernet2022,Bernet2024} track Coma Berenices for $\sim5$ kpc, showing that it is also a large-scale dynamical feature.

There is also significant kinematic substructure in the $R$--$v_{\mathrm{R}}$ space. For example, \cite{Eilers2020} show the large-scale $v_{\mathrm{R}}$ pattern seen in the left panel of Figure~\ref{fig:disc_kinematics} appears wave-like when projected into $R$--$v_{\mathrm{R}}$ space, with the amplitude and wavelength of the pattern increasing with $R$. This is also clearly shown in a combination of \Gaia DR2 and LAMOST \citep{Xu2020} and in the \Gaia Cepheids \citep[e.g.][]{Zhou2024}.

\cite{Friske2019} also mapped the $R$--$v_{\mathrm{R}}$ space over a small range of radii in \Gaia DR2, discovering a wave-like pattern in $v_{\mathrm{R}}$ as a function of guiding radius $R_{\mathrm{G}}$, which has higher wavelength than the wave in $R$. By tracking changes in the wave with Galactic azimuth, they argue that the pattern is consistent with a trailing spiral pattern with an $m=4$ wavenumber. This `radial velocity wave' is also another projection of the classical moving groups (as discussed in Section~\ref{sec:bar_uv_plane}), and aspects of it can be explained by a galactic bar \citep[e.g.][]{Monari2019b} or transient spiral structure \citep{Hunt2020}. There is also a similar `wave-like pattern' in the vertical velocities, and both the radial and vertical patterns are discussed further in Section~\ref{sec:perturbation_waves}.

\subsection{Substructure in the action--angle space}  \label{sec:bar_action}

Alternatively to physical coordinates, one can map substructures in orbit space, using a canonical coordinate system of actions $\boldsymbol J$ and angles $\boldsymbol \theta$, see Section~\ref{sec:observations_orbits_integrals}. The advantage of actions being integrals of motion is somewhat offset by the need to assume a Galactic potential for computing them. The commonly used St\"ackel fudge method \citep{Binney2012} relies on the potential being axisymmetric, produces an approximation to the true actions (which is usually good enough), and is only valid for non-resonant orbits (although actions can be formally defined and calculated for regular orbits around resonances, see e.g. \citealt{Binney2020a}).
For example, \cite{Debattista2024} shows that radial actions computed with the axisymmetric St\"ackel fudge are overestimated in areas of high density compared to the azimuthal average (e.g. in spiral arms) and underestimated in areas of low density (e.g. in the interarm region), with correlated errors up to 100 kpc\,\kms. In addition, the assumption of a potential with a fixed mid-plane can lead to biases in the vertical actions, angles and frequencies when the galaxy is experiencing a vertical perturbation (as seen in the \Gaia data; see Section~\ref{sec:perturbations}), as shown in \cite{Beane2019}.
Nevertheless, they remain a powerful tool for exploring and understanding stellar dynamics in the disc in the \Gaia era even when used to investigate non-axisymmetric perturbations, provided that one is aware and mindful of the caveats and biases involved.

While pre-\Gaia studies have used actions and angles to analyse the orbit structure of stars in the Solar neighbourhood \citep[e.g.][]{Sellwood2010,McMillan2011}, the dramatic increase in the quantity and quality of 6D phase space information with the advent of \Gaia DR2 revealed significantly more detail. \cite{Trick2019} calculated the distribution of orbital actions $(J_{\mathrm{R}}, J_\phi, J_z)$ for almost 4 million stars within 1.5 kpc of the Sun. They show that the action space is highly structured, and that the structures are asymmetric in $v_{\mathrm{R}}$, with the denser regions having significant radial motion. They link many of the `ridges' and `clumps' in action space to the classical moving groups \citep[e.g. see][]{Eggen1996}, and suggest that multiple resonance features are present in the Solar neighbourhood.
\cite{Sellwood2019} and \cite{Hunt2019} showed that the conjugate angles ($\theta_{\mathrm{R}}, \theta_{\phi}, \theta_z$) also contain significant substructure, in line with expectations from some combination of bar and spiral arm resonances and ongoing phase mixing in the Milky Way disc.

\begin{figure*}
\includegraphics[width=\linewidth]{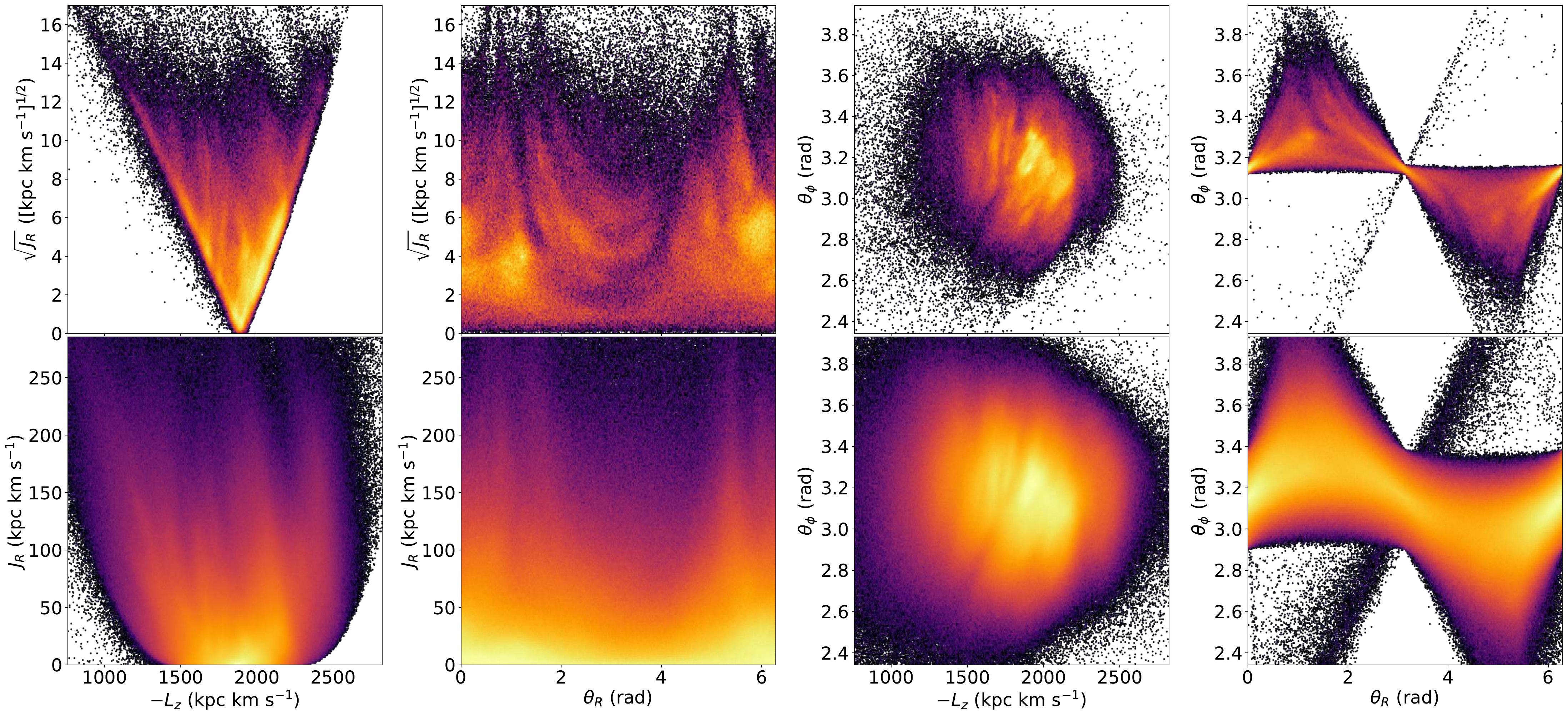}
\caption{Angular momentum vs radial action (left column), radial angle vs radial action (middle left), angular momentum vs azimuthal angle (middle right), and radial angle vs azimuthal angle (right column) for stars with radial velocities and a fractional parallax error of $<10\%$ from \Gaia DR3, within 200 pc (upper row) and within 2 kpc (lower row). The local sample (upper row) displays the square root of the radial action, $\sqrt{J_{\mathrm{R}}}$, to highlight low $J_{\mathrm{R}}$ structure, whereas the lower row displays $J_{\mathrm{R}}$ \citep[following][updated for \Gaia DR3]{Trick2019,Sellwood2019,Hunt2019}.} 
\label{fig:actionangle}
\end{figure*}

Figure~\ref{fig:actionangle} shows the planar actions and angles as calculated from \Gaia DR3 in \texttt{MWPotential2022} from \texttt{gala}, for stars with line-of-sight velocities and a fractional parallax error $\sigma_\varpi/\varpi<10\%$. The top row shows the rich substructure in the orbits of stars in the Solar neighbourhood ($d<200$ pc), and the choice to display the square root of the radial action, $\sqrt{J_{\mathrm{R}}}$, highlights structure at low radial action, in particular the classical moving groups \citep[see][for a detailed explanation of the moving groups in action--angle space]{Trick2019,Hunt2019}.

Note also that the middle right column showing $J_\phi$--$\theta_{\phi}$ looks extremely similar to the classic $v_{\mathrm{R}}$--$v_{\phi}$ plane for the Solar neighbourhood \citep[as explained in][]{Hunt2020}, with the classical moving groups clearly identifiable.

The lower row of Figure~\ref{fig:actionangle} shows the same as the upper row, but for stars within 2 kpc, and displaying the radial action simply as $J_{\mathrm{R}}$. This highlights the more eccentric orbits, and has the benefit that in this projection, resonances will appear as straight lines in $J_\phi$--$J_{\mathrm{R}}$ \citep[see figure 5 of ][]{Sellwood2010}, with the location and slope of such features informing about the properties of the galactic bar or spiral structure.
For example, \cite{Monari2019a} show that the long slow bar model of \cite{Portail2017}, with $\Omega_{\mathrm{b}}=39$ \kmskpc (which contains significant structure beyond the simple $m=2$ quadrupole) induces ridges in the $J_\phi$--$J_{\mathrm{R}}$ distribution caused by the 2:1 (OLR), 3:1, 4:1 and 6:1 resonances, which are qualitatively consistent with the \Gaia DR2 data.

\cite{Trick2021} examine the effect of a purely quadru\-polar bar on the planar action distribution ($L_z$ vs.\ $J_{\mathrm{R}}$), and find that known transition from positive to negative $v_{\mathrm{R}}$ around the OLR owing to the transition between the orientation of the $x_1(1)$ and $x_1(2)$ orbits \citep[see e.g.][]{Contopoulos1989} is clearly seen in this space. They also find a distinctive gradient in $J_z$ around the OLR in their models. By comparing both the `$v_{\mathrm{R}}$ flip' and the $J_z$ gradient to data from \Gaia DR2, they find that the `flip' is present in three places, between Hercules/Hyades, between Hyades/Sirius, and in the `hat'. Thus, they find the data is consistent with bars with pattern speed of 51--52 (short/fast), 45--48 (intermediate), and 33--36 \kmskpc (long/slow), with a weak dependence on the potential.

In a follow-up study, \cite{Trick2022} explore the planar angle distribution ($\theta_{\mathrm{R}}, \theta_{\phi}$), along with a careful analysis of the effect of the selection function, which is particularly important in studies of the conjugate angles. They show that the same orbit orientation flip around the OLR produces an asymmetric feature in the $\theta_{\phi}$--$\theta_{\mathrm{R}}$ distribution, such that differencing the distribution reflected around $\theta_{\mathrm{R}}=0$ and $\theta_{\phi}=\pi$ leaves a coherent ridge, which is resolvable in the data from \Gaia DR2. By examining a range of possible pattern speeds, they find that the strongest candidate from the angle distribution alone is 38.5 \kmskpc, but weaker matches are found for the `slow' and `intermediate' bar models with 33 and 44 \kmskpc. The `fast bar' pattern speed of 51 \kmskpc is not consistent with the \Gaia data.
While the strongest match in angle space is consistent with estimations from the inner Galaxy (see Section~\ref{sec:inner_galaxy}), there is no complementary `$v_{\mathrm{R}}$ flip' in the $L_z$--$J_{\mathrm{R}}$ distribution, as discussed above. The slow and intermediate bar models are reasonable in both of the action and angle studies, although both are messy, and no pattern speed provides a clear and definitive answer. The higher-order resonances (see \citealt{Monari2019a} above) can explain multiple ridges, which was not explored by \cite{Trick2021}, but it remains to be seen how they affects the angle space or the `$v_{\mathrm{R}}$ flip' criterion. Regardless, one would not expect a perfect match from either model, owing to the additional influence of spiral structure.

\cite{Sellwood2019} compares the planar action--angle space from \Gaia DR2 with predictions from three different theories of spiral structure, namely the superposition of transient spiral modes \citep[e.g.][]{Sellwood2014b}, winding material arms \citep[e.g.][]{Toomre1981,Grand2012a,Grand2012b}, and a dressed mass clump model \citep[e.g.][]{Toomre1991}. They find that the \Gaia data appears most consistent with the transient spiral mode model, is somewhat consistent with the material arm model (the model is lacking in particularly narrow features, but otherwise reasonable), and inconsistent with the dressed mass clump model. \cite{Sellwood2019} do not explore the effect of the quasi-stationary classical density wave theory \citep[e.g.][]{Lin1964} citing other flaws with the theory.

\cite{Hunt2019} performed a simultaneous study on the combined effects of bar and spiral models together on the planar action--angle space, and compared them to data from \Gaia DR2. They combine a transient winding spiral arm model and a classical density wave model with simple bar models having different pattern speeds. They are able to \textit{qualitatively} reproduce the Solar neighbourhood planar action--angle distribution for a variety of bar pattern speeds when including transient spiral structure. The addition of the bar coupled with the spirals introduces the fine features missing in the winding material arm model of \cite{Sellwood2019}, removing their main objection to the transient material arm model (although note that \citealt{Hunt2019} do not explore the transient spiral mode model, which was preferred by \citealt{Sellwood2019}, which may also perform well when coupled with a bar). In general, \cite{Hunt2019} do not favour a specific bar pattern speed, but make the point that transient spiral structure can significantly alter the kinematic and orbit space, complicating the inference of bar parameters, commonly `washing out' higher order resonances above 4:1, and occasionally obscuring the stronger ones (e.g. CR and OLR).

\cite{Kawata2021} propose that stars with high radial action are less susceptible to interference from spiral structure (as justified by a comparison to simulations, although see also \citealt{Debattista2024} for a caution). As discussed in Section~\ref{sec:spiralarm_response}, \cite{Palicio2023} also found spiral-like features in the $J_{\mathrm{R}}$ distribution across the disc, where low $J_{\mathrm{R}}$ arcs trace the location of some Milky Way spiral arms. As such, \cite{Kawata2021} explore the ridges in the $L_z$--$J_{\mathrm{R}}$ space at high radial action, and conclude that the resonance-like features at high $J_{\mathrm{R}}$ are equally well reproduced by multiple high-order resonances from a bar with either $\Omega_{\mathrm{b}}\sim34$ or $42$~\kmskpc, similar to the `slow' and `intermediate' pattern speeds from \citet{Trick2022}.

Combining actions with chemistry, \cite{Wheeler2022} examine the effect of bar resonances on the metallicity distribution as a function of $J_\phi$. They make predictions from test particle simulations that there should be a `flattening' of this relation around the CR, and a `wave-crest' around the OLR (and a minor feature around the 4:1 UHR), which appear consistent across slow and fast bar pattern speeds, but are susceptible to selection effects. Comparing their models to data from \Gaia EDR3 and LAMOST DR5, \cite{Wheeler2022} also find that the data are qualitatively consistent with slow, intermediate and fast bar models, but weakly prefer the slow bar model, where the OLR is responsible for the `Hat' moving group. \cite{Wheeler2022} also mention the influence of spiral structure as a confounding factor, but make a strong case for using chemistry in addition to dynamics in future attempts to break the degeneracy.

\subsection{Inference of bar parameters from the halo and streams}  \label{sec:bar_halo}

A full review of streams in the \Gaia era is the subject of another review in this series \citep{StreamsReview}, and the stellar halo is discussed in more detail in Sections~\ref{sec:halo} and \ref{sec:dynamics_halo}. This Section discusses briefly the influence of the Galactic bar on streams and the halo, and what we can learn from features in both.

The bar does not only exert its influence on the stars in the disc, but stars in the halo with the appropriate orbital frequencies can also experience a resonant interaction with the bar, which can leave coherent structure in the stellar halo \citep[e.g.][]{Yang2022,Dillamore2023,Dillamore2024}, or truncate or alter the morphology of stellar streams \citep[e.g.][]{Hattori2016,Pearson2017,Thomas2023}.

\cite{Dillamore2023} use test-particle simulations to show that the Galactic bar can create a prominent ridge in the $E$--$L_z$ plane, which they also observe in the \Gaia data \citep[see also][]{Myeong2018}. The location of the ridge is consistent with the influence of a Galactic bar with pattern speed $\sim$35--40 \kmskpc. \cite{Dillamore2024} then extend the analysis to the $r$--$v_r$ plane, linking the phase space chevrons discovered in \cite{Belokurov2023} to bar resonances rather than the infall of GSE \citep[see also][]{Davies2023}. They find that the two most prominent chevrons are well reproduced by the CR and OLR of a bar with $\sim35$ \kmskpc, without the need for a GSE-like merger.

\cite{Thomas2023} show that the Hyades stream would appear significantly different in a `long slow' barred potential, compared to a `short fast' bar or axisymmetric disc potential. \cite{Thomas2023} are careful to note that the candidate selection for stars in the stream is dependent on the assumed bar potential, making it difficult to draw a conclusion at this stage, but the shape of the Hyades stream could offer an independent test of bar pattern speed in the future, once a more rigorous candidate selection based on unique abundances has been carried out.

\begin{figure}[t]
\includegraphics[width=\linewidth]{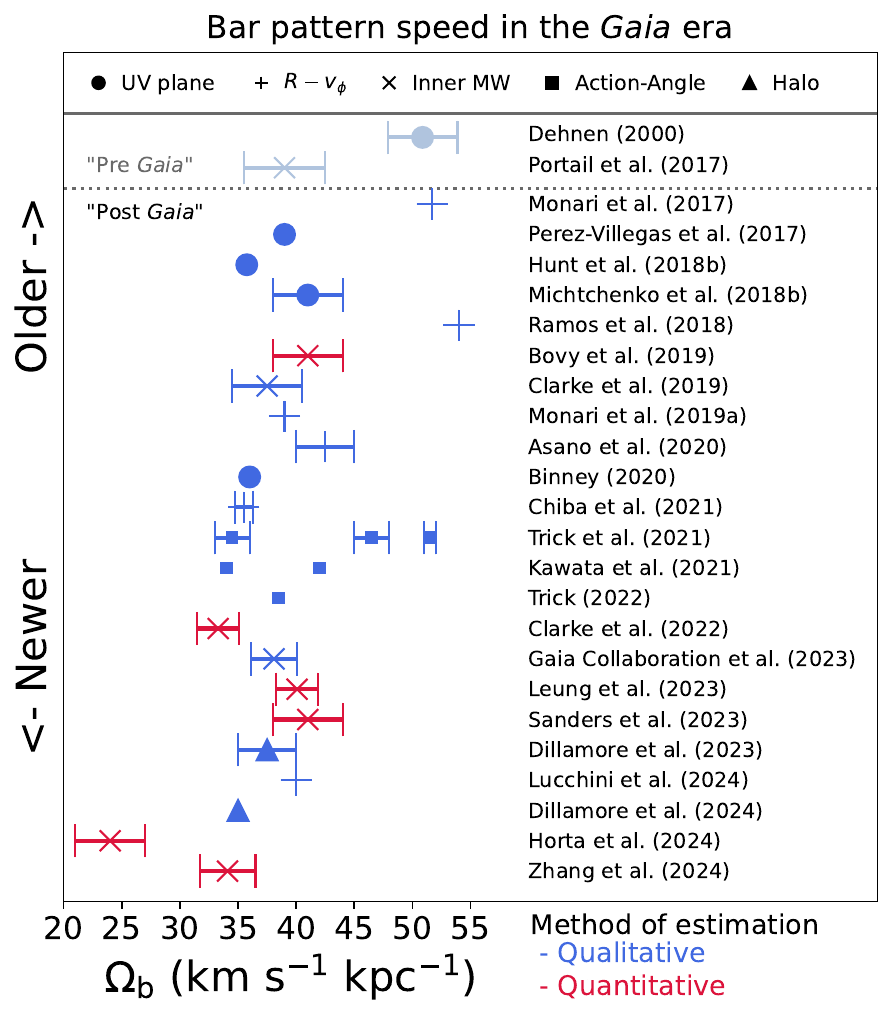}
\caption{Compilation of the pattern speed estimates from the papers discussed in this review, including two historical reference values for the `fast' and `slow' bar solutions. \Gaia papers are ordered vertically by publication date and split by using a qualitative comparison to some kinematic feature (blue) or a `fit' to data (red) for different data sets, split by symbol into `Comparison with the $U$--$V$ plane' (circle), `Tracking moving groups in $R$ and $\phi$' (plus), `Inner Galaxy kinematics' (cross), `Comparison with actions and angles' (square) and `Halo kinematics' (triangles). Error bars are shown only for studies where errors are given.}  \label{fig:BarSpeedsSummary}
\end{figure}

\subsection{Summary of bar parameters}  \label{sec:bar_summary}

While the parameters of the Galactic bar have not (yet) been conclusively measured, and despite a significant scatter in the measurements depending on methods and assumptions, there is a definite trend towards a longer slower bar in literature in the post-\Gaia era. Figure~\ref{fig:BarSpeedsSummary} shows a compilation of the pattern speed estimates from the papers discussed in this review ordered vertically by publication date and split by using a qualitative comparison to some kinematic feature (blue) or a `fit' to data (red) for different data sets, with the symbols showing the method: `Comparison with the $U$--$V$ plane' (circle), `Tracking moving groups in $R$ and $\phi$' (plus), `Inner Galaxy kinematics' (cross), `Comparison with actions and angles' (square) and `Halo kinematics' (triangles). Error bars are shown when errors are given. Note that some papers use multiple data sets, but the most prominent in the paper is used for this loose classification. We include two prominent references from the `pre-\Gaia era (blue-grey) representing the `fast' and `slow' bar dichotomy.

There is a notable clustering in the approximate range of $34\lesssim\Omega_{\mathrm{b}}\lesssim42$ \kmskpc, but no clear consensus within that range. The kinematics in the Solar neighbourhood and nearby disc can be explained by a few different bar pattern speeds in combination with various models of spiral structure, and fitting resonances to gaps and ridges in local data appear to be degenerate between a few different solutions \citep[e.g.][]{Trick2021,Trick2022,Hunt2019,Kawata2021}. While the theoretically expected shape of the resonances in kinematic space is clear \citep[e.g. see][]{Binney2020b}, spiral structure can inflict significant disruption over a local volume \citep{Fujii2019,Hunt2019}, and different studies end up preferring different fits.

Once the radial and azimuthal variation are taken into account, the argument for a long slow bar becomes stronger, with the change in amplitude with radius and azimuth of the Hercules stream matching expectations from the Corotation resonance in studies that consider only the effect of the bar \citep[e.g.][]{Monari2019b,Lucchini2024}, but remaining degenerate in other works that consider the effect of spiral structure \citep[e.g.][]{Hunt2019,Bernet2024}. Interestingly, while the observational appearance of moving groups or ridges is broadly consistent in most studies, their theoretical interpretation significantly varies between papers due to different assumptions or different models.

Measurements of the bar pattern speed from the stellar kinematics in the inner Galaxy prefer lower pattern speeds (e.g. 33--41 \kmskpc; \citealt{Bovy2019,Sanders2019a,Clarke2019,Clarke2022,Leung2023,Zhang2024b}), and inference from the halo (35--40 \kmskpc; \citealt{Dillamore2023,Dillamore2024}) also consistently prefer a slower bar. While not the subject of this review, we note that early estimations from gas kinematics had wide range of estimates covering `fast bar' and `slow bar' solutions (e.g. 52$\pm$10 \kmskpc; see \citealt{Gerhard2011} for a review), more contemporary estimates also favour a slower bar (37.5--42 \kmskpc; e.g.\ \citealt{Sormani2015,Li2022c}), matching the general trend from the stellar dynamics.

In summary, we draw the following, unsatisfactory conclusions:
\begin{itemize}
    \item Kinematic substructure is naturally created by both resonances and phase mixing.
    \item Broadly, studies of the Galactic bar find in favour of one or more of four solutions, with $\Omega_{\mathrm{b}}\sim35, \sim40, \sim45$ or $\sim50$ \kmskpc. These correspond to different bar resonances being responsible for different kinematic substructure.
    \item However, most studies appear to be converging to slower pattern speeds of $\sim35$--40 \kmskpc across multiple tracers and methodologies.
    \item Most studies also now favour a bar length of $R_{\mathrm{b}}\approx 4$--5 kpc, but with disagreement on the `dynamical extent' of trapped orbits or instantaneous alignment with other surface density structures.
    \item Many studies which attempt to `measure' bar parameters from local kinematics ignore spiral structure. It both obscures / disrupts local kinematic signatures, and spiral-bar connection can bias measurements in the inner Galaxy.
    \item Models of a pure quadrupole bar are too simple. Both higher order components of the bar's structure above the quadrupole, and the slow down of the galactic bar are expected and have a significant influence on local kinematics.
    \item There remains disagreement on the angle of the bar $\alpha_{\mathrm{b}}$, primarily driven by stellar distance uncertainties in the inner Galaxy.
    \item On the positive side, stellar labels, such as metallicity and ages, can augment purely dynamical information.
\end{itemize}

The question is unlikely to be fully resolved until we can comprehensively disentangle the bar and spiral structure, both in terms of the kinematic substructure in the disc, but also in the bar region itself.

\subsection{Summary of spiral arms}  \label{sec:spiralarm_summary}

There are now many claimed detections of the Milky Way's spiral structure in the literature across a variety of tracers, both stellar and gaseous. This review has focused on the study of spiral structure in \Gaia data, and thus stellar kinematics, yet there appears to be a disconnect between studies of the Milky Way's spiral structure depending on the tracer used.

For example, the large-scale radial velocity wave of \cite{Eilers2020} shows a different spiral pattern from the UMS sample of \cite{Poggio2021b}, while neither are fully consistent with the maps of young star-forming regions traced by masers \citep[e.g.][]{Reid2019}. However, this is not necessarily a fundamental issue if those tracers are actually expected to trace different dynamical structures. Young stars born in spiral arms on cold orbits and old stars on hot orbits will respond differently to various dynamical perturbations, as will the interstellar medium.

Further work is needed, both observational and through modelling and simulation, to construct a global picture of the Milky Way across stellar populations. Interpreting these varied dynamical signatures across different tracers may also help to better understand the nature of spiral structure itself. In summary;

\begin{itemize}
    \item \Gaia made possible to directly observe local stellar overdensities consistent with the Milky Way's spiral arms \citep[see Section~\ref{sec:spiralarm_density}; e.g.][]{Poggio2021b}.
    \item The distribution of Cepheids and Open Clusters also show spiral-like patterns further across the disc \citep[e.g.][]{Skowron2019a,Gaia2023b}.
    \item Large-scale disc kinematics also show coherent spiral-like patterns in $v_{\mathrm{R}}$ and $v_z$ \citep[e.g.][]{Eilers2020} and higher-order moments of the kinematics \citep[e.g.][]{Akhmetov2024}.
    \item Transient spiral structure reproduces many moving groups and `ridges' in $R$--$v_{\phi}$ and vertex deviation of $v_{\mathrm{R}}-v_\phi$, with or without the presence of a bar \citep[e.g.][]{Hunt2018b}.
    \item Stellar kinematics appear to suggest that the Local and Outer arms are growing, while the Perseus arm is disrupting \citep[see Section~\ref{sec:spiralarm_disruption}, e.g.][]{Funakoshi2024}.
    \item Different traces show different spiral patterns. Is this a problem? Or an opportunity?
\end{itemize}

\section{Vertical perturbations}  \label{sec:perturbations}

The large increase in quantity and quality of data from \Gaia, in particular from DR2 onwards, also revealed clear signatures of vertical perturbation, not only in the outer Galactic disc, but also inside the Solar radius.

It is not an unexpected result that the Milky Way contains such `disequilibrium dynamics' in the planar or vertical distribution and kinematics of stars \citep[e.g. see][etc.]{Minchev2009,Gomez2012b,Widrow2012}, but the large volume of five- and six-dimensional phase-space information made it possible to study known disequilibrium features, such as the classical moving groups (first discovered by Olin Eggen, see \citealt{Eggen1996} for a summary), the vertical asymmetry \citep{Widrow2012}, or the Galactic warp \citep[e.g.][]{Kerr1957,Oort1958}, in new dimensions and in unprecedented detail. \citet{Antoja2018} also showed for the first time a striking phase spiral, or `snail' pattern when plotting the vertical position of stars ($z$) against their vertical motion ($v_z$), further reinforcing the picture that we are in a dynamically active Milky Way.

In this Section we describe vertical perturbations to the Milky Way's disc as have been mapped or detected with \Gaia data. We discuss the large-scale Galactic warp in Section~\ref{sec:perturbations_warp}, the ripples and corrugations which exist on top of the warp in Section~\ref{sec:perturbations_ripples_corrugations_feathres}, and the local vertical disequilibrium feature known as the phase spiral in Section~\ref{sec:perturbations_phase_spiral}. Note that many of these perturbations have planar counterparts, and there will be some overlap with the discussion of disc kinematics from Section~\ref{sec:nonaxi_response}.

\subsection{The Galactic warp}  \label{sec:perturbations_warp}

\begin{figure}
\includegraphics{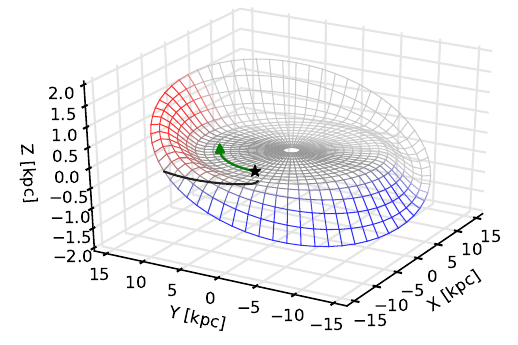}
\caption{
Schematic illustration of the warp in the Milky Way disc. The warp begins beyond Galactocentric radius of 10 kpc, and the black curve shows the line of nodes, at which the mean $z$ crosses zero. The star indicates the Solar position, and the green arrow shows the direction of rotation in the disc, which is also the direction of precession of the warp. The far side of the disc is shown in grey on account of very limited knowledge about the warp geometry in that region; it may differ from the simple reflection of the near side, if modes other than $m=1$ are important. Note that the vertical scale is greatly exaggerated.
}  \label{fig:warp}
\end{figure}

The outer part of the Galactic disc is known to be warped away from the $z=0$ plane. This warp has been observed in both gas and stellar discs, and here we focus only on the latter one, with the associated kinematic signatures becoming available only in the \Gaia era.

To first order, the warp can be described as a sequence of `tilted rings' with a radially varying tilt angle and possibly varying direction of the line of nodes (LoN) where the mean $z$ changes sign. This corresponds to a $m=1$ perturbation in the mean $z$ of disc stars and implies that opposite sides of the disc have opposite deviations; however, observations indicate that the warp amplitude is not exactly antisymmetric with Galactocentric azimuth, necessitating other terms (at least $m=0$ and $m=2$) for a more accurate description. The LoN direction roughly coincides with the line connecting the Sun with the Galactic centre, with the $y>0$ side of the disc (in the direction of Solar motion) having $\overline z>0$, and the warp begins around Galactocentric radius $R\gtrsim 10$~kpc (see Figure~\ref{fig:warp} for an illustration); these features are common to both stellar and gas discs.

The warp can be studied using several types of tracers. Cepheids are relatively young (few hundred Myr) and bright stars with a very good distance calibration using optical (\Gaia, OGLE) or IR (WISE) lightcurves and small amplitudes of random motion, but their number is limited to 1--2 thousand, with very few in the opposite side of the Galactic disc ($x>0$ in the Astropy convention). For UMS stars (spectral types O,B,A), RGB and RC stars, photometric distance estimates have to contend with the extinction and completeness issues, complicating the interpretation of results.

The kinematic signature of the warp is most obvious in the $z$ component of velocity, which for stars near the disc plane mostly depends on the $b$ component of PM. However, for directions other than the anticentre ($l=180^\circ$) and outside the disc plane, and in the absence of \vlos measurements, one needs to subtract the contribution of the perspective motion to $\mu_b$ (see e.g.\ bottom panel of figure 1 in \citealt{Poggio2020} and figure 16 in \citealt{RomeroGomez2019}), typically assuming that stars move on circular orbits with some prescribed radial dependency of the rotational velocity; this assumption is less well justified for older and kinematically hotter populations.

\citet{Chen2019} used 1300 Cepheids from WISE to study the spatial structure of the warp, finding that it starts around 10~kpc and reaches an amplitude of $\overline z\simeq 0.8$~kpc at $R=16$~kpc. They also revealed the twist in the LoN direction, which changes by $\sim 25^\circ$ between 12 and 15~kpc, forming a leading spiral (i.e., in the opposite direction to the spiral arms in the Galactic disc).
\citet{Skowron2019b} examined a larger catalogue of $\sim$2400 Cepheids from OGLE and other surveys, which extends even to the opposite side of the Galactic disc (beyond the bulge). They modelled the spatial distribution by a sum of three angular harmonic terms ($m=0,1,2$) with fixed LoN angles (i.e., without a twist), and found a somewhat higher amplitude of $\overline z\simeq 1.2$~kpc at the outer edge of their sample ($R\simeq 16$~kpc) than the previous study. Combining the distances with the PM measurements from \Gaia DR2, they measured the vertical velocity perturbation $\overline{v_z}\simeq 10$--15 \kms, which is also higher than most other estimates discussed below. A similar conclusion about the warp geometry was obtained by \citet{Lemasle2022} using an updated catalogue of Cepheids from a combination of several surveys.

\citet{Dehnen2023} considered a catalogue of $\sim$2100 Cepheids with full 6d phase-space information from \Gaia DR3 \citep{Gaia2023b}, examining the geometry and kinematics of the warp using the direction of orbital angular momenta as a function of guiding radius (rather than instantaneous location). They confirmed the twist in the LoN found by \citet{Chen2019}, and determined a prograde precession of the warp (in the direction of the Galactic rotation) with an angular velocity decreasing from $\sim$ 15 to 6 \kmskpc between $R=11$ and 14 kpc.  A comparable precession rate of 5 \kmskpc at $R=13$ kpc was found by \citet{Zhou2024} using largely the same data.
Finally, \citet{CabreraGadea2024a} analysed $\sim$2400 Cepheids from OGLE, of which only a third have line-of-sight velocities in \Gaia DR3, using a Fourier series with $m=0,1,2$ in multiple radial bins (due to the limited spatial coverage, it is difficult to constrain higher-order harmonics beyond $m=2$). They also found a twisting LoN between 12 and 15 kpc, and a line of maximum $\overline{v_z}$ with a similar twist, but lagging in azimuth by $\sim$25$^\circ$, from which they inferred a precession rate with a similar radial profile as \citet{Dehnen2023}.

Overall, the amplitude and kinematics of the warp are fairly consistent between different studies that use Cepheids, and both the precession rate and the LoN twist have been well established. However, studies using other tracer populations are less conclusive (see e.g.\ section 5.1.1 and figure 8 in \citealt{CabreraGadea2024a} for a literature comparison), which may be partially attributed to the difficulties in measuring the spatial distribution of stars close to the disc plane caused by the interstellar dust extinction.

Shortly after \Gaia DR2, \citet{Poggio2018} constructed kinematic maps for $1.3\times10^7$ RGB and $6\times10^5$ UMS stars in the range $|y|<8$~kpc and $R$ up to 14~kpc. They found large-scale variations of the mean vertical velocity $\overline {v_z}$ up to a few \kms for both samples, in particular, beyond 12~kpc.
\citet{Poggio2020} used the same sample to measure the precession rate of the warp at a level $11\pm3$ \kmskpc.
\citet{RomeroGomez2019} assembled an even larger catalogue of $1.8\times10^7$ RGB and $2\times10^6$ OB stars, most of which lack \vlos measurements. They found significant differences in the amplitude $\overline z$ and velocity $\overline {v_z}$ between the two populations ($\sim1$~kpc and a few \kms at the radius $R=15$~kpc for RGB stars, and a few times smaller for OB stars), as well as evidence of asymmetry (variations of amplitude between the two sides of the warp) and radial dependence of the LoN direction.

\citet{Cheng2020} constructed a 6d sample of $5.5\times10^6$ stars from \Gaia DR2 RVS and $1.8\times10^5$ stars from APOGEE, using \texttt{StarHorse} photometric distances and age estimates from \citet{Sanders2018}. They found a higher amplitude of the warp ($\overline z\sim$2~kpc at $R=15$~kpc) than in Cepheids-based studies, determined the precession rate of 12--14 \kmskpc, and measured a somewhat higher velocity perturbation in younger population (3--6 Gyr) than in the 6--9 Gyr age bin.
Using nearly $9\times10^6$ RC stars from 2MASS, \citet{Uppal2024} also found a high amplitude of $\sim 1.5$~kpc at $R=15$~kpc, as well as a $\sim$15\% north--south asymmetry.

\citet{CabreraGadea2024b} examined the spatial and kinematic properties of metal-rich RR Lyrae stars from a combined catalogue of \Gaia and other photometric surveys \citep{Mateu2020}. Although these are usually associated with old halo populations, there is now a solid evidence that some of them have intermediate age (few Gyr) and follow disc-like orbits (see references in Section~\ref{sec:observations_distance_measurement}). The warp is detected in RR Lyrae with properties similar to the red clump and intermediate-age populations analysed by \citet{Gaia2021d} and to the Cepheid-based studies described above; its precession speed is estimated to be 10--15~\kmskpc.

\citet{Jonsson2024} updated the analysis of the 6d \Gaia RVS dataset to DR3, with distances from \citet{BailerJones2021}. They find the kinematic line of nodes ($\overline{v_z}$ reaches maximum) shifted more strongly to $y>0$ than in other recent studies and ahead of the geometric line of nodes ($\overline{z}=0$), opposite to the trend found in Cepheids by \citet{CabreraGadea2024a}. They could not robustly constrain its twist, but found the precession rate $\sim 13.5$~\kmskpc to be nearly constant with radius (again in contrast to the declining rate inferred for Cepheids by \citealt{Dehnen2023}). Although they used only the kinematics and not the positions of stars, their model also predicts the warp amplitude on the high end of other estimates, reaching a plateau at $\sim 3$~kpc beyond $R=14$~kpc.

On the other hand, \citet{Wang2020b} and \citet{Li2023c} argued that the warp amplitude decreases with age, based on the sample of RC (and in the second paper also OB and MSTO) stars from LAMOST and \Gaia. The velocity signatures in the second paper were not consistent between the three samples. \citet{Chrobakova2022} used a statistical deconvolution procedure of \citet{LopezCorredoira2019} to infer the distances to young supergiants and older RGB stars from their parallaxes, up to $R=20$~kpc. They also found a lower warp amplitude in the older population, and even the supergiants had a much weaker warp than Cepheids from other studies.

To summarise, the structure and kinematics of the warp are best studied in Cepheids, while the analyses of other tracers do not present a consistent picture -- even the dependence of the warp amplitude on age is not known. Clarifying this dependence would help to discriminate between different theories of warp formation: torques caused by the misalignment between the disc and the triaxial halo, gravitational perturbation from a satellite galaxy, or accretion of intergalactic gas from a direction misaligned with the stellar disc.
The first scenario, although not new, has recently been revived by \citet{Han2023}, who proposed that the debris from the GSE merger create a long-lasting tilt in the halo (see also \citealt{Deng2024} for another scenario involving GSE). On the other hand, \citet{Poggio2020} argued that the observed warp precession rate is much higher than expected in a triaxial halo, and instead advocated the passage of the Sagittarius galaxy through the disc as the driving mechanism. However, simulations of Sagittarius and/or LMC passages \citep{Laporte2018, Poggio2021a} are not precisely tailored to the actual Milky Way system, making a quantitative comparison challenging. \citet{TepperGarcia2022} find that the perturbations in the stellar disc initially trace those of the gas disc, from which the young stars have formed, but eventually drift apart after several hundred Myr.
Finally, a non-gravitational origin of the warp (gas accretion) predicts an absence of precession and a larger amplitude in young populations.

\subsection{Ripples, corrugations and feathers}  \label{sec:perturbations_ripples_corrugations_feathres}

As already mentioned, pre-\Gaia studies uncovered evidence for some external perturbation to the disc of the Milky Way both locally in the planar \citep{Minchev2009} and vertical dynamics \citep{Widrow2012}, and the large-scale warp in the outer disc. It has also been shown that these small-scale perturbations and the large-scale disc warp coexist, with a wave-like oscillations or a `corrugation' pattern extending to the outer disc \citep{Xu2015}, which result in structures such as the Monoceros Ring and Triangulum--Andromeda (TriAnd) overdensity.

While the origin of these features (whether they are accreted structures or a part of the disc) has been a subject of debate historically, more recent data \citep[e.g.][]{Bergemann2018,Sheffield2018} show that they are consistent with being formed out of disc material that has gained kinetic energy following some perturbation (as predicted in simulations of the Milky Way's interaction with the Sagittarius dwarf galaxy, e.g., \citealt{Purcell2011,Gomez2013}). Studies of these collective disc oscillations from the local to global scale, or `Galactoseismology', can provide a window on our Galaxy's history of interactions with its satellites, and on the nature of dark matter \citep[see e.g.][]{Johnston2017}.

We note that many of the features discussed in this Section are related, and the split by subsections is somewhat dependent on how the results are presented in the literature, while the waves, ripples, corrugations and feathers themselves are likely part of the same large-scale disc dynamical phenomena seen in different projections and datasets.

\subsubsection{Velocity waves and corrugations across the disc}  \label{sec:perturbation_waves}

There has long been evidence for a vertical wave-like oscillation in the Solar neighbourhood \citep{Widrow2012}. \cite{Bennett2019} expand upon this original detection of vertical asymmetries and disequilibrium in the Solar neighbourhood to larger $|z|$ and to higher accuracy. Using \Gaia DR2 in combination with APOGEE and GALAH, they find a wave-like oscillation in the local stellar disc, consistent across all stellar colours and ages, implying that they were all excited by the same perturber (simulations of a Sagittarius-like merger can reproduce Solar neighbourhood asymmetry \textit{qualitatively} similar to the data, e.g., \citealt{Laporte2018}, \citealt{Bennett2022}, but are unable to find an exact match). Accounting for the asymmetry, \cite{Bennett2019} then make a precise measurement of the Sun's height above the disc plane of $z_\odot=20.8\pm0.3$ pc.

\cite{Schoenrich2018} uncover a wave-like feature further across the disc in both $R$--$v_z$ and $L_z$--$v_z$ in data from \Gaia TGAS for approximately $7\lesssim R\lesssim9$ or $4\lesssim R_{\mathrm{G}}\lesssim11$ kpc. As discussed in Section~\ref{sec:bar_action}, using $R_{\mathrm{G}}$ (equivalently, $L_z$) instead of $R$ groups stars on similar orbits regardless of their current epicyclic phase, and makes possible to use a physically local sample to explore further across the disc in orbit space. While the wave steadily increases in $v_z$ with $R$ and $R_{\mathrm{G}}$, which they link to the Galactic warp, the shorter wavelength oscillations on top of the warp are likely linked to phase mixing perturbation following interaction with the Sagittarius dwarf galaxy.

This wave is also seen in \cite{Kawata2018}, \cite{Gaia2021d} in $R$--$v_z$, and in \cite{Friske2019}, \cite{Antoja2022} in $R_{\mathrm{G}}$--$v_z$. Such vertical waves are not easily explainable by a bar or spiral structure in an isolated `flat' Galaxy. However, it is also possible that such corrugations, along with generic bending and breathing modes, can arise from the interaction of a planar spiral wave and the large-scale warp \citep{Masset1997,Khachaturyants2022a,Khachaturyants2022b}. Here we also remind the reader of the discussion of a wave-like pattern in the radial velocities in Section~\ref{sec:maps}. It is natural that some external perturbation would induce coherent kinematic substructure in both the planar and vertical direction. As such, many studies link the wave-like patterns in the planar and vertical velocities.

Such wave patterns are also present in the energy of stars, as shown in \cite{Khanna2019}. They find a wave-like pattern in the dimensionless energy, defined as [$E-E_{\mathrm{circ}}(R_\odot)]/V^2_{\mathrm{circ}}(R_\odot)$, vs.\ normalised density, $v_{\mathrm{R}}$, $v_z$ and $z$, where the peaks are approximately evenly spaced, except for the peak corresponding to the Hercules stream, which is an extra peak between the other regular pattern.

As discussed above in Section~\ref{sec:bar_action}, actions are orbit labels, and examining dynamics in action space sharpens features by grouping stars on similar orbits. \cite{Cheng2020} show a comparison of the radial and vertical velocity waves in $R$ and $L_z$ using a combination of data from \Gaia and APOGEE, for approximately $6\lesssim R\lesssim16$ kpc and $1000\lesssim L_z\lesssim4000$ kpc\,\kms. They show that the short wavelength pattern is clear as a function of angular momentum, present on top of the large wavelength pattern as a function of $R$. They propose an origin of either secular Milky Way spiral arms, or the influence of a perturbing satellite galaxy, and find the data is consistent with both potential explanations when also superimposed on top of the distinct influence of the large scale warp, as discussed in Section~\ref{sec:perturbations_warp}.

\begin{figure}
\includegraphics[width=\linewidth]{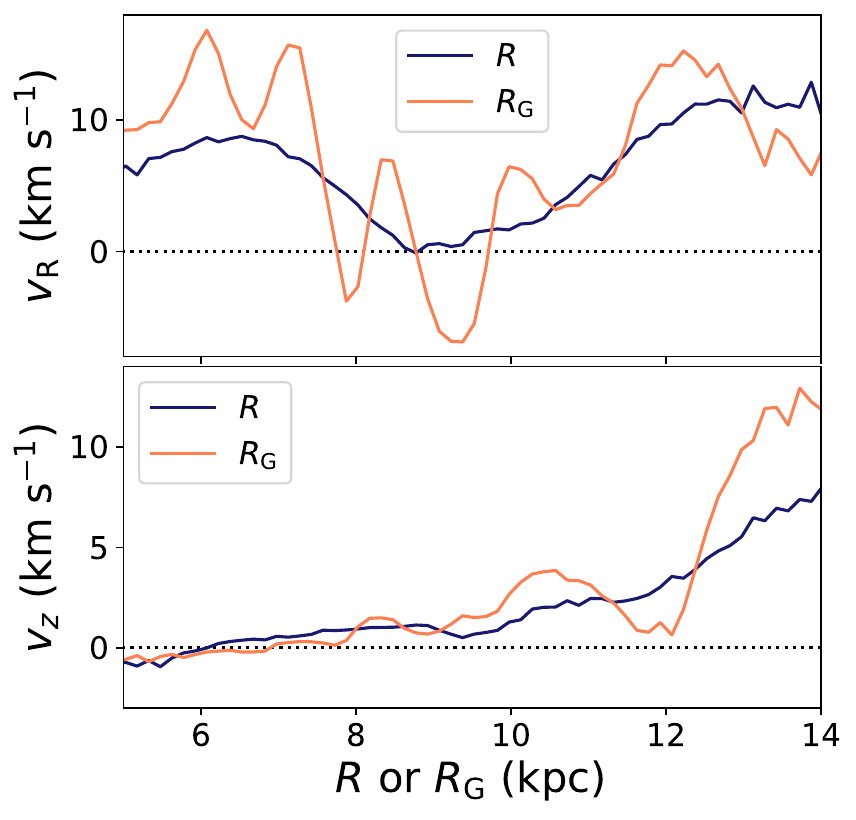}
\caption{Radial (upper panel) and vertical (lower panel) velocity waves as a function of $R$ (blue) and $R_{\mathrm{G}}$ (orange) for stars in \Gaia DR3 with fractional parallax error $<10\%$, and $|\phi|<0.1$ rad from the Sun Galactic centre line \citep[see also e.g.][]{Schoenrich2018,Friske2019,Khanna2019,Xu2020,Hunt2020,Cheng2020,Zhou2024,Poggio2024,Sun2024}.}  \label{fig:velwaves}
\end{figure}

These features are illustrated in Figure~\ref{fig:velwaves}, which shows the radial velocity (upper panel) and vertical velocity (lower panel) as a function of $R$ (blue) and $R_{\mathrm{G}}$ (orange), for stars in \Gaia DR3 with fractional parallax error $<10\%$ and located $|\phi|<0.1$ rad from the Sun--Galactic centre line. Note that the exact amplitude of the pattern varies with choices made in the sample selection, but the overall pattern itself remains robust. The `long-wavelength' radial velocity pattern as a function of $R$, and the `shorter-wavelength' $v_{\mathrm{R}}$ pattern as a function of $R_{\mathrm{G}}$ are readily apparent. It is not yet clear whether the pattern in $R$ is merely a smoothed version of the pattern in $R_{\mathrm{G}}$ owing to `blurring' from the stars epicyclic motions, or if they are two distinct dynamical signals. The vertical velocity increases steadily with $R$ over this range, while the vertical velocity pattern as a function of $R_{\mathrm{G}}$ oscillates around the signal in $R$.

\cite{Poggio2024} examine two samples from \Gaia DR3; one of young giant stars, and one of Cepheid variables. Both samples show a large-scale corrugation on top of the warp signature, with coherent radial and vertical motion, which they interpret as a large-scale wave propagating radially outwards across the disc. They also find that when subtracting the signature of the warp, the residual vertical velocity fluctuations are systematically positive and coupled to the radial motion, such that the stars in the wave crest experience `outward' and `upward' motion of $\sim10$--15 \kms as the wave propagates, while stars outside the wave crest move inwards. Thus, \cite{Poggio2024} claim the discovery of a large-scale Rayleigh-like vertical wave propagating outwards, with correlated vertical radial motion.

\subsubsection{Rings and Feathers in the outer disc}  \label{sec:perturbations_feathers}

These large-scale corrugations can lead to the appearance of other morphological features in the outer disc above and below the disc plane. Here we refer to corrugations that remain part of the disc as `rings', and to structures formed out of disc material that has been pulled away from the disc as `feathers' (also sometimes called tidal spiral arms), when there is a gap between the feather and the edge of the disc, even if it is connected at some other Galactic azimuth.

\cite{Laporte2020a} use \Gaia DR2 to examine the Anticentre stream (ACS) and the Monoceros stream and find them to be kinematically and spatially separate structures. With LAMOST and SEGUE data they show that ACS is both $\sim0.1$ dex more metal poor, and comprised of older stars than Monoceros. They argue that this can be explained if ACS is a tidal spiral arm which was excited from the first pericentric passage of the Sagittarius dwarf galaxy, while Monoceros is a ring / corrugation which arises from the later extended interactions. \cite{Ramos2021} show that the ACS and Monoceros can in fact both be detected `blind' purely from \Gaia DR2 astrometry, and track them to low latitudes using a wavelet transform. They again find the properties of both Monoceros and ACS to favour a disc origin, where the ratio of RR Lyrae stars and M giants are inconsistent with coming from a disrupted satellite. The disc origin of the Monoceros ring is also advocated by \citet{Borbolato2024} based on the kinematics and chemistry of M giants from \Gaia DR3 and several spectroscopic surveys.

\cite{Laporte2022} further explore the Galactic anticentre with data from \Gaia EDR3 in combination with LAMOST, revealing several new feathers in the outer disc by selecting stars with low parallaxes ($\varpi<0.1$~mas) and similar PM. They show that these new structures have disc-like kinematics ($v_\phi$ between 170 and 230 \kms) and distance gradients, with a small spread in energy. \cite{Laporte2022} find the most likely explanation that these structures are feathers lifted away from the disc following interaction with a satellite, but note that they may also be the corrugations of the disc seen in projection. As with the corrugations, such rings and feathers are naturally reproduced through interaction with a large satellite such as the Sagittarius dwarf galaxy. For example, \cite{Laporte2018} demonstrate that Monoceros and TriAnd-like structures are created in a simulation of a Sagittarius-like merger. \cite{Laporte2019a} then use this simulation to show that these feathers can be used both to constrain the galactic potential and to infer the mass of the perturber.

\subsubsection{Modelling the corrugations}  \label{sec:perturbations_corrugations}

Both the `velocity waves' and the rings and feathers in the outer disc appear to be part of the same coherent ripples and corrugation pattern seen across the disc, yet our location within the Milky Way makes it difficult to get a Galaxy-wide and complete picture of them. However, large numerical simulations help to explore their formation and evolution in a variety of scenarios. For example, the large-scale ripples and corrugations are naturally reproduced through interaction with some external perturbation, as shown in numerous studies \citep[e.g.][]{Purcell2011,Gomez2012a,Gomez2013,Laporte2018,BlandHawthorn2021,Hunt2021,Antoja2022,TepperGarcia2022,Stelea2024}.

\cite{BlandHawthorn2021} trigger an impulsive perturbation in a high-resolution $N$-body disc and show that the response is a complex superposition of two modes; a density wave and a bending wave that both wind up at different rates. The bending waves creates corrugations in the disc as it winds up, and they show the interaction of the two waves can result in the phase spiral described in Section~\ref{sec:perturbations_phase_spiral}. \cite{TepperGarcia2022} then revisit this scenario with the addition of a gas disc, and find that the corrugations settle faster in the gaseous disc compared to the stellar, providing another potential method of age-dating the perturbation.

However, more realistic satellite interactions would consist of a series of repeated pericentres and disc crossings, which superimpose subsequent waves on top of older interaction signatures. \cite{Antoja2022} show how tidally induced spiral arms create the corrugations (and $v_{\mathrm{R}}$, $v_\phi$ waves) both from single and multiple perturbation test-particle models. While the single-perturbation models have a simple well-behaved wave pattern, the multiple-perturbation models show more complex behaviour and a superposition of wave modes, which provides a better match to the \Gaia data. However, this approach neglects self-gravity of the disc and assumes the perturbations to be instantaneous, which is likely a poor approximation for later interactions between the satellite and the disc.

On the other hand, \cite{Poggio2021a}, \cite{GrionFilho2021} and \cite{Antoja2022} examine the self-consistent $N$-body simulations of \cite{Laporte2018}, and show that there are different interaction `regimes' as the satellite merges into the Milky Way-like host. The acceleration felt by the disc comes from a complex combination of the direct force of the satellite, the force from the dark halo distortion, and the self-gravity of the disc response itself.

\cite{GrionFilho2021} find that the direct force is most important in a small window around disc crossings, when Sagittarius is either massive (at early times) or closer to the inner disc (at late times), driving long-lived slowly decaying modes, while the dark halo distortion has the largest effect on the outer disc. The disc self-gravity becomes important at later times, where the self-gravity of the bending wave grows to rival the influence of the satellite, and has a significant effect on the inner disc. The exact contributions from each mechanism in the real Milky Way will depend on the orbit and the mass loss history of the Sagittarius dwarf galaxy. Regardless, early stages of the interaction with Sagittarius should clearly have left a lasting imprint on the Milky Way's disc.

However, it is very difficult to construct a simulation of the entire Sagittarius--Milky Way interaction, starting from the time before its first pericentre, that would match all present-day constraints. In contrast to the aforementioned `Sagittarius-like' simulations, which resemble the actual satellite's structure and trajectory qualitatively but not in detail, other works focused on a more quantitative fit to the properties of the Sagittarius remnant and its stream. For instance, \citet{Vasiliev2020} compared a series of simulations of the disrupting satellite to the \Gaia DR2 kinematic maps of the remnant and argued that its present-day total mass is of order $4\times 10^8\,M_\odot$, and even at the previous pericentre passage ($\sim$1 Gyr ago) it could not exceed a few times $10^9\,M_\odot$. \citet{Vasiliev2021a} presented simulations of the Sagittarius disruption in the presence of the LMC, which matched well the properties of the entire stream. These simulations did not start from the first infall of the satellite, but only covered the last 3 Gyr of its evolution, with an initial mass already significantly reduced from the pre-infall one (estimated to be up to a few times $10^{10}\,M_\odot$). Nevertheless, \cite{Stelea2024} found that even this last stage of evolution leaves some large-scale ripples in the Milky Way disc, but no significant tidal arms or feathers, consistent with the conclusion of \citet{Laporte2020a} that they are induced by the first pericentre passage. Notably, a recent passage of Sagittarius naturally reproduces the large-scale radial velocity wave / spiral feature of \cite{Eilers2020}, although it is also well reproduced from secular spiral structure (see Section~\ref{sec:spiralarm_response}), and correlation does not necessarily imply causation.

\subsection{The phase spiral}  \label{sec:perturbations_phase_spiral}

\begin{figure}
\includegraphics[width=\linewidth]{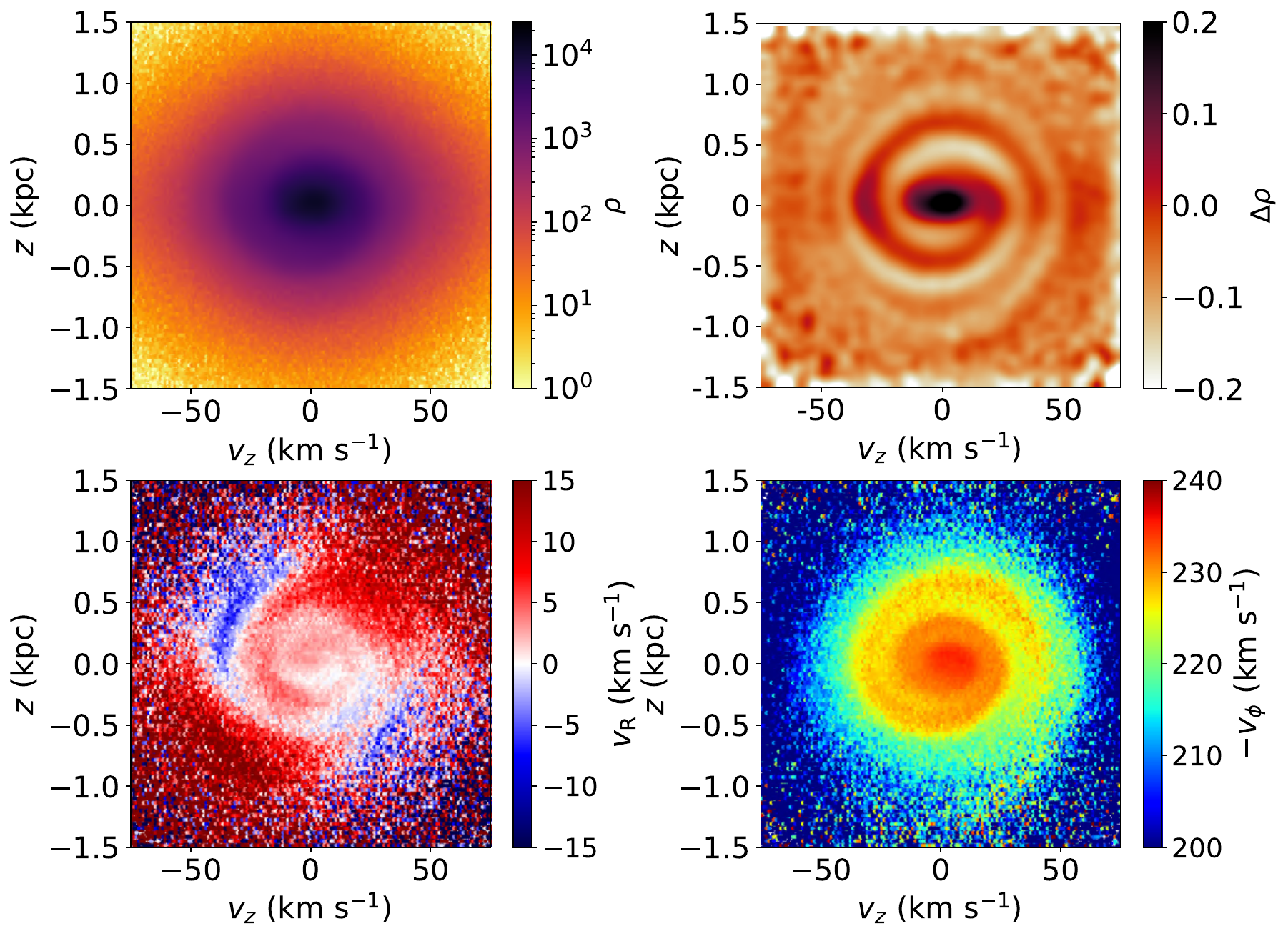}
\caption{Phase spiral as seen by \Gaia DR3 data within a local 1 kpc cylindrical volume from the Sun, shown in logarithmic number density (upper left), normalised density (upper right), radial velocity (lower left) and azimuthal velocity (lower right), following several studies as discussed in the text \citep[e.g.][and many other references in Section~\ref{sec:perturbations_phase_spiral_mapping}]{Antoja2018,Antoja2023}.}  \label{fig:phasespirallocal}
\end{figure}

The \zvz phase spiral is another indication that there has been some vertical perturbation to the Galactic disc, which is now mixing back towards equilibrium. In the discovery paper with \Gaia DR2, \cite{Antoja2018} reproduce the \zvz phase spiral with a simple model of phase mixing following a perturbation, and estimate that the perturbation occurred 300 to 900 Myr ago. Since that initial work, many groups have attempted to reproduce and explain the phase spiral through a mix of theory and simulation, and to map it further across the Galactic disc. It is also highly likely to be related to the corrugations in the disc (see Section~\ref{sec:perturbations_feathers}), as discussed below.

In this Section we describe the morphology of the phase spiral across the disc, and discuss potential origin scenarios. The phase spiral can also be used to measure the vertical potential and dark matter density, as discussed in Section~\ref{sec:dynamics_phase_spiral}. While other dynamical processes in the Milky Way have been the subject of their own reviews, the phase spiral is a relatively new area of study, thus we give a more detailed overview of the literature, which is all in the \Gaia era.

\subsubsection{Mapping the phase spiral}  \label{sec:perturbations_phase_spiral_mapping}

The initial detection of the phase spiral in \Gaia DR2 was restricted to a local volume \citep[$d \lesssim 0.2$ kpc;][]{Antoja2018}. They showed that the phase spiral is also clear when visualised as a function of mean azimuthal velocity $\overline{v_\phi}$ \kms, or mean radial velocity, $\overline{v_\mathrm{R}}$. Figure~\ref{fig:phasespirallocal} shows a reproduction of the local phase spiral as seen in \Gaia DR3 data for a local cylindrical selection of stars with $d_\mathrm{cyl}<1$ kpc from the Sun, where $d_{\mathrm{cyl}}=\sqrt{(x-x_{\odot})^2+(y-y_{\odot})^2}$. The phase spiral is faintly visible in the raw number density (upper left panel), but becomes clear in normalised density (upper right panel), $v_{\mathrm{R}}$ (lower left panel)\footnote{although note that there will always be a $\overline{v_\mathrm{R}}$ quadrupole in the outer part of the \zvz plane, which arises from the tilt of the velocity ellipsoid as a function of $z$ regardless of the phase spiral itself, as shown in \cite{BlandHawthorn2019}.} and $v_{\phi}$ (lower right panel). Note that there is a split in the literature whether the phase spiral is displayed with $v_z$ or $z$ on the horizontal axis, altering the appearance of a clockwise or anti-clockwise rotation of the spiral; regardless of this choice, the phase spiral is always a trailing spiral.

Subsequent studies have mapped the phase spiral across Galactic radius and azimuth, across action--angle space (see Section~\ref{sec:perturbations_phase_spiral_action}), and as a function of planar velocities and stellar labels (see Section~\ref{sec:perturbations_phase_spiral_chemistry}). Depending on the geometry and nature of the perturbation, one may also expect coherent changes in properties of the phase spiral (e.g. amplitude, pitch angle, phase angle) across the disc, which in turn may provide clues on the nature of the interaction.
This mapping can be done in physical space ($x$--$y$) or action--angle space (e.g., $J_\phi$--$\theta_\phi$). The phase spiral itself can also be examined in physical space (\zvz), or in action space (e.g., $\sqrt{J_z}\cos\theta_z$--$\sqrt{J_z}\sin{\theta_z}$, as in \citealt{BlandHawthorn2019}). While mapping in physical space is more intuitive, dust extinction in the midplane leaves significant selection effects in phase spirals away from the Solar neighbourhood, and a selection in physical space blends stars on many different orbits.

Conversely, one can split the local volume into an action-based grid, minimising selection effects from dust extinction in the mid-plane and grouping stars with shared orbital histories, which should have experienced similar perturbations. However, such an action-based split instead has selection effects by orbital phase (see \citealt{Hunt2022}, \citealt{DarraghFord2023} for a description and illustration), and requires assumption of a potential in which to calculate actions (or assumption of a rotation curve if splitting just by $L_z$), which adds different biases. For example, \cite{Guo2022} and \cite{Lin2024} show that the long known wave-like North--South vertical asymmetries in density and $\sigma_{z}(z)$ \citep[e.g.][]{Widrow2012,Bennett2019} are a projection (or a consequence) of the phase spiral. The pattern is consistent between $R$ and $R_{\mathrm{G}}$, but it is clearer in guiding radius.

\cite{Hunt2024} also found an analogous phase spiral in $\Delta R$--$v_{\mathrm{R}}$, which may be expected for any perturbation that influences both planar and vertical motions, and is likely another projection of residuals seen in the $\Omega_{\mathrm{R}}$--$\theta_{\mathrm{R}}$ plane, as shown in \cite{Li2023b}. A combination of selection effects, planar kinematic substructure, and a weak radial frequency gradient with radial action make the $\Delta R$--$v_{\mathrm{R}}$ spiral harder to resolve, and the rest of this Section will focus on the \zvz phase spiral, although future studies may benefit from modelling both the vertical and planar disequilibrium in tandem.

\cite{Laporte2019b} demonstrated that the phase spiral changes with Galactic radius, using bins centred around $R=6.6,\ 8$ and 10 kpc. They show that the phase spiral is much clearer in relative density $\delta\rho(v_z,z)$, as adopted by many subsequent studies, and that the aspect ratio of the phase spiral and the pitch angle changes with $R$ owing to the change in the restoring force from the vertical potential across the disc (note that this allows measurement of the vertical potential owing to its anharmonic nature; see Section~\ref{sec:dynamics_phase_spiral}). They compare the data to an $N$-body simulation of a Milky Way--Sagittarius merger \citep{Laporte2018}, linking the local phase spiral to disc-wide corrugations (as described in Section~\ref{sec:perturbations_feathers}) and estimating an impact time of 500--800 Myr. \cite{Xu2020} extended the mapping of the phase spiral in physical space out to $7\lesssim R\lesssim17$ kpc using a combination of \Gaia DR2 and LAMOST data, where the phase spirals are partially visible as a function of azimuthal velocity.

Extinction in the midplane can become a significant issue for studies that map the phase spiral in physical space; however, this can be modelled out. For example, \cite{Widmark2022a} map the phase spiral across $\sim3$ kpc in the disc with \Gaia EDR3, while masking out areas of high extinction around $z=0$. The paper is primarily concerned with using the phase spiral to map the vertical potential, and it is discussed in more detail in Section~\ref{sec:dynamics_phase_spiral}.

\Gaia DR3 significantly increased the 6D stellar sample and enabled maps of the phase spiral to extend further in physical space, as shown nicely in the DR3 science demonstration paper \citep{Gaia2023c}, which maps the spiral in three coarse bins stretching across $5.25<R<11.25$ kpc. The change in aspect ratio and pitch angle is clear from the inner to outer Galaxy in normalised density, $\hat{v}_{\mathrm{R}}$, $\hat{v}_\phi$ and $\Delta$[M/H] (see Section~\ref{sec:perturbations_phase_spiral_chemistry} for a discussion of the chemistry). \cite{Antoja2023} detect the phase spiral in a similar range ($6<R<11$ kpc) with DR3, but split it into finer bins, again resolving a clear and consistent trend in phase spiral pitch angle. They also extend the analysis to Galactic azimuth, finding more subtle variation over $-20^\circ<\phi<20^\circ$. \cite{Alinder2023} also map the phase spiral in physical space in the outer disc with DR3. They measure the phase angle of the phase spiral, and show that there is a clear and consistent change of phase angle with Galactic azimuth, where the phase spiral shows a change of approximately three degrees in phase angle per degree of Galactic azimuth.

\subsubsection{The phase spiral in action space}  \label{sec:perturbations_phase_spiral_action}

\cite{Li2020a} examine the phase spiral for stars in different classical moving groups, i.e.\ Hercules, Sirius, Hyades, Pleiades etc., and find that the phase spiral varies between groups. They show that this variation is linked to the radial action, $J_{\mathrm{R}}$, where stars with higher radial action show weaker phase spirals, almost disappearing for $J_{\mathrm{R}}>40$ kpc\,\kms. They suggest that phase mixing happens quicker for hotter orbits, and by comparing with test particle models argue that the perturbation must be older than 500 Myr.

\cite{BlandHawthorn2019}, \cite{Khanna2019}, \cite{Li2021a} and \cite{Gandhi2022} split local samples of \Gaia stars as a function of angular momentum (or $R_{\mathrm{G}}$), and show that the phase spiral is significantly enhanced when grouping stars by their actions instead of their physical location, along with mitigating selection effects. They also show that the phase spirals changes clearly with $J_\phi$, both in phase and pitch angle, finding more wound spirals at lower $|J_\phi|$. The vertical stellar frequencies $\Omega_z$ are higher for lower $|J_\phi|$ \citep[see e.g.\ figure 2 of][]{Gandhi2022}, so this is expected from stellar dynamics following a single vertical perturbation. \cite{BlandHawthorn2019} also show that the phase spiral changes with $J_{\mathrm{R}}$ at a fixed $J_\phi$, further highlighting the complexity of the phase spiral morphology in a physically local volume. All three studies compare the data to test particle simulations, confirming the trends with $J_\phi$. \cite{Li2021a} estimates a perturbation time of $\sim500$--700 Myr by comparing the pitch angle of the simulations with the data, and \cite{BlandHawthorn2019} estimate an interaction time of $\lesssim500$ Myr for a $\sim3\times10^{10}\ M_{\odot}$ Sagittarius perturber.

With \Gaia DR3, \cite{Antoja2023} extend the range of $J_\phi$ in which the phase spiral is observable to $1300<|J_\phi|<2600$ kpc\,\kms, again finding clearer phase spirals than when binning in $R$. They calculate timings for the perturbation for individual phase spirals from the difference in vertical frequencies between wraps, with
\begin{equation*}
    T=\frac{2\pi}{\Omega_z(L_z/Z_{c_1})-\Omega_z(L_z/Z_{c_2})},
\end{equation*}
where $\Omega_z(L_z)$ is the vertical frequency for a given $L_z$ and $z_{\mathrm{max}}$ (see section 4 of \citealt{Antoja2023} for full description of the method and frequency calculation), and $c_1,c_2$ are different wraps of the phase spiral as they cross the $v_z=0$ line. From this method, \cite{Antoja2023} find a clear and consistent trend of increasing time since interaction with $J_\phi$, with some wave-like variation on the scale of $\sim100$--200 kpc\,\kms. They recover times ranging from $0.2<T<1.2$ Gyr, with an average time of 0.5 Gyr. The exact time measured is dependent on the choice of Milky Way potential, which sets the frequencies, but the overall trend of increasing with increasing $|J_\phi|$ persists for all common Milky Way potential models.

\cite{Frankel2023} model the amplitude and impact time of the phase spiral over $1250\lesssim |J_\phi| \lesssim2400$ kpc\,\kms, finding inferred times in the range 0.2--0.6 Gyr. Their time estimates also increase with $|J_\phi|$ for $|J_\phi|>1750$ kpc\,\kms but find a decline with $|J_\phi|$ for $|J_\phi|<1750$ kpc\,\kms. \cite{Frankel2023} finds a general increase in amplitude with $|J_\phi|$, but both the amplitudes and the inferred times show significant variation on $\sim100$ kpc\,\kms scale, leading to local peaks around $|J_\phi|\sim1750$ kpc\,\kms in amplitude, and $|J_\phi|\sim1900$ kpc\,\kms in inferred time. \cite{Alinder2023} also model the amplitude of the phase spiral over $2000<|J_\phi|<2600$ kpc\,\kms, and find an increase in amplitude from $2000\lesssim |J_\phi| \lesssim2350$ kpc\,\kms, before a decrease towards a dip at $|J_\phi|\sim2500$ kpc\,\kms. The general increase in amplitude is qualitatively consistent between \cite{Frankel2023} and \cite{Alinder2023}, but details differ. This is not surprising as \cite{Frankel2023} expands a local sample as a function of $J_\phi$ while \cite{Alinder2023} explores a sample with a wider range of physical $R$ across the outer disc, meaning the two samples contain stars with different orbital eccentricities even for the same $J_\phi$.

Both \cite{Antoja2023} and \cite{Frankel2023} assume that a single interaction should create phase spirals with a single inferred interaction time. However, \cite{Gandhi2022} show that a single satellite passage should not create phase spirals with a single, uniform interaction time for realistic interactions, and different stars in the disc experience the maximum force from the satellite at different times over a period of hundreds of Myr, which should lead to smoothly varying phase spiral parameters as a function of galactic position or actions \citep[as seen e.g.\ in][]{DarraghFord2023,Alinder2023}.

This is a key point when interpreting `impact time' measurements from phase spirals and can explain some of the variation within other studies which search for a single impact time (see also Section~\ref{sec:perturbations_phase_spiral_self_gravity} for further complications arising from self-gravity). Conversely, models which fit the phase spirals frequently find structured residuals in the fits \citep[e.g.][]{Frankel2023,Alinder2023}, hinting that the simple single spiral models are incomplete, and the data likely contain multiple perturbations.

\begin{figure}
\includegraphics[width=\linewidth]{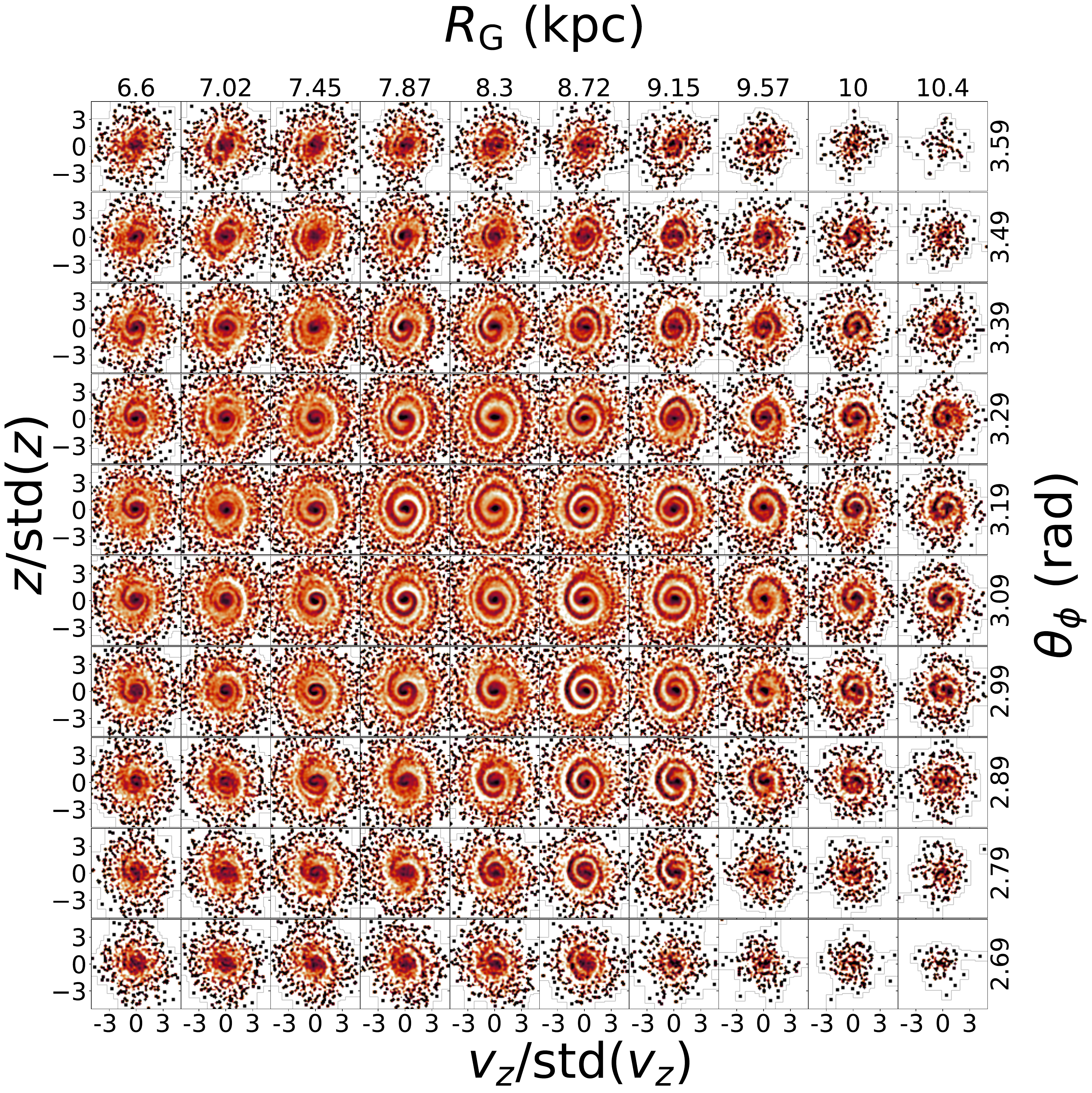}
\caption{The phase spiral in \Gaia DR3 splitting a local sample of stars ($d_{\mathrm{cyl}}<1$ kpc) by guiding radius, $R_{\mathrm{G}}$, and azimuthal phase angle, $\theta_\phi$, reproduced following \cite{Hunt2022}. Note that this is a split of the same sample in Figure~\ref{fig:phasespirallocal}, and that each panel is normalised by the standard deviations of coordinate ($\sqrt{\langle z^2 \rangle}$) and velocity ($\sqrt{\langle v_z^2 \rangle}\equiv\sigma_z$) in the corresponding spatial region.}  \label{fig:phasespiralsplit}
\end{figure}

\cite{Hunt2022} extended the split of the local sample to include splitting the sample by both the azimuthal action, $J_\phi$ and the azimuthal phase angle, $\theta_{\phi}$ (the conjugate phase angle to the azimuthal action; see Section~\ref{sec:observations_orbits_integrals}). This `two dimensional' expansion showed \textit{qualitatively} that the phase spiral varies not only as a function of $J_\phi$, but also in $\theta_{\phi}$, as seen in Figure~\ref{fig:phasespiralsplit}, which is a reproduction of figure 2 from \cite{Hunt2022}. While this enables them to probe orbit space away from the Solar neighbourhood, it introduces a bias in epicyclic phase and radial action, as discussed in \cite{Hunt2022} and \cite{DarraghFord2023}. In brief, the panels in the centre of the figure will contain stars with low $J_{\mathrm{R}}$ and the panels towards the edge of the figure will contain stars with high $J_{\mathrm{R}}$. Thus, \cite{Hunt2022} still find phase spirals present in groups or stars with higher $J_{\mathrm{R}}$, in contrast to \cite{Li2020a}, likely owing to a combination of the improved data quality and quantity in \Gaia DR3, and the dissection of the sample by $J_\phi$ and $\theta_\phi$.  

Using this dynamical projection, \cite{Hunt2022} also discovered two-armed \zvz phase spirals in the Solar neighbourhood stars with low $L_z$ (seen on the left-hand side of Figure~\ref{fig:phasespiralsplit}), indicating a breathing mode interior to the Solar radius \citep[as previously measured in, e.g.][]{Williams2013,Carrillo2018}. They argue that the transition from one- to two-armed phase spirals imply that more than one perturbation is present, which must be taken into account in both the mapping of phase spiral parameters across the disc, and also the modelling of the phase spiral when attempting to uncover the origin of the perturbations. \cite{Hunt2022} linked this change to the transition from internally to externally excited phase spirals, as discussed below in Section~\ref{sec:perturbations_phase_spiral_origins}.

\begin{figure}
\includegraphics[width=\linewidth]{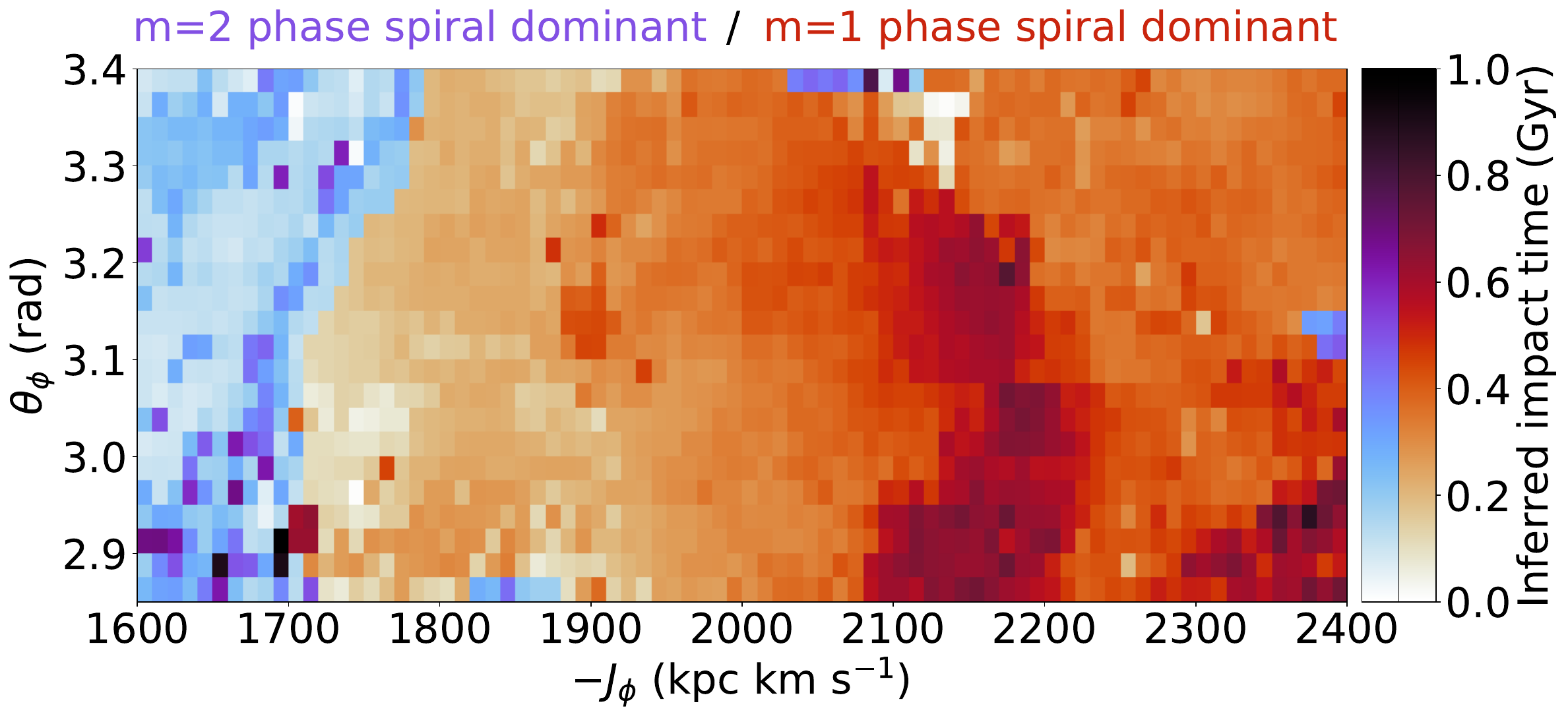}
\caption{`Impact time' inferred from the \texttt{Escargot} algorithm of \cite{DarraghFord2023} as applied to \Gaia DR3 data, loosely following their figure 8, over a wider angular momentum range and separating impact time estimates from one-armed (red) and two-armed (blue) phase spirals. While there is a clear and coherent variation with $J_\phi$ and $\theta_\phi$, these specific time estimates neglecting self-gravity should be treated with caution (See Section~\ref{sec:perturbations_phase_spiral_self_gravity} for discussion of the accuracy of inferred time values when neglecting self-gravity).}  \label{fig:phasespiraltimes}
\end{figure}

\cite{DarraghFord2023} followed this up with a measurement of the phase angle and pitch angle of the phase spirals seen when splitting a local volume of stars as a function of $J_\phi$ and $\theta_\phi$, by extracting the density ridgeline of the spiral projected into $\sqrt{J_z}\cos\theta_z$--$\sqrt{J_z}\sin\theta_z$. In this projection, the phase spiral is given by
\begin{equation*}
\theta_z = (\theta_0+\Omega_z t_0)\,\mathrm{mod}\frac{2\pi}{m},
\end{equation*}
\citep{Binney2018}, allowing them to fit for an inferred impact time from each phase spiral. \cite{DarraghFord2023} find coherent variation in both interaction time and phase angle. Figure~\ref{fig:phasespiraltimes} shows an updated reproduction of their figure 8, illustrating the inferred interaction time for individual phase spirals in each pixel (calculated using their \textsc{Escargot} method), as a function of $J_\phi$ and $\theta_\phi$, where $m=1$ spirals are shown in red, and $m=2$ spirals are shown in blue, finding a transition in the dominance of one-armed to two-armed phase spirals around $|J_\phi|\sim1700$ kpc\,\kms.

The two-armed phase spiral was subsequently seen in radial velocities in \cite{Antoja2023} for $|J_\phi|\lesssim1600$ kpc\,\kms, but is not resolved in the density, potentially owing to midplane selection effects, or the lack of a $\theta_{\phi}$ selection. \cite{Li2023d} subsequently show that the two-armed spirals appear clearly as a function of radial velocity, $v_{\mathrm{R}}$ (km s$^{-1}$) for $\sim1200<|J_\phi|<1800$ kpc\,\kms, while \cite{Alinder2024} find that two-armed spirals are only visible for a narrow range of $7.5<R<8.5$ kpc and $1400<|J_\phi|<1000$ kpc\,\kms. Both studies also link the two-armed phase spirals to a breathing motion (as do the more theoretical studies of \citealt{Banik2023} and \citealt{Widrow2023}). \cite{Alinder2024} also show that two-armed phase spirals are significantly clearer when stars are selected by azimuthal phase angle, $\theta_\phi$, rather than the physical azimuthal angle, $\phi$, and that the phase-angle of the two-armed phase spiral increases with $\theta_{\phi}$ by $\sim2.25\deg$ per degree of $\theta_{\phi}$.

\subsubsection{The `Iron snail'}  \label{sec:perturbations_phase_spiral_chemistry}

\cite{Tian2018} show that the phase spiral is present, and similar in morphology, for stars with all ages $\tau<6$ Gyr, using data from \Gaia and LAMOST. \cite{Laporte2019b} also showed that the phase spiral has the same shape across all ages \citep[for a given $R$ and $\phi$, using astrometry from \Gaia DR2, and stellar ages from][]{Sanders2018} suggesting that all stars responded to the same recent perturbation, while \cite{BlandHawthorn2019} showed that the phase spiral is more clearly visible in the low-$\alpha$ disc, using data from \Gaia and GALAH. \cite{Gaia2023c} and \cite{Alinder2023} then showed that the phase spiral is also directly visible when visualising it as a function of stellar chemistry, using the \Gaia DR3 GSP-Spec catalogue. The one-armed phase spiral appears in the residual $\Delta$[M/H] once subtracting off the running median metallicity from 100 pc bins. \cite{Horta2024a} also see the phase spiral in the residuals for [Mg/Fe] when subtracting their Orbital Torus Imaging model (see Section~\ref{sec:dynamics_chemistry}). \cite{Alinder2024} also show that the two-armed phase spiral is (faintly) visible in [M/H] for a limited range of $R$ and $J_\phi$.

\cite{Frankel2024} show that this is expected for any stellar label with a vertical gradient (such as [Fe/H], [$\alpha$/Fe], stellar age etc., but also $\overline{v_\phi}$). If there is a phase spiral present in the density distribution, there will also be a phase spiral in the label distribution. Note that this also has implications for the spiral arms, as discussed in Section~\ref{sec:spiralarm}.

The vertical gradient in [Fe/H] leads to a clear phase spiral in the residuals, $\Delta$[Fe/H], hence referred to as the `Iron Snail' (and similarly for other labels). Tracking the phase spiral in label space can be less influenced by selection effects, such as extinction in the Galactic plane, or the survey selection function, providing that it does not depend on the given stellar label.

\cite{Frankel2024} fit the amplitude and `dynamical age' from the `Iron snail' (using [Mg/Fe]) as a function of $L_z$ and compare it to their fits of the phase spiral in the density from \cite{Frankel2023}. While broadly consistent in the overall dynamical time found, there are notable differences in the oscillations in dynamical age as a function of $L_z$. The amplitude of the phase spiral increases with $L_z$ in both density and [Mg/Fe], yet while there are significant oscillations in the amplitudes (as calculated from the density phase spiral), the increase is smooth in [Mg/Fe].

Considering the expected agreement between the density and stellar labels, as above, this may seem unexpected. However, the measurements from the phase spiral in the density are from `local stars' separated into fine resolution $L_z$ bins, while the `Iron snail' sample has a much wider spatial coverage, with wider bins, compensating for fewer stars. In addition, the fits to the phase spiral in \cite{Frankel2023} are calculated assuming the \texttt{MWPotential2014}\ from \texttt{galpy}, while the fits from \cite{Frankel2024} make use of empirically derived parameters derived from Orbital Torus Imaging \citep{PriceWhelan2021,PriceWhelan2024}, see Section~\ref{sec:dynamics_chemistry}. As such, the differences between the amplitudes and dynamical ages inferred from density-based phase spiral and the `Iron Snail' may simply reflect the different nature of the samples, and the picture should become clearer with future surveys. What is clear, is that stellar labels are again a powerful complement to purely dynamical studies, with different advantages and disadvantages.

\subsubsection{Phase spiral origins}  \label{sec:perturbations_phase_spiral_origins}

There are now several proposed origins for the \zvz phase spiral, as discussed below.

\textbf{The direct influence of the Sagittarius dwarf galaxy:} Perhaps the most obvious contender is the Sagittarius dwarf galaxy, which is currently merging into the Milky Way following multiple pericentric passages and disc crossings over several Gyr. Numerous theoretical models \citep[e.g.][]{Binney2018,Bennett2021,Banik2022,Banik2023}, test-particle simulations \citep{BlandHawthorn2019,Li2021a,Gandhi2022} and live $N$-body simulations \citep[e.g.][]{Haines2019,Laporte2019b,Khanna2019,Hunt2021,BlandHawthorn2021,Bennett2022} show that a dwarf galaxy will induce a \zvz phase spiral as it merges into the Milky Way. However, attaining a quantitative reproduction of the phase spiral in the \Gaia data in a live simulation (e.g. matching pitch angle, amplitude etc.) in a simulation while also matching the properties of the actual Sagittarius galaxy (mass, orbit, etc.) has proven more difficult.  

One issue is that the low mass of the present day Sagittarius remnant ($\sim 4\times10^8\,M_\odot$; \citealt{Vasiliev2020}) appears inconsistent with the amplitude of the phase spiral and with the larger-scale vertical perturbation and asymmetry in the disc. For example, \cite{Bennett2022} run a series of $N$-body models tailored to the last Gyr or so of the Sagittarius interaction, and find that the shape and wavelength of the disc asymmetry do not match the data. However, it is expected that the Sagittarius progenitor was significantly more massive in the past \citep[e.g.][]{Gibbons2017}, and \cite{BlandHawthorn2021} showed that it is possible to reconcile the amplitude of the perturbation and the low mass of the present day remnant if Sagittarius has experienced significant mass loss during its last orbit, but no models have currently been able to $quantitatively$ explain all aspects of the data at the same time.

\textbf{The direct influence of a different Milky Way satellite galaxy:} \cite{Banik2022,Banik2023} use linear perturbation theory to investigate the disc response to Sagittarius along with the Milky Way's other larger satellites, namely the LMC, Hercules, Leo II, Segue 2, Bootes 1, Draco 1 and Ursa Minor. They conclude that Sagittarius dominates the Solar neighbourhood response at the present day by an order of magnitude. While the LMC is currently perturbing the Milky Way (see Section~\ref{sec:halo_perturbation}) and has a significant effect on the disc \citep[e.g.][]{Laporte2018}, there has not been sufficient time to form phase spirals. It does not appear possible that any current Milky Way satellite other than Sagittarius could be responsible for the \zvz phase spiral (although they do contribute small torques, which may contribute to the `everything' option below).

\textbf{Buckling of the galactic bar:} \cite{Khoperskov2019} proposed the first internal explanation for the phase spirals. They showed that the buckling of a bar in an $N$-body simulation creates long lived vertical bending waves which give rise to one-armed phase spirals (owing to the breaking of vertical symmetry) across the disc. Other studies have not found evidence of a recent bar buckling event in the Milky Way \citep[e.g.][]{Hey2023}, but the bending waves themselves can last for several Gyr following the buckling phase owing to the presence of self-gravity.

\textbf{A slowing bar and transient spiral structure:} \cite{Hunt2022} reproduced the two-armed `breathing' phase spirals (and no one-armed `bending' spirals) in an isolated self-consistent $N$-body barred spiral galaxy. \cite{Monari2016} have previously shown that bars and spiral arms together create significant breathing modes, in a non-linear superposition of the isolated effects of both components. The bar in the model of \cite{Hunt2022} does not buckle, and the two-armed spirals appear to correlate with the transient spiral structure. \cite{Banik2023} also used their linear perturbation theory framework to show that transient spiral structure $can$ create phase spirals, but it requires a quick growth and decay of the spiral potential. \cite{Li2023d} also reproduce two-armed phase spirals in a test-particle model with a growing and slowing bar, and a two-armed spiral pattern, but only for a limited range of times. They note that self-gravity is likely important in maintaining the two-armed phase spirals (see also \citealt{Widrow2023} and Section~\ref{sec:perturbations_phase_spiral_self_gravity}). Thus, the combination of a slowing bar and transient spiral structure appears to be able to explain the presence of the two-armed breathing spirals well, but is likely not responsible for the main one-armed phase spiral in the Solar neighbourhood and outer disc, which require a separate origin. For example, \cite{Hunt2022} also use their model as a host galaxy for a satellite interaction simulation, where the satellite induces one-armed phase spirals, leading to the transition between one- and two-armed phase spirals as seen in the \Gaia data.

\textbf{Stochastic heating from dark matter subhaloes:} \cite{Tremaine2023} propose that instead of one single large perturber, the phase spiral can be caused by Gaussian noise from many small perturbations, such as from subhaloes in the dark matter halo. They also show that the phase spirals originating from this mechanism should have an age of approximately 0.5 Gyr, which is consistent with the average of the data, but lacks the variation in age as a function of angular momentum and azimuth. Whether the stochastic background is dominant compared to single large perturbers will depend on the subhalo mass distribution function. \cite{Tremaine2023} suggest that for $\Lambda$CDM the largest haloes will dominate, but it will be a continuum, rather than a single force from a single satellite. \citet{Gilman2024} explore this scenario further and find that while the perturbation from dark (starless) haloes are individually an order of magnitude weaker than those from luminous satellite galaxies, the large number predicted from $\Lambda$CDM would create a significant and consistent perturbation to the Solar neighbourhood, although this alone remains insufficient to match the observed amplitude of the phase spiral.

\textbf{Torque from a distorted dark matter halo:} \cite{Grand2023} propose that the dark matter wake induced by a large satellite can be more important than the direct force from the satellite. They show that in a cosmological simulation where a $10^{10}$ M$_{\odot}$ satellite falls into a disc galaxy about 6 Gyr before their present day, it is the torque from the dark matter wake that creates \zvz phase spirals around the `Solar neighbourhood' in their present-day disc which are qualitatively similar to those seen in the \Gaia data. It is worth nothing that Sagittarius may have experienced its first interaction with the MW approximately 6 Gyr ago, leaving the possibility that Sagittarius did indeed cause the phase spirals, just indirectly. Another contender is the earlier GSE merger, if it did indeed lead to a tilted triaxial halo, as advocated by \cite{Han2023}. However, \cite{Laporte2018} find that in their model, the force from the dark matter halo wake is comparable to the direct influence of the satellite only at early times. Thus, whether the `halo wake' is a larger influence than Sagittarius itself depends on the true evolutionary history and properties of Sagittarius, the Milky Way, and the geometry of the merger.

\textbf{A cosmological formation history and gas accretion:}
\cite{GarciaConde2022,GarciaConde2024} show that phase spirals are prevalent in a cosmological simulation with a complex formation history, with satellite impacts, a misaligned dark matter halo, dark subhaloes as above, and they also add a misaligned gas disk and accretion to the list of perturbative influences as opposed to a singular large impact event \citep[see also][for an exploration of bending and breathing waves in a warped disc, and how misaligned gas accretion can lead to bending waves]{Khachaturyants2022a,Khachaturyants2022b}.

\textbf{The `everything' option:}
In practice, the Milky Way is interacting with Sagittarius, has a bar (whether buckled or not), has spiral arms, likely has dark matter subhaloes and a lumpy triaxial halo (depending on the true nature of dark matter), is interacting with other smaller satellites even if their effect is minor, and has some cosmological formation history which includes gas accretion. All these dynamical mechanisms cause some level of vertical perturbation to the disc, resulting in phase spirals that are ubiquitous, yet difficult to relate to a single cause.

Thus, the question becomes: Are the \zvz phase spirals in the Milky Way caused by one dominant pertuber? Or do they arise from a complex combination of the above? There are now many proposed mechanisms that qualitatively match the observed properties of the phase spiral, but one needs to make clear predictions to differentiate between these scenarios and quantitatively compare them to current and future observations.

However, the picture is further complicated by the fact that phase spirals may be erased by scattering over shorter timescales than their mixing time in a smooth potential. \cite{Tremaine2023} show that small-scale kicks, such as from interaction with GMCs in the disc, can erase phase spirals on timescales of $\sim$ 1 Gyr \cite[or even $\sim0.6$ Gyr;][]{Gilman2024}, and as discussed above, many studies consistently find `interaction times' of $\lesssim1$ Gyr \citep[][]{Antoja2023,Frankel2023,DarraghFord2023}. This then calls into question any interpretation of the phase spirals that links their origin with some singular perturbation further in the past. However, the density of GMCs decreases beyond the Solar neighbourhood, and orbital timescales are longer, so future surveys aimed at mapping the phase spirals across the outer disc may be needed to quantitatively link the disc response to a specific perturbation.

\subsubsection{The role of self-gravity}  \label{sec:perturbations_phase_spiral_self_gravity}

This is not a theory focused review, yet in this Section we discuss briefly the effect of self-gravity on the formation and evolution of the phase spirals, to help contextualise results and interpretation of the \Gaia data. We direct the reader to the papers discussed in this Section for a more thorough review of the underlying theory.

As mentioned above, studies of the phase spiral that attempt to `unwind' it to infer interaction times, rely on some form of the argument that the spiral is caused by a displacement, which then phase mixes in a well-behaved manner, following the gradient of vertical frequency $\Omega_z$ as a function of vertical action $J_z$. In practice, this is not so straightforward. In an `ideal' case, for an impulsive interaction in a fixed potential, these methods work well. However, as discussed in \cite{Binney2018}, these methods make use of `unperturbed frequencies', while in reality when the disc is perturbed, regardless of the origin of this perturbation, the period of the vertical oscillations of stars can become significantly longer than $2\pi/\Omega_z$.

\cite{Darling2019} show nicely how a one-armed phase spiral emerges naturally from a bending perturbation, and how $R$--$z$ coupling is important in order to reproduce the $\overline{v_\mathrm{R}}$ and $\overline{v_\phi}$ dependence of the phase spiral. They also show that the winding speed of the phase spiral differs between two simulations of an induced bending wave that include or do not include self-gravity. Namely, the phase spiral in the self-gravitating disc shows approximately half the number of wraps compared to the test-particle disc, and the bending wave is sustained over the 1 Gyr run time of the simulation in the self-gravitating case. This is also seen in \cite{BlandHawthorn2021}, whose $10^8$ particle self-gravitating disc model shows a density wave and a bending wave following a single impact. The bending wave winds up slowly, leading to a corrugated wave that persists for $\sim1.5$~Gyr. They also find that it takes $\sim380$~Myr for a phase spiral to form at the Solar radius following the impact, a delay time which challenges recent impact time measurements of a few hundred Myr derived from the models currently being applied to the \Gaia data. \cite{BlandHawthorn2021} suggest that the phase spiral seen in the \Gaia data instead originates from an earlier Sagittarius impact, approximately 1--2 Gyr before the present day. This behaviour is also seen in the time evolution of the self-gravitating simulation of \cite{Hunt2021}, where phase spirals only appear several hundred Myr following the first impact\footnote{An animation of the phase spiral evolution across the disc is available at \url{https://zenodo.org/records/8402668}. While the first impact occurs at $t=2.4$ Gyr, the phase spirals appear closer to $t=3$ Gyr. See \cite{Hunt2021,Hunt2024} for model description and discussion.}.

\cite{Widrow2023} further explores the role of self-gravity on the formation and evolution of both one- and two-armed (or bending and breathing) phase spirals in the context of swing amplification in a shearing box. \cite{Widrow2023} shows that swing amplification can enhance the amplitude of a phase spiral before it starts to wind, and that stationary phase spirals can form following a co-rotating mass. They also show that the winding rate of the phase spiral is significantly slower when self-gravity is included in their model. In one example, their self-gravitating model produces a phase spiral with a single `wrap' over the same timescale as it takes the non-self-gravitating phase spiral model to wrap three times. This has obvious and significant implications for estimating impact times from phase spirals in the \Gaia data, namely that one may expect the times inferred to be underestimates from the true impact times \textit{if} the phase spirals arise from a single perturbation.

\subsubsection{Phase spiral summary}  \label{sec:perturbations_phase_spiral_summary}

So far, efforts to map the phase spiral across the Milky Way disc in $R$, $\phi$, $L_z$ and $\theta_{\phi}$ show coherent variation in amplitude, pitch angle and phase-angle, as discussed in Section~\ref{sec:perturbations_phase_spiral_mapping} and as predicted by single perturbation models \citep[e.g.][]{Antoja2023,Frankel2023,DarraghFord2023,Alinder2023,Frankel2024,Alinder2024}. It also appears clear that the local phase spirals are linked to the disc-wide corrugations as discussed in Section~\ref{sec:perturbations_feathers} \citep[e.g.][]{Laporte2019b,BlandHawthorn2021}, as any large perturbation will influence the planar motion of stars \citep[and potentially create $\Delta R-v_{\mathrm{R}}$ phase spirals;][]{Hunt2024} as well as their vertical motion.

It remains to be seen whether a `complex combination', or the `Stochastic heating' scenario can lead to such coherent signals, or whether one would expect a more uniform set of phase spirals in this case. The Milky Way appears to have experienced a comparatively quiet history \citep[e.g.][]{Fragkoudi2020}, and thus it may be that a single recent perturbation is the dominant cause of the currently visible phase spiral(s), on top of a weaker stochastic background.

Alternatively, future studies may show that the coherent pattern in phase spiral properties is to be expected even in the `complex combination' scenario, making it difficult or impossible to use the phase spirals to learn about a specific perturbative event in our Galaxy's past. However, even if there is no single clear origin, the phase spirals are still useful for studies of the Milky Way's potential (see Section~\ref{sec:dynamics_phase_spiral}).

Inferred interaction times range from a few hundred Myr to just over 1 Gyr, with significant scatter, but most current models neglect self-gravity, likely biasing such estimates towards shorter times. The study of the phase spiral is a comparatively young field, and there are many unanswered questions. Future works will need to satisfy the following constraints:
\begin{itemize}
    \item Winding models that infer interaction times need to include the effect of self-gravity and the large-scale disc response \citep[e.g.][]{Darling2019,Widrow2023}
    \item Models must be able to explain the coherent change in phase spiral properties with $R,\phi,L_z,\theta_\phi$ \citep[][]{Antoja2023,Frankel2023,Alinder2023,DarraghFord2023}, and link local phase spirals to the disc-wide corrugations \citep[e.g.][]{Laporte2019b,BlandHawthorn2021,Hunt2021}.
    \item Models must account for the transition and overlap from two-armed to one-armed phase spirals in the inner disc, or at least explain both separately \citep{Hunt2022,Li2023d,Alinder2024}.
    \item Models must be able to retain strong phase spirals at the present day despite the fact that the phase spirals can be erased by scattering from giant molecular clouds \citep{Tremaine2023,Gilman2024} on dynamically short timescales.
\end{itemize}

\section{Halo}  \label{sec:halo}

\Gaia brought transformative changes in our knowledge of the structure and dynamics of the stellar halo.
Since the halo is believed to be one of the oldest components of the Galaxy, and also contains various remnants of accreted stellar systems, its study to a large extent belongs to the fields of Galactic archaeology and streams, and is covered in other reviews in this series (\citealt{ArchaeologyReview}, \citealt{StreamsReview}, see also \citealt{Helmi2020}). Here we summarise the findings relevant for stellar dynamics.

\subsection{Kinematics}  \label{sec:halo_kinematics}

Two complementary ways of studying the halo population are the selection of stars on halo-like orbits locally in the velocity space (i.e., the Toomre diagram showing the distribution in $v_\phi$--$v_R$ or $v_\phi$--$v_\bot$ spaces, where $v_\bot \equiv \sqrt{v_R^2+v_z^2}$), and a more global view offered by the integrals-of-motion space, such as $E$--$L_z$ or actions \citep{Myeong2018,Lane2022}.

\citet{Bonaca2017} analysed $\sim10^3$ kinematically selected stars in the local stellar halo using \Gaia DR1 astrometry (TGAS) in combination with RAVE line-of-sight velocities, and found two chemically distinct populations -- metal-rich and metal-poor, the first of which was rather unexpected and conjectured to have in-situ origin.
At the same time, \citet{Belokurov2018} used a much larger catalogue of PM obtained by combining \Gaia DR1 and SDSS astrometry with a baseline of 10--15 years \citep{Deason2017}, with a similar precision as TGAS, but for fainter stars and therefore a larger volume. This sample of MS stars with photometric distance estimates, and metallicity and line-of-sight velocity from SDSS, was split into several bins in metallicity and $z$ coordinate. They discovered that the more metal-rich component ([Fe/H]${}\gtrsim -1.5$) displayed a clearly bimodal distribution in $v_R$--$v_\phi$ space, with the disc stars occupying the region around $v_\phi\sim 200$~\kms and the halo stars having much higher dispersion in $v_R$ than in $v_\phi$ (i.e., a strong radial anisotropy), the hints of which were already seen in \citet{Bonaca2017}, but not identified as such. \citet{Belokurov2018} suggested that this metal-rich, strongly radially-anisotropic component of the halo must originate from a single accretion of a massive enough satellite galaxy early in the history of the Milky Way. This population was independently discovered in the \Gaia DR2 RVS dataset, with metallicity provided by APOGEE, by \citet{Haywood2018} and \citet{Helmi2018}, who also advocated a similar accreted origin for it. It became known as GSE (see \citealt{Evans2020} for a brief historical overview of the concept and name).

The spatial distribution and kinematic properties of the GSE debris and other halo populations have been extensively studied using \Gaia astrometry and complementary spectroscopic surveys.
\citet{Bird2019} examined the velocity dispersion of K giants in the halo with photometric distances and line-of-sight velocities from LAMOST and PM from \Gaia DR2. They found that the velocity anisotropy $\beta\equiv 1-\sigma_{\text tan}^2/(2\sigma_{\text rad}^2)$ of the entire sample stays at the level 0.8 up to 20--25~kpc and then rapidly drops to 0.4--0.5, while selecting a metal-rich subset ([Fe/H]${}\gtrsim -1.8$) gives an even higher $\beta\simeq 0.9$ in the same radial range.
\citet{Lancaster2019} performed a similar analysis using BHB stars from SDSS, and found that the velocity distribution is much better described by two rather than one Gaussian components, with the radially anisotropic one (GSE) having $\beta\simeq 0.9$ and rapidly disappearing beyond 25--30~kpc.
\citet{Bird2021} reanalysed both samples, using somewhat different selection criteria and sample cleaning procedures, and found lower values of $\beta$ but in a larger radial range, with BHB stars having systematically lower $\beta$ than K giants, which they attribute to the difference in their metallicity. \citet{Vickers2021} constructed another sample of halo BHB stars from LAMOST, and found $\beta\simeq 0.6$--0.7 up to 20~kpc, with more metal-rich stars having slightly higher values. \citet{Han2024} used the chemokinematic sample from the H3 survey, splitting it into in-situ and accreted components in the [$\alpha$/Fe] vs.\ [Fe/H] plane, and confirmed the high radial anisotropy for the latter.

With full 6d phase-space information and a reasonable choice of Galactic potential, one can compute orbital parameters (notably, peri- and apocentre radii) and thus extrapolate a relatively local sample of stars to the entire halo (subject to an obvious constraint that stars with pericentres above the maximum Galactocentric distance in the sample will be missed).
Using this approach, \citet{Deason2018} found that the apocentre radii of the metal-rich halo population, associated with the GSE merger, have a concentration around 20~kpc. A similar orbit-based analysis by \citet{Sato2022} confirmed that this metal-rich component is dominant up to 20~kpc, and has a more triaxial and boxy shape than the outer, more metal-poor halo.

\citet{Iorio2019,Iorio2021} analysed the density profile and (in the second paper) kinematic properties of RR Lyrae stars identified by \Gaia. The vast majority of these stars have no line-of-sight velocities, but by fitting multivariate Gaussian profiles to PMs in different sky areas, one can infer the 3d velocity distributions. Using the same two-component Gaussian mixture model as \citet{Lancaster2019}, they found that the radially-anisotropic population dominates within 25~kpc and has $\beta \simeq 0.9$, whereas the remaining (isotropic) halo population is slightly counter-rotating, and a few percent of stars have disc-like kinematics and spatial distribution. Using a similar approach, but in a smaller volume and treating the entire sample as a single population, \citet{Wegg2019} determined its anisotropy to be $\beta\simeq 0.6$--0.8 up to 20~kpc (the outer limit of their dataset).
\citet{Liu2022} examined a much smaller 6d subset of RR Lyrae with line-of-sight velocities and metallicity from LAMOST and SDSS and updated PM from \Gaia EDR3. They confirm that the metal-rich population ([Fe/H]${}>-1.8$) is highly radially anisotropic ($\beta\simeq 0.9$) and dominates between 4 and 30~kpc, whereas the remaining metal-poor population is still radially anisotropic with $\beta\simeq 0.5$.
 
To summarise, it is clear from multiple tracers that the stellar halo in the inner 20--30 kpc is dominated by a highly radially anisotropic and metal-rich component (GSE) with $\beta$ reaching 0.9 and metallicity peaking around $-1.5$. The general consensus is that it represents the material accreted from a relatively massive satellite that merged with the Milky Way $\sim10$~Gyr ago, and was the last major merger in its history (not counting the ongoing interaction with the LMC, which may also be marginally considered as a major merger).

In many of these studies, significant efforts have been spent on cleaning the sample from known unmixed structures, such as globular clusters, satellites (primarily Magellanic Clouds) and streams (chiefly the Sagittarius stream). Of course, one can turn the table and examine the spatial distribution and kinematics of different dynamically coherent populations. This topic is reviewed in detail in \citet{ArchaeologyReview}; here we only mention that the GSE is dominant between 5--10 and 20--30 kpc (see \citealt{Mackereth2019}, \citealt{Naidu2020}, \citealt{An2021a}, \citealt{Ablimit2022}, \citealt{Wang2022b}, \citealt{Wu2022}, in addition to the studies described above), and Sagittarius takes over further out (e.g., figure 19 in \citealt{Naidu2020}).

Similarly to the Galactic disc, the stellar halo can be studied with the help of DF-based dynamical models, which link the spatial and kinematic properties via the collisionless Boltzmann equation.
\citet{Hattori2021} and \citet{Li2022a} applied the DF-fitting approach described in Section~\ref{sec:disc_df} to the RR Lyrae stars, which have precise distances and relatively well measured PM, but no line-of-sight velocities, using individual stars rather than binned data. Although in this case the RR Lyrae are tracer populations, whose mass is negligible for the overall potential, one still has a choice of keeping the potential fixed or letting it vary during the fit. The former study took the second option and used the data to constrain the Galactic potential, particularly the shape of the dark halo. However, both papers considered only a single DF for the entire population, which may be an oversimplification given the spatially varying mix of in-situ and accreted stars.

\subsection{Spatial structure}  \label{sec:halo_structure}

The full-sky coverage of the \Gaia catalogue and the addition of astrometric information for a cleaner selection of halo stars makes it much easier to study the density profile and the shape of the stellar halo.

The halo is often modelled as a broken power-law profile with inner and outer indices and a break radius, sometimes allowing for different shape and/or orientation in the two asymptotic regimes (see table 6 in \citealt{BlandHawthorn2016} for a compilation of pre-\Gaia measurements, and figure 15 in \citealt{Li2022a}, figure 12 in \citealt{Han2022b} and table 6 in \citealt{Medina2024} for more recent compilations).

\citet{Iorio2018,Iorio2019} studied the halo using RR Lyrae stars from \Gaia DR1 and DR2, respectively. They found the halo to be close to spherical in the inner Galaxy (within a few kpc), and triaxial further out, with the intermediate axis tilted from the equatorial plane by some 20$^\circ$, forming two roughly opposite structures at Galactocentric radii 10--20~kpc: Virgo Overdensity (VOD) and Hercules--Aquila Cloud (HAC). \citet{Simion2019} demonstrated that both VOD and HAC are dominated by stars on very radial orbits, and thus could be linked to the GSE merger debris, which are expected to produce localised shell-like structures. \citet{Perottoni2022} confirmed these findings with more detailed analysis of chemistry and kinematics based on \Gaia EDR3, APOGEE and SEGUE. This interpretation is corroborated by $N$-body simulations of the GSE merger by \citet{Naidu2021}, who found that the merger debris have an elongated triaxial shape and that the two overdensities are produced by successive apocentres of the progenitor orbit (however, in their model it is the major, not intermediate axis, that connects the two structures). The association between VOD and debris from a nearly-radial merger is also advocated by \citet{Donlon2019}; however, this group argues for a more recent merger event (a few Gyr instead of 8--10, as commonly assumed for the GSE merger), based on the mixing rate of debris \citep{Donlon2020,Donlon2024}.
As a counter-argument, \citet{Han2022a} suggested that rapid mixing rate may be an artifact caused by the assumption of a spherical potential; if the potential of the dark halo itself is tilted with respect to the disc plane, the stellar halo may keep its misalignment much longer.

\citet{Han2022b} modelled the spatial distribution of RG stars from the H3 survey combined with \Gaia. The accreted halo population was defined by a kinematic and chemical selection, and by construction, dominated by the GSE debris, and modelled as a triaxial density profile with three power-law segments (i.e., a `doubly-broken' power law). Their best-fit model was tilted by some 25$^\circ$ from the disc plane, and had break radii around 12 and 28 kpc, which they interpret as the last two apocentre radii of the GSE progenitor orbit.

\citet{Balbinot2021} used a similar approach to \citet{Deason2018}, integrating the orbits of nearby halo stars selected from the 6d subset of \Gaia EDR3, and confirmed that their apocentres concentrate around the VOD and HAC regions on the sky, as well as a couple of other locations predicted to contain GSE-related overdensities.
\citet{Belokurov2023} examined a much larger 6d catalogue of \Gaia DR3 within a few kpc, and identified several chevron-like structures in the radial phase space ($r$ vs.\ $v_r$), which are essentially kinematic signatures of apocentre pile-up. \citet{Wu2023} proposed that additional chevrons should be located at larger distances, up to 25--30~kpc, based on the orbit integration of chemically and kinematically selected local halo stars.

Further out, \citet{Ye2023} identified additional break radii at $\sim 50$ and 90~kpc in the density of RR Lyrae, associating them with more distant shell-like structures remaining from earlier orbital apocentres of the GSE progenitor, although without any kinematic signatures, as the PM uncertainties for RR Lyrae in \Gaia DR3 are too large (Figure~\ref{fig:velocity_error_rrl}); the situation will considerably improve in future data releases. \citet{Chandra2023} instead used significantly brighter RG stars with photometric distance estimates using metallicity determined from BP/RP spectra calibrated against H3 spectroscopy, and adding line-of-sight velocity measurements from existing catalogues and dedicated follow-up spectroscopic observations. They identified several structures beyond 40~kpc, and also interpreted them as more distant apocentres of the GSE progenitor.

The total mass of the stellar halo is estimated to be (1--1.5)${}\times10^9\,M_\odot$ \citep{Deason2019b,Mackereth2020}, somewhat higher than pre-\Gaia estimates summarised in section 6.1 of \citet{BlandHawthorn2016}. According to various studies \citep{Helmi2018,Naidu2020,Han2022b,Kurbatov2024}, GSE debris contributes roughly a half of the stellar halo mass, i.e., (5--8)${}\times 10^8\,M_\odot$. Using typical relations between stellar and halo masses at redshifts 1.5--2, this implies a fairly major merger with a mass ratio as high as 1:2.5 \citep{Naidu2021}. This also conforms with the expected orbital history of the GSE progenitor: in order to sink deep into the Galactic potential before being disrupted, it needs to have experienced strong dynamical friction, which also makes its orbit more eccentric \citep{Amorisco2017,Vasiliev2022}. However, some studies \citep{Mackereth2019,Lane2023} advocate for a lower mass of GSE debris in the halo, at the level (1.5--3)${}\times10^8\,M_\odot$, which would translate to a more minor merger with the total mass of the GSE progenitor in the range (1--3)${}\times10^{10}\,M_\odot$ \citep[e.g.][]{Mackereth2020}.  A low \textit{stellar} (not \textit{halo}) mass ratio for the GSE merger ($\lesssim 1$:20) is also advocated by \citet{Fragkoudi2020} based on the relatively undisturbed kinematics and high amount of rotation in the bulge region.
So the matter is not settled yet.

\subsection{Perturbations in the halo}  \label{sec:halo_perturbation}

In addition to the complicated structure and kinematics of the stellar halo resulting from a superposition of multiple accreted components, it has also been affected by the more recent events -- namely, the ongoing interaction with the LMC. This topic is covered in detail in \citet{Vasiliev2023} and section 5.2 of \citet{ArchaeologyReview}; here we summarise the key points.

It is now well established that the LMC is fairly massive, with most estimates being in the range (1--2)${}\times 10^{11}\,M_\odot$ (see figure 1 and section 2.2.1 in \citealt{Vasiliev2023}), which makes it only 5--10 times less massive than the Milky Way (and only 2--4 times smaller than the enclosed Galactic mass within its current distance of 50~kpc). Moreover, it has just passed its pericentre with a relative velocity exceeding 300 \kms, and being part of a binary system, our Galaxy has acquired a fraction of this velocity in the centre-of-mass reference frame. However, the Milky Way is not a rigid body, and outer parts of the Galactic halo have not experienced the same gravitational acceleration from the LMC as the inner part. As a result, the outer halo (beyond $\sim 30$~kpc) appears to move preferentially `up' (towards positive $z$, i.e. `Galactic North') in the Galactocentric reference frame. This displacement manifests itself as a dipole signal in density (higher in the Northern hemisphere), line-of-sight velocity (positive/negative towards North/South poles respectively), and $\mu_b$ (positive everywhere). The kinematic signatures have been detected in various tracers (halo stars, globular clusters and satellite galaxies) by \citet{Erkal2021} and \citet{Petersen2021}. The density perturbations proved to be more elusive: \citet{Conroy2021} reported a dipole signal with the expected orientation, but twice stronger than predictions from simulations (e.g., \citealt{GaravitoCamargo2019}), whereas \citet{Amarante2024} find no statistically significant signal in the distribution of BHB stars. The orientation of the velocity dipole (sometimes called `the apex of the reflex motion') is also somewhat inconsistent between studies, see \citet{Yaaqib2024}, \citet{Chandra2024b} and \citet{Bystrom2024} for recent updates and comparisons.
Despite these discrepancies, it is clear that the overall effect of the LMC on the kinematics and structure of the outer halo cannot be neglected (the velocity perturbations are at the level of few tens \kms).

\subsection{Hypervelocity stars}  \label{sec:halo_hypervelocity_stars}

The fastest stars in the Milky Way are not gravitationally bound to it, and can be produced by two mechanisms. The first is the tidal disruption of a stellar binary by a supermassive black hole (SMBH), in which one of the components of the binary remains bound to the SMBH and the other is ejected \citep{Hills1988}, or a three-body scattering of a single star by a binary SMBH \citep{Yu2003}; in both cases the orbit of the ejected star originates from the Galactic centre. A few of these hypervelocity stars (HVS) have been discovered in the pre-\Gaia era (see \citealt{Brown2015} for a review). The second mechanism produces (hyper-)runaway stars originating from the stellar disc as a result of a disruption of a binary stellar system when one of its components explodes as a supernova.

HVS are interesting for Galactic dynamics in several ways: first, they probe the environment around the central SMBH \citep[e.g.,][]{Evans2022,Verberne2024}, and second, their trajectories constrain the shape of the Galactic potential \citep[e.g.,][]{Gnedin2005}, although quite a large sample would be needed to resolve degeneracies \citep{Contigiani2019,Gallo2022}, and the gravitational influence of the LMC cannot be ignored \citep{Kenyon2018,Boubert2020,Liao2023}.
Historically, HVS candidates were mostly selected in the outer halo as young blue giants, which are unlikely to be formed at their present location, and since their velocities are almost entirely radial, they could be confirmed using only spectroscopy.
\Gaia DR2 made it possible to search for objects with high three-dimensional velocities among the 6d RVS dataset, but since true HVS are expected to be very rare, the sample of candidates is overwhelmingly dominated by objects whose measurement errors scatter them to higher distances and/or PM, even though the actual velocity is below the escape speed.
Several studies \citep{Hattori2018,Bromley2018,Marchetti2019,Du2019} have identified a few dozen possible candidates, but only a few stars remaining after applying various quality cuts. In addition, it was discovered that the reported line-of-sight velocity measurements can be off by more than 500~\kms in case of contamination of the RVS spectra by a nearby bright star \citep{Boubert2019}; this problem has been rectified in EDR3. For the majority of candidates, the updated astrometry in EDR3 brought down the estimated sky-plane velocity \citep{Scholz2024}, highlighting the difficulty of finding a needle in a haystack. \citet{Marchetti2021} and \citet{Marchetti2022} updated the catalogue of high-velocity stars using EDR3 and DR3, respectively, but did not find any new high-quality candidates whose orbits could be unambiguously traced back to the Galactic centre (i.e., true HVS), although \citet{Erkal2019a} found one high-velocity star that could originate from the LMC centre.
In the meantime, \citet{Koposov2020} serendipitously discovered a star with a total Galactocentric velocity exceeding 1700~\kms in the S$^5$ spectroscopic survey, which can be traced back to the Milky Way nuclear star cluster with high precision. However, its extremely fast motion means that its orbit is a nearly straight line and provides no constraints on the Galactic potential. To date, this remains the only genuinely HVS found in the \Gaia dataset, but future data releases are expected to uncover more objects \citep{Marchetti2022}.

\subsection{Summary of stellar halo}  \label{sec:halo_summary}

Like in the rest of the Milky Way, the advent of \Gaia uncovered new levels of complexity in the stellar halo:
\begin{itemize}
\item The most dramatic discovery is undoubtedly the GSE \citep{Belokurov2018,Helmi2018}, which is generally acknowledged to be the last major merger in the Milky Way history. Some of its properties are known rather well, e.g., high radial anisotropy ($\beta \sim 0.9$, e.g., \citealt{Bird2019,Lancaster2019,Liu2022}), relatively high metallicity (median [Fe/H]$\simeq -1.2$, \citealt{Mackereth2019,Naidu2020}), accretion time (8--10 Gyr ago according to most estimates), and to some extent its spatial structure (e.g. association of VOD and HAC with its apocentres, \citealt{Simion2019,Naidu2021}).
However, its mass estimates vary by a factor of a few, and even some basic kinematic features (e.g. whether it is prograde or retrograde at low binding energies) disagree between studies.
\item The effects of the GSE merger on the subsequent Milky Way evolution are debated: it may have triggered the formation of the bar \citep{Merrow2024}, caused the precession of the disc plane \citep{Han2023}, or created chevrons in the $r$--$v_r$ phase-space \citep{Belokurov2023}. However, all these features have alternative interpretations unrelated to GSE.
\item Other accreted structures in the halo have been proposed in many studies, but there remains much confusion about their properties, selection criteria, etc.
\item Undoubtedly, our ability to identify these structures has been affected by the ongoing interaction with the LMC, which creates velocity perturbations at the level of a few tens \kms in the outer halo, but there is some disagreement about the orientation of the main dipole component of this kinematic signal \citep{Yaaqib2024,Chandra2024b}, and the detection of corresponding density perturbations is still debated \citep[e.g.,][]{Amarante2024}.
\item In view of the complexity of substructures in the halo, one may question the utility of approximating it with simple profiles (single or broken power laws in density, or smooth DF models), or using it for constraining the Galactic potential, as discussed in the next Section.
\end{itemize}

\section{Global dynamical modelling}  \label{sec:dynamics}

\subsection{Basics}  \label{sec:dynamics_intro}

One of the main science goals of the \Gaia mission is to provide kinematic measurements across a large part of the Galaxy, which would be used to constrain its gravitational field and mass distribution -- including, of course, any unseen matter. This inference on the gravitational potential is called dynamical modelling, and is discussed in detail, e.g., in \citet{Binney2013}, \citet{Wang2020a}, \citet{Gardner2021}, \citet{deSalas2021}; here we recall its basic principles.

Galaxies are usually considered to be collisionless stellar systems, ignoring two-body relaxation and any dissipative processes related to gas. Such systems are described by the distribution function(s) of one or more components (different populations of stars, dark matter, etc.) in the 6d phase space $\{\boldsymbol x, \boldsymbol v\}$. Each DF $f_{\text c}$ satisfies the collisionless Boltzmann equation (CBE)\footnote{see \citet{Beraldo2019} for discussion of the CBE (in)applicability on timescales longer than a few dynamical times.}, which in the most general (time-dependent) case reads
\begin{equation}  \label{eq:CBE}
\frac{\partial f_{\text c}}{\partial t} + \boldsymbol v\,\frac{\partial f_{\text c}}{\partial \boldsymbol x} - \frac{\partial \Phi}{\partial \boldsymbol x}\,\frac{\partial f_{\text c}}{\partial \boldsymbol v} = 0.
\end{equation}

By taking moments of the CBE over velocities at a given position $\boldsymbol x$, one obtains the set of Jeans equations; in practice, these are almost always used under the assumption of stationarity and in spherical or axisymmetric geometry. In the latter case, there are two equations:
\begin{eqnarray}
\frac{\partial\Phi}{\partial R} &\!\!=\!\!&
\frac{\overline{v_\phi^2}}{R} - \frac{\overline{v_R^2}}{R}
-\frac{1}{\rho} \frac{\partial\big(\rho\,\overline{v_R^2  }\big)}{\partial R}
-\frac{1}{\rho} \frac{\partial\big(\rho\,\overline{v_R v_z}\big)}{\partial z},  \label{eq:jeans_radial} \\
\frac{\partial\Phi}{\partial z} &\!\!=\!\!&
-\frac{1}{\rho} \frac{\partial\big(\rho\,\overline{v_z^2  }\big)}{\partial z}
-\frac{1}{\rho} \frac{\partial\big(\rho\,\overline{v_R v_z}\big)}{\partial R}
-\frac{\overline{v_R v_z}}{R}.  \label{eq:jeans_vertical}
\end{eqnarray}
Here $\rho$ and $\overline{v_i v_j}$ are the density and second moments of velocity of the given tracer population, while $\Phi$ is the total potential, which is of course related to the sum of densities of all components via the Poisson equation.
In both equations, the first term in the r.h.s. is the dominant one, but other terms can also contribute non-negligibly. In particular, $\overline{v_R v_z}$ is directly related to the tilt of the velocity ellipsoid discussed in Section~\ref{sec:disc_velocity_ellipsoid_orientation}, and can be quite significant at large $|z|/R$.

The vertical Jeans equation (\ref{eq:jeans_vertical}) is commonly used to probe the gravitational field near the disc plane, often using the one-dimensional approximation (i.e., keeping only the first term in the r.h.s.).
On the other hand, the radial Jeans equation (\ref{eq:jeans_radial}) usually appears in studies that probe the circular-velocity curve in the disc plane $v_\text{circ}(R) \equiv \sqrt{R\,\partial\Phi/\partial R}\big|_{z=0}$.
Finally, combining the two equations under certain assumptions about the shape and orientation of the velocity ellipsoid, one arrives at a global dynamical model of the Galaxy. It should be noted that the Jeans equations use only the first two velocity moments and ignore the information contained in the shape of the velocity distribution (particularly in $f(v_\phi)$, which is significantly non-Gaussian). Moreover, they require the knowledge of not only the kinematics, but also the spatial density profile, and the latter is usually more difficult to measure (Sections~\ref{sec:observations_selection}, \ref{sec:disc_spatial_structure}).

Aside from Jeans equations, another popular approach is based on the Jeans theorem, which states that in the equilibrium state, the DF may only depend on the integrals of motion (usually at most three, instead of six phase-space coordinates). Two of these integrals (energy $E$ and $z$-component of angular momentum $L_z$) are readily available in the axisymmetric case, while the third integral can be derived using the St\"ackel approximation for the potential \citep[e.g.,][]{Bienayme2015}. Alternatively, one may use the triplet of radial, vertical and azimuthal actions $\{J_r, J_z, J_\phi\equiv L_z\}$, which are also usually computed using the St\"ackel approximation \citep{Binney2012}. The often used assumption of separable motion in the $R$ and $z$ dimensions is less accurate than the St\"ackel approximation, but is still suitable near the disc plane ($\lesssim 1$~kpc). In the context of DF-based modelling, the one-dimensional DF $f(E_z)$, where $E_z\equiv \Phi(R,z)+v_z^2/2$ is an approximate integral of motion, provides a complementary approach to the vertical Jeans equation \citep[e.g.,][]{Kuijken1989}.
More realistically, the Galaxy contains multiple stellar populations, which differ in metallicity, $\alpha$-element abundances, ages, etc., and this complexity can be taken into account in two ways: either considering multiple mono-age, mono-abundance populations described by separate DFs (e.g., \citealt{Bovy2013,Ting2013}), or introducing an extended DF, in which the dynamical parameters (e.g., velocity dispersions) are linked to stellar properties (e.g., \citealt{Sanders2015,Das2016}). In any case, the constraints on the gravitational potential from the dynamics of multiple stellar populations are stronger. A conceptual advantage of DF-based models over Jeans equations is that they may be meaningfully constrained by the shape of the velocity distribution at different locations throughout the Galaxy, without the detailed knowledge of the density profile.

Finally, the \citet{Schwarzschild1979} orbit-superposition approach and the Made-to-Measure (M2M) method of \citet{Syer1996} still rely on the assumption of dynamical equilibrium, but can deal with a more complex geometry (e.g., a triaxial rotating bar). In the context of recent Milky Way studies, only the latter method has seen some practical applications, so far limited to constraining the mass distribution in the bar region and its pattern speed \citep{Portail2017,Clarke2022}. \citet{Khoperskov2024a,Khoperskov2024b} developed a variant of the Schwarzschild method for transferring the age and chemical information from a limited footprint of the APOGEE survey to the entire disc and bar, but kept the gravitational potential fixed throughout their analysis.

\subsection{Dynamics in the stellar disc}  \label{sec:dynamics_disc}

\subsubsection{Vertical dynamics in the Solar neighbourhood}  \label{sec:dynamics_disc_vertical}

Using either the vertical Jeans equation or the 1d DF $f(E_z)$, one can measure the vertical force $K_z \equiv \partial\Phi/\partial z$, or equivalently the surface mass density in a slab $\mbox{$\Sigma(<\!z)$} \equiv \int_{-z}^z \rho_\text{total}(z)\;\text{d}z$, which is related to the vertical force by $K_z(z) \approx 2\pi\,G\,\mbox{$\Sigma(<\!z)$}$; the approximation neglects the $R$--$z$ tilt term and the radial gradient of the circular velocity (aka the `rotation curve term'). Here the total mass density $\rho_\text{total} = \rho_\star + \rho_\text{gas} + \rho_\text{DM}$ contains contributions from stars, gas and dark matter; the first two components are dominant near the equatorial plane, but their density rapidly decreases with $z$, unlike that of the dark matter, which is usually assumed not to be strongly concentrated near the equatorial plane. Thus, in order to constrain the latter, one needs to examine the dynamics of tracers sufficiently far from the plane; on the other hand, the 1d approach becomes increasingly inaccurate at large $|z|$. The measurements of $K_z$ or $\Sigma(<\!z)$ are often quoted at a certain compromise value of $z$, e.g., 1.1~kpc (dating back to \citealt{Kuijken1989}).

\citet{Hagen2018} used RC stars with PM from \Gaia DR1 (TGAS) and line-of-sight velocities from RAVE, splitting them into thin and thick components by metallicity and constructing maps of velocity dispersions within $\sim 1$~kpc from the Sun. They then applied the 1d vertical Jeans equation (\ref{eq:jeans_vertical}) for both tracer populations simultaneously, ignoring the tilt term, to derive the total mass density and constrain the scale heights of both discs. After subtracting the baryonic contribution, they deduced $\rho_\text{DM} = 0.018\pm0.002\,M_\odot\text{pc}^{-3}$, which is 1.5--2$\times$ higher than most recent estimates (see figure 1 in \citealt{deSalas2021} for a compilation).

\citet{Widmark2019a} also used the TGAS sample in a much smaller volume ($\lesssim 150$~pc) to build a 1d DF $f(E_z)$ and measure the total mass density $\rho_\text{total} \approx 0.12\,M_\odot\text{pc}^{-3}$ with a 10\% uncertainity. \citet{Widmark2019b} updated the analysis for the \Gaia DR2 kinematics, and inferred an excess of mass density near the disc plane that was significantly larger than the adopted model for the baryonic distribution. This could be attributed to a quite dramatic underestimate of the gas disc density, or to some form of dissipative dark matter concentrated near the plane. However, a more likely explanation proposed in the follow-up study \citep{Widmark2021a} that covered a larger spatial volume is that the measurements are systematically biased by non-equilibrium effects. \citet{Schutz2018} and \citet{Buch2019} employed a similar DF-based method and obtained only an upper limit on the dark disc density, though both studies found significant variation of results depending on the selected samples, and also admitted that systematic errors from the equilibrium assumption likely affect their conclusions -- a problem compounded by the small spatial region of their datasets.

\citet{Guo2020} used a combination of \Gaia DR2 and LAMOST, selecting $9\times10^4$ metal-rich dwarfs that represent the thin disc population in the range $0.2 \le (z/\text{kpc}) \le 1.3$ and a narrow range of Galactocentric radii. From the vertical Jeans equation with or without the tilt term, and placing a strong prior on the stellar surface density $\Sigma_\star$, they measured $\rho_\text{DM}=0.013\pm0.002\,M_\odot\text{pc}^{-3}$; however, with a non-informative prior on $\Sigma_\star$, the inferred $\rho_\text{DM}$ had much larger uncertainties and was nearly consistent with zero. A model with an additional thick disc component in the potential was likewise underconstrained, largely due to the limited vertical coverage of the dataset. Moreover, estimates of $\rho_\text{DM}$ from the separate analysis of northern and southern Galactic hemispheres differed by more than a factor of three -- a discrepancy attributed to disequilibrium effects.

\citet{Salomon2020} constructed vertical density profiles for $\sim\! 4\times10^4$ RC stars from \Gaia DR2 (though they derived distances from inverse parallax rather than photometry), of which $\sim$3/4 had \vlos measurements in the RVS dataset. They found global density variations between the two hemispheres, as well as more localised oscillations at a level $\lesssim10$\% (similar fluctuations have been found by \citealt{Bennett2019}), and associated variations in the velocity dispersion. After approximating both $\rho(z)$ and $\sigma_z(z)$ by smooth analytic functions, they determined $\Sigma(<\!z)$ from the vertical Jeans equation, including the tilt term and correcting for the circular velocity gradient with radius. At $|z|\ge 2$~kpc, both hemispheres yielded consistent estimates of $\rho_\text{DM} \simeq 0.010$--0.013\,$M_\odot\,\text{pc}^{-3}$. However, at lower altitudes there were significant fluctuations, especially in the northern hemisphere, which they also attributed to disequilibrium. With a different modelling method based on the three-integral DF and jointly fitting both hemispheres, they found a similar range of values for $\rho_\text{DM}$. This method is detailed in the follow-up study of \citet{Bienayme2024}, which used an updated DR3 sample of RC stars and a DF $f(E,L_z,I_3)$ (with the third integral $I_3$ computed in the St\"ackel approximation) to determine the gravitational field up to $|z|=3.5$~kpc from the midplane. At such large altitudes, the mass density is dominated by dark matter, and disequilibrium effects are less important; the inferred $\rho_\text{DM}(z=0)\simeq 0.013$\,$M_\odot\,\text{pc}^{-3}$ is consistent with other estimates, but is still sensitive to the assumed contribution of baryons, which was taken from the latest iteration of the Besan\c con model.

Recognising that a single-component model is clearly an oversimplified description of the Milky Way disc, \citet{Cheng2024} used the chemical tagging (metallicity and $\alpha$-abundance) to separate the \Gaia DR3 + APOGEE DR17 sample of $\sim2\times10^5$ giants into $\alpha$-poor and $\alpha$-rich discs and the stellar halo. They determined the 3d velocity dispersion profiles within a few kpc from the Sun, fitted them with simple linear or exponential trends, and applied the vertical Jeans equation (with the tilt term estimated from the data, and density profiles taken from the literature) to determine the profiles $\Sigma(<\!z)$ at several Galactocentric radii, separately from both disc populations. The results were significantly different between the two populations, with the $\alpha$-poor one being most discrepant from the literature; they attributed the disagreement to the shortcomings of the model, as well as the disequilibrium features in the data (more apparent in the colder $\alpha$-poor population).

\citet{Li2021b} used a sample of $10^5$ giants with 6d phase-space information to construct the vertical density and velocity dispersion profiles and measure the potential using a variant of the \citet{Kuijken1989} method with a more general type of the DF $f(E_z)$ describing a superposition of isothermal components. The novel feature of their approach is that they determined the density of stars in the $z$--$v_z$ space directly from the observations, and after subtracting the best-fitting model, found clear signatures of the phase spiral in the residuals, from which they estimated the perturbation to occur $\sim 0.5$~Gyr ago.
They found some differences between hot and cold populations, which they attribute to the simplifications of the 1d approximations. In a follow-up paper, \citet{Li2023b} used a $2.5\times$ larger dataset, and fitted it with a sum of two 3d quasi-isothermal DFs $f(E, E_z, L_z)$, with $E_z$ being a somewhat less well conserved quantity than $I_3$, but still useful to overcome the limitations of the 1d approximation. Despite the obvious disequilibrium signatures in the residuals, the inferred surface mass density was comparable to other studies.

\subsubsection{Dynamics across the entire disc}  \label{sec:dynamics_disc_global}

\begin{figure*}
\includegraphics{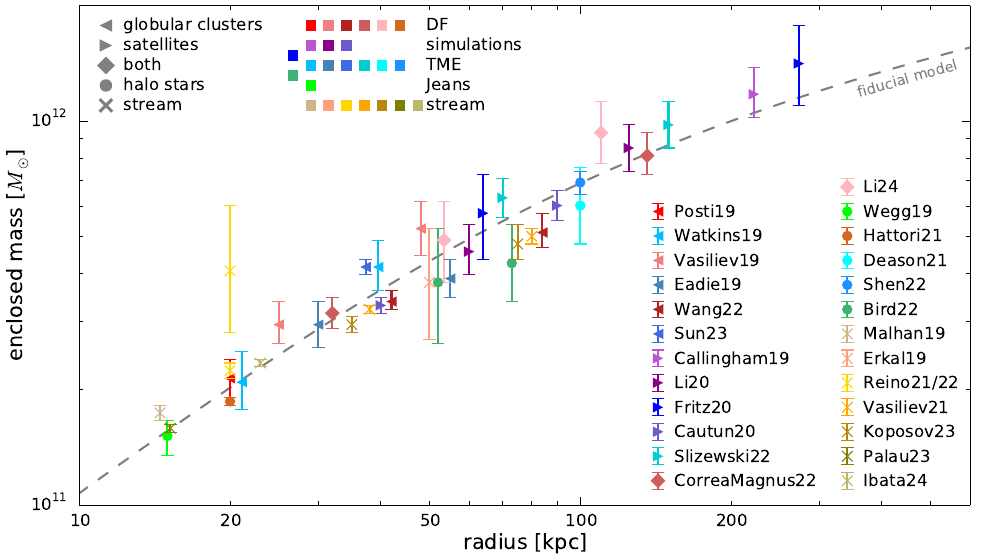}
\caption{A compilation of measurements of the MW enclosed mass profile. Marker styles indicate the kinematic tracers, while colours indicate the modelling method, as detailed in the top left corner. Studies that provide an entire mass profile \citep{Vasiliev2019,Eadie2019,Wang2022a,Li2020b,Cautun2020,Slizewski2022,CorreaMagnus2022,Vasiliev2021a,Koposov2023} are reduced to two points selected to minimise overlap with other points, while most other studies are shown by a single point at a fiducial radius, except a few cases \citep{Watkins2019,Fritz2020,Bird2022,Reino2021,Li2024b} where two tracer populations were used to determine the mass at two values of radii. Dashed line shows the fiducial Milky Way potential model from \texttt{gala} \citep{PriceWhelan2017}.
}  \label{fig:enclosedmass}
\end{figure*}

\begin{figure*}
\includegraphics{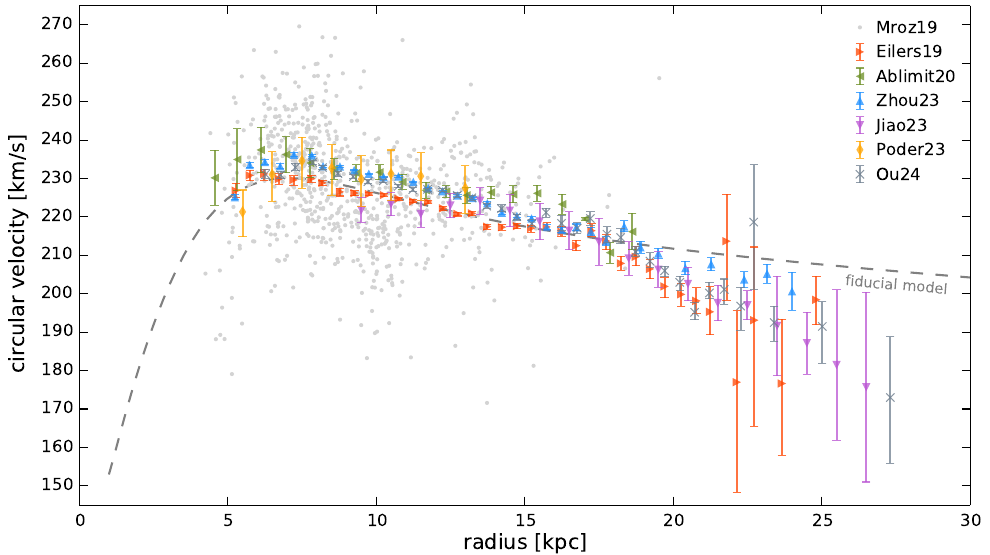}
\caption{A compilation of measurements of the Milky Way circular-velocity curve in the radial range 5--30 kpc. The dashed line shows the same fiducial model as in the previous plot.
}  \label{fig:vcirc}
\end{figure*}

\begin{figure}
\includegraphics{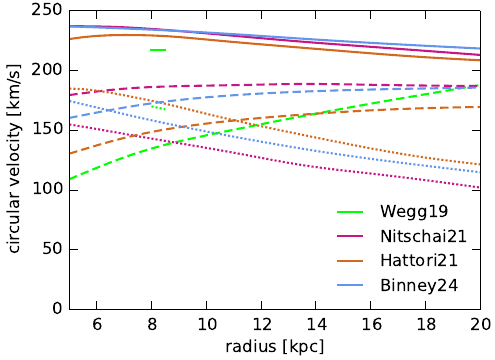}
\caption{Contributions of baryons (dotted lines) and dark halo (dashed lines) to the total circular velocity curve (solid lines), according to several dynamical modelling studies: \citet{Wegg2019} (quotes only the local value of circular velocity), \citet{Nitschai2021}, \citet{Hattori2021}, \citet{Binney2024}.
}  \label{fig:vcirc_decomposition}
\end{figure}

In one of the first dynamical studies to rely on \Gaia DR2 data, \citet{Wegg2019} constructed velocity dispersion maps of RR Lyrae in the radial range 2--20 kpc, using only their PM and distances. Then they employed the full 2d Jeans equations (including the tilt term measured observationally) to determine the total force field, and after subtracting the contribution of baryons from the \citet{Portail2017} model, constrained the mass distribution and shape of the dark halo. The shape turned out to be close to spherical, while the contribution of the halo to the circular-velocity curve within the Solar radius was relatively low (smaller than the baryonic one).

\citet{Nitschai2020} constructed 2d maps of mean $v_\phi$ velocity and three velocity dispersions in the meridional plane within a few kpc from the Sun and $|z|\le 2.5$~kpc, using $2\times10^6$ stars with precise 6d phase-space coordinates (a subset of \Gaia DR2 RVS sample). They then applied the axisymmetric Jeans Anisotropic modelling method of \citet{Cappellari2008,Cappellari2020} to determine the gravitational potential and the anisotropy parameter of the stellar distribution (using a single-component model describing the entire population, which is dominated by disc stars). \citet{Nitschai2021} updated the observational sample to \Gaia EDR3 PM and complemented it with distances and line-of-sight velocities from APOGEE, extending the radial range to 5--20~kpc. In both papers, the density profile of stars was assumed to follow the thin and thick discs with parameters taken from \citet{Juric2008}, and in the second paper they additionally considered a flared disc model, with little impact on the results. Their best-fitting dark halo had a considerably higher normalisation and a steeper density profile in the central region than in the \citet{Wegg2019} study.

Many studies restricted the analysis to the stars within the Galactic disc.
\citet{Mroz2019} and \citet{Ablimit2020} used $\mathcal O(10^3)$ Cepheid stars, whose distances can be measured rather precisely, in combination with \Gaia DR2 PM and \vlos, and inferred the circular velocity assuming that these stars move on nearly-circular orbits. Several other studies \citep{Eilers2019,Ou2024,Zhou2023} used spectrophotometrically estimated distances and line-of-sight velocities for $\mathcal O(10^4$--$10^5)$ red giants from APOGEE and LAMOST (in the latter paper) and \Gaia PM (DR2 in the first of these papers, EDR3 in the latter two). \citet{Poder2023} used photogeometric distances for $\sim 6\times10^5$ red giant stars from \Gaia DR3 RVS with good parallax measurements. Alternatively, \citet{Wang2023} applied the Lucy deconvolution method \citep{LopezCorredoira2019} to the entire \Gaia DR3 RVS sample ($\sim 3\times10^7$ stars), estimating the distance and 3d velocity distributions statistically even for stars with large relative errors in parallax. Since the old stellar populations are kinematically hotter and therefore have smaller rotational velocities, these studies (plus \citealt{Jiao2023}, who used the same sample as \citealt{Wang2023}, but with a more complete analysis of systematic uncertainties) compensated for this asymmetric drift using the 1d radial Jeans equation~(\ref{eq:jeans_radial}), which can be recast as follows:
\begin{equation*}
R\,\frac{\partial\Phi}{\partial R} =
\overline{v_\phi^2} - \overline{v_R^2} - \overline{v_R^2}\,\frac{\partial \ln \rho}{\partial \ln R} - R\,\frac{\partial \overline{v_R^2}}{\partial R} - \frac{R}{\rho}\,\frac{\partial \big(\rho\,\overline{v_R v_z}\big)}{\partial z}.
\end{equation*}
The left-hand side of this equation evaluated in the disc plane ($z=0$) is by definition $v_\text{circ}^2$, but of course the sample of stars covers a finite range of $z$ (typically 5--10 degrees from the disc plane). Neglecting the last term in the above equation ignores this fact and introduces a systematic error that is usually fairly small. The third term containing the logarithmic derivative of tracer density by radius is the most problematic one, since as discussed in Section~\ref{sec:observations_selection}, determining the density profile of tracers is much more difficult than their kinematic properties. It is usually assumed that the disc follows an exponential profile in radius, with a scale length between 2 and 3 kpc, which is one of main sources of systematic uncertainty. Finally, the above equation assumes a steady state, which is again questionable especially in the outer parts of the disc (Section~\ref{sec:perturbations}). These complications notwithstanding, there is a general agreement between $v_\text{circ}$ profiles derived in different studies (Figure~\ref{fig:vcirc}) up to $R\simeq 20$~kpc, although with systematic offsets of a few \kms. $v_\text{circ}$ steadily declines by $\sim 20$~\kms between 8 and 20~kpc. At larger radii, measurement uncertainties and systematic errors become larger, and the circular-velocity profile appears to decline even faster in some studies \citep{Eilers2019,Ou2024,Wang2023,Jiao2023}, though not in \citet{Zhou2023}; the \citet{Poder2023} analysis was limited to the radial range 5--14~kpc. Based on this accelerated decline, \citet{Jiao2021,Jiao2023}, \citet{SylosLabini2023} and \citet{Ou2024} argued for a rapidly declining DM halo profile (e.g., the \citealt{Einasto1965} model) with a virial mass of only a few times $10^{11}\,M_\odot$. However, this is inconsistent with the enclosed mass profile at larger radii found in many other studies (Figure~\ref{fig:enclosedmass}), as discussed below.
While the most natural resolution of this discrepancy would be that the circular-velocity profile derived from Jeans analysis of disc stars is biased low at large Galactocentric distances, the above papers performed a detailed analysis of systematic errors and concluded that they are unlikely to explain the magnitude of this bias, so this mismatch remains a puzzle (see \citealt{Koop2024} for a discussion).

As discussed in Section~\ref{sec:disc_df}, some (though not all) dynamical modelling studies based on DFs are also designed to constrain the Galactic potential. With the action-based iterative approach, \citet{Binney2023,Binney2024} measured the circular-velocity curve in agreement with other studies, although only up to Galactocentric radius of 15 kpc. Although the overall amount of mass in this region is well constrained, there is still a significant freedom in redistributing it between baryons (stars and gas) and dark matter, as illustrated in Figure~\ref{fig:vcirc_decomposition} (see also figure 4 in \citealt{deSalas2019}).

\subsection{Dynamics in the halo}  \label{sec:dynamics_halo}

The Galactic disc `runs out of stars' beyond 20--25 kpc, so any dynamical mass measurement further out must rely on other tracer populations. Globular clusters are well suited for this purpose: almost a half of $\gtrsim 150$ currently known clusters had their PM measured by \citet{Gaia2018c}, and PM of nearly all remaining ones have soon been measured by \citet{Baumgardt2019} and \citet{Vasiliev2019}, while \citet{Vasiliev2021b} and \citet{Vitral2021} updated these values using EDR3 astrometry. Together with precisely measured distances \citep{Baumgardt2021} and line-of-sight velocities from various studies, this gives full 6d phase-space coordinates up to distances beyond 100 kpc, though the majority of clusters are located within the Solar radius and only a handful beyond 50 kpc.
Dwarf galaxies orbiting the Milky Way are less numerous but extend to even larger distances, and several groups independently measured their PM from DR2 \citep{Fritz2018,Kallivayalil2018,Massari2018,Simon2018,Pace2019,McConnachie2020a} and EDR3 \citep{McConnachie2020b,Li2021c,Battaglia2022,Pace2022}.
Individual stars such as K giants, BHB stars and RR Lyrae are also suitable for this purpose, although their distances and especially PM are much less precise than those of entire stellar systems. In addition, the outer stellar halo is full of incompletely relaxed substructures, the most important being the Sagittarius stream, which needs to be excised from the dataset (usually by applying a spatial mask). The cleaned samples contain $\mathcal O(10^2)$ stars beyond 50~kpc, being the largest population of tracers.

These tracer populations were used in numerous studies using several dynamical modelling approaches. Distribution function-based methods evaluate the likelihood of each object in the given model and then use the Bayes theorem to compute the likelihood of the model itself, given the data. They require that the observed sample is complete or at least has a well-determined selection function, and thus are mainly applied to globular clusters \citep{Posti2019,Vasiliev2019,Wang2022a} or satellite galaxies \citep{CorreaMagnus2022,Li2024b}, although \citet{Hattori2021} used it with RR Lyrae satisfying a well-defined spatial selection (excluding stars at low latitudes or beyond 25~kpc). Another popular approach is the Tracer Mass Estimator (TME, \citealt{Evans2003,Watkins2010}), which assumes that both the potential and tracer density follow power-law profiles and that the velocity anisotropy is constant, and produces an estimate of the enclosed mass at the outermost data point. It has been applied to halo stars \citep{Deason2021,Shen2022,Bird2022}, globular clusters \citep{Watkins2019,Eadie2019,Sun2023a} and satellite galaxies \citep{Fritz2020,Slizewski2022}.
However, the more distant is the tracer population, the less reliable is the assumption of dynamical equilibrium and phase-mixedness, which underly the classical dynamical modelling approaches. Recognising this limitation, some studies opted for simulation-driven potential inference methods, comparing the positions and velocities of observed satellite galaxies with appropriately scaled populations of haloes from cosmological simulations \citep{Callingham2019,Li2020b,Cautun2020,RodriguezWimberly2022}, although sometimes they provide only the total mass estimate and not the entire profile, and in most cases are limited to specific parametric forms of the dark halo potential (e.g., NFW). The most significant perturbation in the outer halo comes from the ongoing interaction with the LMC (Section~\ref{sec:halo_perturbation}), but it can be approximately accounted for in the context of equilibrium models \citep{Deason2021,CorreaMagnus2022}.

\subsection{Modelling of stellar streams}  \label{sec:dynamics_streams}

Last but not least, modelling of stellar streams offers another opportunity to constrain the Galactic potential across a range of scales. Currently around 100 streams have been discovered (see \citealt{Mateu2023} for a compilation), although the number of `useful' ones is much smaller: the stream needs to be long enough to provide substantial constraints on the potential.

\citet{Malhan2019} used orbit fitting for $\sim 100$ stars from the GD-1 stream with full 6d phase-space information, and measured the mass between 8 and 20~kpc (the radial extent of the stream), as well as found a marginally significant preference for an oblate halo.
\citet{Palau2023} fitted three streams  with known globular cluster progenitors (NGC 3201, M~68 and Pal~5) between 10 and 20 kpc, separately and all together, using 5D astrometry from \Gaia DR2 plus line-of-sight velocities for some stars in Pal~5. The results were consistent between streams, and slightly favoured a prolate halo shape.
\citet{Ibata2024} analysed a much larger catalogue of 29 streams with updated \Gaia DR3 astrometry, augmented by various line-of-sight velocity measurements from literature and dedicated follow-up observations. They performed fits of the Galactic potential with a dozen free parameters, and found a rather strong preference for an oblate halo with axis ratio $q\simeq 0.75$.
The last two studies also used the circular-velocity measurements from \citet{Eilers2019} and other external observational constraints in fitting the potential, while \citet{Malhan2019} fixed the stellar disc parameters and varied only the halo mass and flattening. Thus the inference about the Galactic mass distribution from streams is not entirely independent from other studies.

The above described studies used streams within $\sim 20$--25~kpc from the Galactic centre. There are only a handful of streams at larger distances, and two of them are particularly instrumental in constraining the Galactic potential. The first one is the Orphan--Chenab stream (initially discovered as two independent streams in the Northern and Southern hemispheres), located between 15 and 60 kpc and spanning over half of a great circle on the sky. Soon after \Gaia DR2, \citet{Koposov2019} and \citet{Fardal2019} independently discovered a misalignment between the stream track and the Solar reflex-corrected direction of PM of stream stars. Such a misalignment cannot occur in a static Galactic potential, regardless of its geometry, and \citet{Erkal2019b} demonstrated that it could be explained by the gravitational perturbation from the LMC, which passes within 10~kpc from the stream at the closest point. By modelling the stream in a time-dependent potential of the Milky Way and LMC moving under mutual gravitational forces, they were able to constrain the LMC mass and the Galactic potential simultaneously; in a subsequent paper, \citet{Koposov2023} confirmed these results with a much improved observational dataset and an updated modelling method. \citet{Shipp2021} used the same approach to measure the mass of the LMC from the perturbation inflicted on several other streams in the Southern hemisphere, and found a result consistent with the above studies of the Orphan--Chenab stream, although they assumed a fixed Milky Way potential and only varied the LMC properties.

The most prominent tidal structure in the outer halo is the Sagittarius stream, which has been extensively mapped out with \Gaia DR2 \citep{Antoja2020,Ramos2020,Ibata2020} and EDR3 \citep{Ramos2022} across its more than 360$^\circ$ track on the sky and the range of Galactocentric distances from 20 to over 100~kpc. \citet{Vasiliev2021a} discovered a similar misalignment between the stream track and the PM direction, and demonstrated that it also can be explained by the effect of the LMC. The constraints on the Galactic mass distribution from this study and from the analysis of the Orphan--Chenab stream are very similar, and in both cases the Milky Way dark halo needs to have a rather complicated shape with radially varying axis ratios and orientation, in order to adequately reproduce the properties of the streams, although the inferred potentials do not match in detail. By contrast, \citet{Wang2022c} modelled the Sagittarius stream in a much lighter Milky Way, but their simulations displayed quite significant deviations from the observed line-of-sight velocity profile along the stream track. Finally, \citet{Valluri2021} proposed the figure rotation of the triaxial dark halo of the Milky Way as another mechanism capable of producing misalignment between PM and the stream track.

Another approach to constraining the Galactic potential with tidal streams is to maximise the clustering of debris in the space of integrals of motion (e.g., actions, \citealt{Sanderson2015}). Its advantage is that, in principle, one does not need to assign membership labels for multiple structures beforehand, although a contamination by a smooth background population would reduce the efficiency of the method. \citet{Reino2021} applied this approach to three `classical' tidal streams (Pal~5, GD-1 and Orphan), as well as the Helmi stream, which is an older and more phase-mixed structure. The enclosed mass constraint from three streams was comparable to other recent studies, while adding the Helmi stream significantly increased both the value and its uncertainty\footnote{The original paper had an error that was corrected in \citet{Reino2022}; Figure~\ref{fig:enclosedmass} shows the updated result.}.
\citet{Yang2020} introduced a related method based on the two-point correlation function in action space and applied it to the \Gaia DR2 RVS dataset, although their derived circular-velocity curve was systematically lower than most other measurements, possibly due to the effect of selection cuts that are yet to be fully understood and mitigated.

\subsection{Escape speed}  \label{sec:dynamics_escape_speed}

\begin{figure}
\centering
\includegraphics{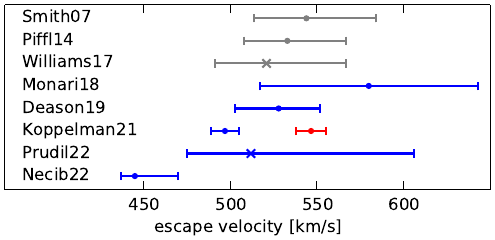}
\caption{Compilation of escape velocity determinations in the Solar neighbourhood. Grey points with error bars show pre-\Gaia studies based on RAVE and SDSS \citep{Smith2007,Piffl2014a,Williams2017}, blue -- those after \Gaia DR2 \citep{Monari2018,Deason2019a,Koppelman2021b,Necib2022,Prudil2022}. Circles show papers that used local samples (within a few kpc from the Sun), crosses -- papers that mapped the escape velocity across a range of Galactocentric radii. The study of \citet{Koppelman2021b} estimated that their measurement is biased low by $\sim 10\%$, and the value after this correction is shown in red.
}  \label{fig:v_escape}
\end{figure}

One of the more direct ways of constraining the Galactic potential is through the measurement of the escape velocity $v_\mathrm{esc} \equiv \sqrt{2(\Phi_\infty-\Phi_\odot)}$, where $\Phi_\odot$ is the potential at the Solar location and $\Phi_\infty$ is the potential at infinity (or rather, at the virial radius of the Galaxy or its small multiple). Under the assumption of dynamical equilibrium, one should not expect any stars with velocities above $v_\mathrm{esc}$, except for a handful of ejected hypervelocity stars (Section~\ref{sec:halo_hypervelocity_stars}) that may be indeed leaving the Milky Way.
\citet{Leonard1990} introduced the now standard method, in which the velocity distribution of stars near the escape boundary is assumed to follow a power law, $f(v) \propto (v_\mathrm{esc}-v)^k$, and developed practical approaches for dealing with observational uncertainties and missing dimensions (at the time and until \Gaia DR2, the method could only be used with line-of-sight velocities). Unfortunately, there is a strong degeneracy between the power-law index $k$ and the escape velocity (the higher the index $k$, the more sparsely populated is the tail of the velocity distribution and the higher is the escape velocity). Earlier studies based on a relatively small RAVE catalogue \citep{Smith2007,Piffl2014a} had to introduce strong priors on $k$ to obtain any constraints on $v_\mathrm{esc}$. With a significantly larger sample from SDSS, \citet{Williams2017} were able to constrain both parameters, as well as the radial profile of the escape velocity up to 50~kpc. They also used a more statistically robust analysis allowing for outliers in the dataset -- an improvement shared by subsequent studies.

\citet{Monari2018} explored the full 3d velocity distribution of the \Gaia DR2 RVS sample, using $\sim 3000$ stars on retrograde orbits. They again used a restricted prior on $k$, and arrived at a somewhat larger value of $v_\mathrm{esc}$, but still compatible with pre-\Gaia estimates that used only line-of-sight velocities. However, all these studies relied on the velocity distribution being isotropic -- an assumption clearly disproved by the discovery of highly radial GSE debris in the halo.

\citet{Deason2019a} analysed a large suite of simulated Milky Way analogues and found that those whose assembly history and kinematic signatures are most similar to our Galaxy (dominated by a single, old, radially anisotropic merger), have somewhat lower values of $k$ (at the level 1--2.5) than assumed by earlier studies (2--4). With this revised prior on $k$, and using largely the same subset of retrograde stars from \Gaia DR2 RVS, they obtained a lower estimate of $v_\mathrm{esc}$ than \citet{Monari2018}, but very close to pre-\Gaia studies (which used higher ranges of $k$).
The escape speed can be translated into the total mass of the Galaxy, assuming some parametric form for the dark halo (typically NFW); the above papers all converged on the range (1--1.5)${}\times10^{12}\,M_\odot$. However, by applying the method to a suite of cosmological simulations, \citet{Grand2019} found that the mass determined from $v_\mathrm{esc}$ is underestimated by some 20\%.

\citet{Koppelman2021b} examined a large dataset of main-sequence stars with photometric distance estimates and having PM but not necessarily line-of-sight velocity measurements. Their estimates from 5D and 6D samples were consistent at 2$\sigma$ level, and both favoured higher values of $k\sim3$--3.5 than advocated by \citet{Deason2019a}, but nevertheless had lower inferred $v_\mathrm{esc}$, with much smaller uncertainties than earlier studies because of greatly increased sample size. However, by calibrating the method on simulations, they estimated that the escape velocity is biased low by $\sim 10\%$; after applying the correction, the Milky Way mass was found to be (1--1.2)${}\times10^{12}\,M_\odot$.

\citet{Prudil2022} examined RR Lyrae stars with 6D phase-space coordinates from \Gaia EDR3 and complementary spectroscopic surveys, and using a fairly broad prior on $k$, obtained a similarly broad range of $v_\mathrm{esc}$, consistent with all other studies. Their sample spanned a large range of Galactocentric distances, and the enclosed mass within 20 and 30 kpc agreed well with other studies shown in Figure~\ref{fig:enclosedmass}.

\citet{Necib2022} generalised the \citet{Leonard1990} estimator to the case of multiple populations with different values of $k$, as well as a (small) admixture of outliers, and applied it to the 6d dataset from \Gaia EDR3. With two- or three-component fits, they obtained a considerably lower $v_\mathrm{esc}$ than previous studies that assumed a single halo population, and thus their estimated Milky Way mass was also more than twice lower, $\lesssim 5\times 10^{11}\,M_\odot$, although they also admitted that the inferred $v_\mathrm{esc}$ might be biased low by $\sim 10\%$. \citet{Roche2024} applied the method to a larger sample of stars across a range of Galactocentric radii, and updated the Milky Way mass estimate to $(6.4\pm1.5)\times10^{11}\,M_\odot$.

In summary, the estimates of $v_\mathrm{esc}$ in the \Gaia era are mostly consistent with the earlier studies (except the last two papers, which used a more flexible model for the velocity distribution and obtained substantially lower values). A caveat mentioned in several papers is that the presence of unrelaxed populations may lead to an underestimate of the escape velocity and the inferred Milky Way mass by 10--20\%. Although the masses inferred from $v_\mathrm{esc}$ measurements are in the same range as other literature estimates, they are inherently based on local samples of stars and extrapolated to the virial radius, and for this reason we do not show them on Figure~\ref{fig:enclosedmass}.

\subsection{Other methods}  \label{sec:dynamics_other}

\subsubsection{Phase spiral}  \label{sec:dynamics_phase_spiral}

Most dynamical modelling methods rely on the assumption of equilibrium, and are therefore adversely affected by the perturbations discussed in Section~\ref{sec:perturbations}. However, given a specific model for these non-equilibrium features, one can use them for constraining the potential; an example would be the modelling of stellar streams, which are not phase-mixed, but originate from a compact clump in the integrals space. In a series of papers, \citet{Widmark2021b,Widmark2021c,Widmark2022a} introduced a novel method that exploits the phase spiral (Section~\ref{sec:perturbations_phase_spiral}) to measure the vertical mass distribution in the Galactic disc. The idea is that if it formed due to a single impulsive perturbation, the way it winds up in the vertical phase space ($z$--$v_z$) depends on $\Phi(z)$ being anharmonic; see figure 2 in the first of these papers for an illustration. By applying the method to the \Gaia DR2 6d sample in the Solar neighbourhood (up to $|z|\le 600$~pc), \citet{Widmark2021c} determined the value of local dark matter density with a precision comparable to the traditional methods, though it was mostly limited by the uncertainties in the baryon distribution. \citet{Widmark2022a} used updated \Gaia EDR3 PM and extended the analysis to more distant regions, mostly in the outer disc (the results were deemed unreliable in most subregions in the inner Galaxy). They mapped out the surface mass density in the radial range 8--11~kpc, complementing the pre-\Gaia work by \citet{Bovy2013} based on equilibrium action-space DFs.

\citet{Guo2024} used a similar idea, but fitting only the intersections of the phase spiral with the $z=0$ and $v_z=0$ axes. Applying their method to \Gaia DR3 RVS, they obtained a somewhat higher local density of dark matter, but still in general agreement with the literature (see their figure 14).

\subsubsection{Link between chemistry and dynamics}  \label{sec:dynamics_chemistry}

\citet{PriceWhelan2021, PriceWhelan2024} introduced an approach, which they call `orbital torus imaging', for constraining the potential using joint distribution of stars in chemical and phase spaces. Specifically, under the assumption of phase-mixedness, elemental abundances may depend only on the integrals of motion, but not on orbital phases. When colouring the phase space (e.g., $z$--$v_z$) by mean abundance, its foliation by orbital tori should coincide with isolines of abundances, ideally for all elements (see their figure 4 for illustration). Since the tori depend on the potential, maximising the similarity between these two labellings of phase space should reveal the correct potential. This idea is somewhat similar to the use of the phase spiral for the same purpose, except that here one looks at abundance patterns, not over-/underdensities in the vertical phase space (and the approach is formulated more generally for a St\"ackel potential, not just the 1d vertical dynamics). The key advantage of this method is that it is very insensitive to the selection function of the dataset, provided that it remains purely spatial and independent of kinematics and chemistry.
\citet{Horta2024a} put this formalism to practice, using $\sim 10^5$ thin-disc stars with \Gaia DR3 astrometry and APOGEE velocities and abundances. They measured the surface mass density within 1.1~kpc from the midplane, local density of dark matter, and several other quantities, all in agreement with previous estimates using independent methods.
\citet{Frankel2024} demonstrated that the phase spiral from the previous Section can also be traced in the abundances of iron and other elements, opening the way to extending the orbital torus imaging to non-equilibrium structures.

A related approach was employed by \citet{Eilers2022} to calibrate the chemical abundances in the APOGEE survey across a range of stellar types, assuming that the kinematic properties of a given stellar population may depend only on its chemical pattern, but not on the stellar mass, temperature, surface gravity and other parameters. Though their calibration only used kinematic, not dynamical arguments, the idea is similar to orbital torus imaging.

\subsubsection{Direct application of the collisionless Boltzmann equation}  \label{sec:dynamics_cbe}

The CBE (\ref{eq:CBE}) is a continuity equation for the DF in the 6d phase space, which relates the (gradient of the) potential to the derivatives of the DF in position, velocity and time.
In principle, given sufficient information about the DF in the phase space, one may apply it directly to determine the force field. Of course, if both $\partial\Phi/\partial \boldsymbol x$ and $\partial f/\partial t$ are unknown, there is no way to measure both, so one still has to invoke the equilibrium assumption, discarding the time derivative, or somehow assume its value. Moreover, the DF is not known in a closed form, but only through a finite set of stars sampling it.

To construct a smooth non-parametric DF, suitable for computing its derivatives, from discrete samples, several groups, starting from \citet{Green2020}, proposed to use normalising flows -- a method based on deforming a known analytic function (e.g., a Gaussian) through a series of invertible and differentiable nonlinear transformations represented by a neural network. 
\citet{Green2020}, \citet{Green2023} then train a second neural network to represent the gravitational potential, by requiring it to minimise the r.h.s.\ of the CBE and to have a non-negative Laplacian. Alternatively, \citet{An2021b} propose to determine the acceleration field directly from the CBE, which is actually an overdetermined equation (the potential gradient at a given point $\boldsymbol x$ must be independent of $\boldsymbol v$), but in this approach, the acceleration is not guaranteed to have zero curl or to correspond to a non-negative total density. \citet{Naik2022}, \citet{Green2023}, \citet{Buckley2023} demonstrate that these methods remain applicable in realistic situations (including observational uncertainties and non-stationarity), while \citet{Kalda2024} extend the formalism to the rotating frame. Finally, \citet{Putney2024} address the complications arising from dust extinction and the associated spatial selection function.

The caveats of this approach is that one needs to estimate the DF in 6d using noisy and incomplete samples, and then take derivatives of this estimate, and that it still relies on the time-independence of the DF for a lack of better alternative. These complications notwithstanding, \citet{Lim2025} applied the method to the absolute magnitude-limited subset of \Gaia DR3 RVS and inferred the local dark matter density in a pleasant agreement with other estimates.

In another variation of this idea, \citet{Han2016} proposed an `orbital PDF' method for inferring the acceleration field from a discrete kinematic sample without explicitly constructing the DF, but only assuming its stationarity (i.e., relying on the Jeans theorem). Namely, the observed 6d sample of tracers is integrated in time in various trial potentials, and select those that keep the spatial distribution invariant (as much as possible). \citet{Li2024b} applied this method to satellite galaxies, while \citet{Kipper2020,Kipper2021} developed a similar `orbital arc' approach to measure the tangential acceleration in the Solar neighbourhood caused by the bar and to estimate the non-stationary perturbations.

Besides normalising flows, there are other methods for reconstructing the velocity field as a smooth differentiable function. \citet{Nelson2022} used Gaussian process regression to determine a smooth map of mean velocity in a $\sim$2~kpc Solar neighbourhood from the \Gaia DR2 RVS sample, while \citet{Akhmetov2024} expanded the velocity field into Taylor series up to 2nd order at a grid of $\sim 3\times10^4$ points within 10~kpc from the Sun, using roughly half of the \Gaia DR3 RVS sample. These studies also revealed perturbations in the velocity field, but did not attempt to constrain the potential.

\subsection{Summary of potential constraints}  \label{sec:dynamics_summary}

\begin{figure}
\centering
\includegraphics{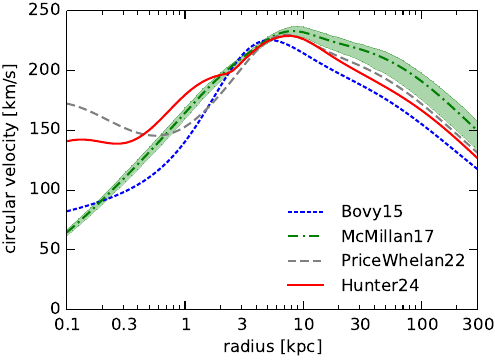}
\caption{Circular-velocity curves of several commonly used Milky Way potentials: \texttt{MWPotential2014} from \texttt{galpy} \citep[blue dotted]{Bovy2015}, the fiducial potential and its $1\sigma$ uncertainty from \citet[green dot-dashed]{McMillan2017}, \texttt{MilkyWayPotential2022} from \texttt{gala} \citep[grey dashed, same as the fiducial potential in Figures~\ref{fig:enclosedmass}, \ref{fig:vcirc}]{PriceWhelan2017}, and the potential from \citet{Hunter2024} as implemented in \texttt{agama} \citep[red solid]{Vasiliev2019}.
}  \label{fig:vcirc_compilation}
\end{figure}

It is fair to say that the precision of Milky Way potential measurements has improved in recent years, with many studies shown in Figure~\ref{fig:enclosedmass} quoting formal uncertainties at the level of 10\% or even less, although the scatter between them does exceed this value. At the same time, it became apparent that the equilibrium assumption underpinning many modelling methods might systematically bias the results by a larger amount than the statistical uncertainty. This may explain the discrepancy between the circular-velocity curves derived from disc kinematics (Figure~\ref{fig:vcirc}) and the larger-scale enclosed mass measurements from streams, satellites and other tracers in the outer halo (Figure~\ref{fig:enclosedmass}). To highlight this discrepancy, we plot a fiducial \texttt{MilkyWayPotential2022} from the galactic dynamics package \texttt{gala} \citep{PriceWhelan2017}, which appears to be a relatively good representation of the Milky Way mass distribution as we know it now. It fits both the circular-velocity curve within $\sim$15~kpc and the larger-scale constraints, but cannot reproduce the declining $v_\mathrm{circ}$ curve in the range 20--30~kpc; any attempt to do so would violate the constraints in the outer Galaxy. This potential has a local escape velocity around 500 or 550~\kms depending on whether one uses the virial radius or infinity for $\Phi_\infty$, which also satisfies most observational constraints shown in Figure~\ref{fig:v_escape}.

Figure~\ref{fig:vcirc_compilation} shows the circular-velocity curves of this fiducial potential together with two other commonly used ones that were tuned to various observational constraints in the pre-\Gaia era. \texttt{MWPotential2014} from \citet{Bovy2015} has been calibrated to have $v_\text{circ}(R_\odot)=220$~\kms, a value that is now considered too low, and has a lower enclosed mass at large scales than inferred by most recent studies. On the other hand, \citet{McMillan2017} has a higher local circular speed of 233~\kms and a higher enclosed mass at large scales; this study provides not only the best-fit potential, but an entire Monte Carlo Markov Chain, shown by a green uncertainty interval. The \texttt{MilkyWayPotential2022} from \texttt{gala} has been tuned to a variety of recent constraints from \Gaia, including the circular-velocity profile from \citet{Eilers2019} and the vertical force inferred from the \Gaia phase spiral \citep{DarraghFord2023}; it has a slightly lower local circular speed, and its mass profile at large scales sits between the other two models, better matching various constraints shown in Figure~\ref{fig:enclosedmass}, where it is shown as the fiducial potential. It uses only a few simple analytical components and may serve as a good general-purpose potential for Galactic studies. All three potentials are axisymmetric and are not well suited for studying the dynamics in the inner few kpc. Finally, the potential from \citet{Hunter2024} has been tailored to match both gas and stellar kinematics in the inner Galaxy, incorporating the boxy bar model from \citet{Portail2017} and the nuclear stellar disc model from \citet{Sormani2022}, and satisfies a variety of observational constraints in the Solar neighbourhood and outer Milky Way; its axisymmetrized version is shown by a red curve in Figure~\ref{fig:vcirc_compilation}.

Even though the overall mass distribution appears to be well constrained at all scales, with the total mass of the Milky Way converging around $10^{12}\,M_\odot$ with a conservative uncertainty of 20--30\% (see recent reviews by \citealt{Wang2020a} and \citealt{Bobylev2023}), this cannot be said about the shape of the potential. Studies based on stellar streams in the inner Galaxy, both pre-\Gaia \citep[e.g.,][]{Kuepper2015, Bowden2015, Bovy2016} and more recent \citep{Malhan2019,Palau2023}, as well as Jeans analysis of RR Lyrae \citep{Wegg2019} and \Gaia DR2 RVS sample \citep{Nitschai2020}, are consistent with a spherical halo shape (see figure 8 in \citealt{Palau2023} and figure 14 in \citealt{Woudenberg2024} for a compilation of literature measurements). An exception is the analysis of multiple streams by \citet{Ibata2024}, who determined an oblate shape with high confidence (axis ratio $z/R\simeq0.75\pm 0.03$). The orbit of a hypervelocity star presumably ejected from the Galactic centre \citep{Hattori2020}, as well as DF-based dynamical modelling of globular clusters \citep{Posti2019} and RR Lyrae \citep{Hattori2021}, even favour prolate shapes of the inner halo\footnote{Note however that the action estimation method used in \citet{Posti2019} is not applicable for prolate potentials.}. This is in a mild tension with the expectations from cosmological simulations, in which the halo usually becomes flattened in the inner part due to the baryonic compression \citep[e.g.,][]{Chua2019}.
On the other hand, fitting of the Orphan--Chenab stream in the outer Galaxy (15--60~kpc) \citep{Erkal2019b,Koposov2023} preferred either oblate or prolate haloes, but not aligned with any principal axes, and the Sagittarius stream required a radially varying triaxial shape and orientation \citep{Vasiliev2021a}. Such a tilted halo could be a leftover of the GSE merger \citep{Han2022a} and could be responsible for the coherent orientation of Galactic satellites \citep{Shao2021} or the warp in the Galactic disc \citep{Han2023}. Obviously, the measurement of the outer halo shape is greatly complicated by the ongoing interaction with the LMC, and it is not clear if this concept even makes much sense.

In addition, many of the studies based on different methods and tracers also invoke additional  observational constraints unrelated to their primary dataset, or overly tight priors on the potential, making it difficult to interpret the results and possibly even contributing to the emergence of a consistent picture of the Galactic mass distribution. Some of these external constraints may be well justified, e.g., the Solar distance and velocity with respect to Sgr A$^\star$ precisely measured by \citet{Gravity2019} and \citet{Reid2020}. On the other hand, the commonly used circular-velocity curve from \citet{Eilers2019} is not a primary observational constraint, but is derived using a particular dynamical modelling method (Jeans equations). While its use could be appropriate in studies that fit a parametric potential model to a variety of constraints from the literature \citep[e.g.,][]{deSalas2019,Karukes2020,Jiao2021}, it should better be avoided when fitting dynamical models directly to observations, e.g., using stellar streams or DF-based methods.

As a final remark, in the \Gaia era we are blessed with enormous quantities of data, so in most cases, the statistical uncertainty on the gravitational potential is smaller than possible systematic errors arising from model assumptions and limitations of the method. It is therefore important to test the methods on realistically complex mock datasets mimicking the Milky Way \citep[e.g.,][]{Grand2018,Sanderson2020} to get a more reliable estimate of their overall accuracy.

\section{Conclusions}  \label{sec:conclusions}

It goes without saying that \Gaia has dramatically advanced our knowledge of the Milky Way structure and dynamics. Some of these improvements were expected, others came as a surprise, as usually happens with major new observational facilities. Even more impactful is the synergy between \Gaia and several ground-based spectroscopic surveys, discussed in Section~\ref{sec:observations_spectroscopic_surveys}, which on the one hand, extend the line-of-sight velocity measurements to much fainter stars than are available from \Gaia alone, and on the other hand, complement the catalogues with chemical information, allowing one to better distinguish different Galactic components. In addition, the rapid progress in advanced machine learning techniques makes it possible to extract much more information from the \Gaia data (particularly, BP/RP spectra, see Section~
\ref{sec:observations_distance_measurement}) than originally foreseen.

Below we summarise the key features of Galactic dynamics, the knowledge of which can be directly attributed to or facilitated by \Gaia.

\begin{itemize}
\item The pattern speed of the Galactic bar has been revised downward from $\sim 50$--60~\kmskpc typically quoted two decades ago to $\sim 35$--40~\kmskpc in most recent studies (with some exceptions), see Figure~\ref{fig:BarSpeedsSummary}. At the same time, it became clear that most methods can be confounded by multiple overlapping dynamical phenomena (e.g., combination of bar and spiral arms), making measurements and their interpretation more difficult. Moreover, there is evidence that the bar slows down, but its formation time and evolution are still highly uncertain.
\item Various deviations from equilibrium, although relatively small (at a level $\sim$10--20\%), are now reliably measured. These include the phase spiral, north-south asymmetries in star counts and velocities, warp of the outer disc, LMC-induced dipole perturbation in the outer halo, and numerous unrelaxed structures (streams, shells) throughout the halo. Yet their interpretation is often challenging because of the complexity of underlying dynamical mechanisms, and more than one of these may play important roles in any given phenomenon.
\item Spiral-like structure has been seen both directly in the stellar density contrast and via the imprint left in the kinematics. However, details differ between studies and tracers used. Evidence appears to point towards a more flocculent pattern, with arms that are actively growing (e.g.\ the Local arm) and disrupting (e.g.\ the Perseus arm), but the full picture is not yet clear.
\item The local stellar halo of the Milky Way is dominated by high-eccentricity debris from a single disrupted satellite (GSE), most likely accreted 8--10 Gyr ago. There is still some dispute about its fractional contribution to the total mass, the spatial distribution of debris, and its orbital evolution. As for other, smaller accretion events, the current state of knowledge can be characterised as a delightful patchwork of proposed structures, which partly overlap but have different names in different studies -- in other words, there is no established classification yet, see the review by \citet{ArchaeologyReview} for details.
\item The current estimates of the Milky Way mass converge around $10^{12}\,M_\odot$ with an uncertainty $\sim 20$--30\% (accounting for the scatter between different studies); significantly higher or lower values are likely excluded by the combination of available dynamical evidence. However, there is a notable disagreement between the circular-velocity curve estimates at $r\simeq 20$--30 kpc based on the Jeans analysis of disc kinematics, some of which prefer much lower total masses, and other methods using more distant tracers (see Figures~\ref{fig:enclosedmass} and \ref{fig:vcirc}). The origin of this discrepancy has not been conclusively established, although the above mentioned perturbations of the Galactic disc likely play a role in it.
\item The shape of the dark halo remains largely unknown, despite many conflicting measurements. Given the disequilibrium state induced by the current interaction with the LMC, as well as the echoes of distant accretion events, the very question about the halo shape may be ill-posed.
\item Despite the obvious transient phenomena, global equilibrium models of the Galaxy based on the Jeans theorem still have considerable value, offering a baseline `synoptic' view of the Milky Way at large scales. However, there is currently no such model that would adequately describe the joint distribution of stars in the space of position, velocity, chemistry and age; existing models miss one or more of these dimensions.
\end{itemize}

It is clear from the above list that despite considerable progress, there are outstanding questions in the Milky Way dynamical studies, which should be addressed in the forthcoming years. The \Gaia journey is not over, and the upcoming DR4 in 2026, alongside with further increase in data quantity and quality, will undoubtedly bring a new round of discoveries. But even the interpretation of the existing data remains a challenge, and a considerable progress on the theoretical side is required to fully exploit the legacy of the \Gaia mission.

\section*{CRediT authorship contribution statement}

Jason A.S. Hunt: sections 4 and 5. Eugene Vasiliev: remaining sections.

\section*{Declaration of competing interest}

The authors declare that they have no known competing financial interests or personal relationships that could have appeared to influence the work reported in this paper.

\section*{Acknowledgements}

We thank Leandro Beraldo e Silva, Gregory Green, and the referees for valuable comments and suggestions, and Adrian Price-Whelan for providing an observational dataset used in several figures. EV is supported by an STFC Ernest Rutherford fellowship (ST/X004066/1).
This work, as well as most of the referenced articles, has made use of data from the European Space Agency (ESA) mission \Gaia (\href{https://www.cosmos.esa.int/gaia}{www.cosmos.esa.int/gaia}), processed by the \Gaia Data Processing and Analysis Consortium (DPAC). 
This work also used data from the Sloan Digital Sky Survey IV (\href{http://www.sdss.org}{www.sdss.org}). Funding for SDSS-IV has been provided by the Alfred P. Sloan Foundation, the U.S. Department of Energy Office of Science, and numerous Participating Institutions. SDSS-IV acknowledges support and resources from the Center for High Performance Computing at the University of Utah.

\section*{Data availability}

Python scripts and notebooks for reproducing all Figures except Figure~\ref{fig:Poggio2021spiralarms} are available at \url{https://github.com/JASHunt/NARev_Gaia_Dynamics_Review}, with some large data files available at \url{https://zenodo.org/records/14609059}. The data and notebook for the UMS stars and Cepheids in Figure~\ref{fig:Poggio2021spiralarms} are available at \url{https://github.com/epoggio/Spiral_arms_EDR3} \citep{Poggio2021b}, and the open clusters from \citet{Hunt2023} are available at \url{https://cdsarc.cds.unistra.fr/viz-bin/cat/J/A+A/673/A114}.

\bibliographystyle{model2-names-astronomy}
\bibliography{paper}
\end{document}